\documentclass[fleqn,usenatbib]{mnras}

\usepackage{newtxtext,newtxmath}
\usepackage[T1]{fontenc}

\DeclareRobustCommand{\VAN}[3]{#2}
\let\VANthebibliography\thebibliography
\def\thebibliography{\DeclareRobustCommand{\VAN}[3]{##3}\VANthebibliography}

\newcommand{\pd}[2]{\frac{\partial #1}{\partial #2}}

\usepackage{graphicx}	% Including figure files
\usepackage{amsmath}	% Advanced maths commands
\usepackage{bm}
\usepackage{xcolor}
\title[galactic ionizing output]{On the rise and fall of galactic ionizing output at the end of reionization}

\author[Cain et. al.]{
Christopher Cain$^{1,2}$,\thanks{E-mail: clcain3@asu.edu}
Anson D'Aloisio$^{2}$,
Garett Lopez$^{2}$,
Nakul Gangolli$^{2}$,
Joshua T. Roth$^{2}$
\\
$^{1}$School of Earth and Space exploration, Arizona State University, Tempe, AZ 85281, USA\\
$^{2}$Department of Physics and Astronomy, University of California, Riverside, CA 92521, USA
}

\date{Accepted XXX. Received YYY; in original form ZZZ}

\pubyear{2015}

\begin{document}
\label{firstpage}
\pagerange{\pageref{firstpage}--\pageref{lastpage}}
\maketitle

\begin{abstract}
Quasar absorption spectra measurements suggest that reionization proceeded rapidly, ended late at $z \sim 5.5$, and was followed by a flat ionizing background evolution.  Simulations that reproduce this behavior often rely on a fine-tuned galaxy ionizing emissivity, which peaks at $z \sim 6 - 7$ and drops a factor of $1.5-2.5$ by $z \sim 5$.  This is puzzling since the abundance of galaxies is observed to grow monotonically during this period.  Explanations for this include effects such as dust obscuration of ionizing photon escape and feedback from photo-heating of the IGM.  We explore the possibility that this drop in emissivity is instead an artifact of one or more modeling deficiencies in reionization simulations.  These include possibly incorrect assumptions about the ionizing spectrum and/or inaccurate modeling of IGM clumping.  Our results suggest that the need for a drop could be alleviated if simulations are underestimating the IGM opacity from massive, star-forming halos.  Other potential modeling issues either have a small effect or require a steeper drop when remedied.  We construct an illustrative model in which the emissivity is nearly flat at reionization’s end, evolving only $\sim 0.05$ dex at $5 < z < 7$.  More realistic scenarios, however, require a $\sim 0.1-0.3$ dex drop.  We also study the evolution of the Ly$\alpha$ effective optical depth distribution and compare to recent measurements.  We find models that feature a hard ionizing spectrum and/or are driven by faint, low-bias sources most easily reproduce the mean transmission and optical depth distribution of the forest simultaneously.  

\end{abstract}


%%%%%%%%%%%%%%%%%%%%%%%%%%%%%%%%%%%%%%%%%%%%%%%%%%

%%%%%%%%%%%%%%%%% BODY OF PAPER %%%%%%%%%%%%%%%%%%

\section{Introduction}
\label{sec:intro}

The past decade has seen great progress towards constraining the timing of cosmic reionization.  Measurements of the Thomson scattering optical depth by Planck have localized the midpoint of reionization to $z \approx 7.5 \pm 0.8$~\citep{Planck2018}.  QSO damping wing measurements have suggested that the IGM was significantly neutral at $z \sim 7-8$~\citep{Davies2018,Wang2020}, while Ly$\alpha$ emitter surveys have hinted at large ionized bubbles at the same redshifts~\citep{Ouchi2018,Hu2019,Endsley2022b}.  The tail end of reionization has been probed by the Ly$\alpha$ forest of high-redshift QSOs~\citep{Fan2006,Becker2015,Bosman2018,Eilers2018,Bosman2021,Zhu2022}, and measurements of the ionizing photon mean free path~\citep[MFP,][]{Worseck2014,Becker2021,Zhu2023}.  These observations support a relatively late end to reionization at $z \approx 5-5.5$~\citep{Kulkarni2019,Keating2019,Nasir2020,Qin2021,Cain2021,Davies2021b}.  

Despite our improving understanding of reionization's timing, relatively little is known about the properties of the sources that drove it.  Measurements of the high-redshift UV luminosity function~\cite[UVLF,][]{Bouwens2015,Finkelstein2019,Bouwens2021} and simulations~\cite[e.g.][]{Ocvirk2018,Kannan2022} suggest that galaxies with physically reasonable ionizing properties could have completed reionization.  Conversely, measurements of the quasar luminosity function~\citep{Willott2010,McGreer2013,Georgakakis2015,Matthee2023}, simulations~\citep{Trebitsch2021,Kannan2022} and constraints on the temperature of the IGM~\citep{Daloisio2017} disfavor quasars as the dominant sources.  However, there remains ongoing debate as to which galaxies were the dominant ionizing photon producers~\citep{Robertson2015,Finkelstein2019,Naidu2020}, and how the ionizing properties of these galaxies evolved over time.  

\begin{figure}
    \centering
    \includegraphics[scale=0.215]{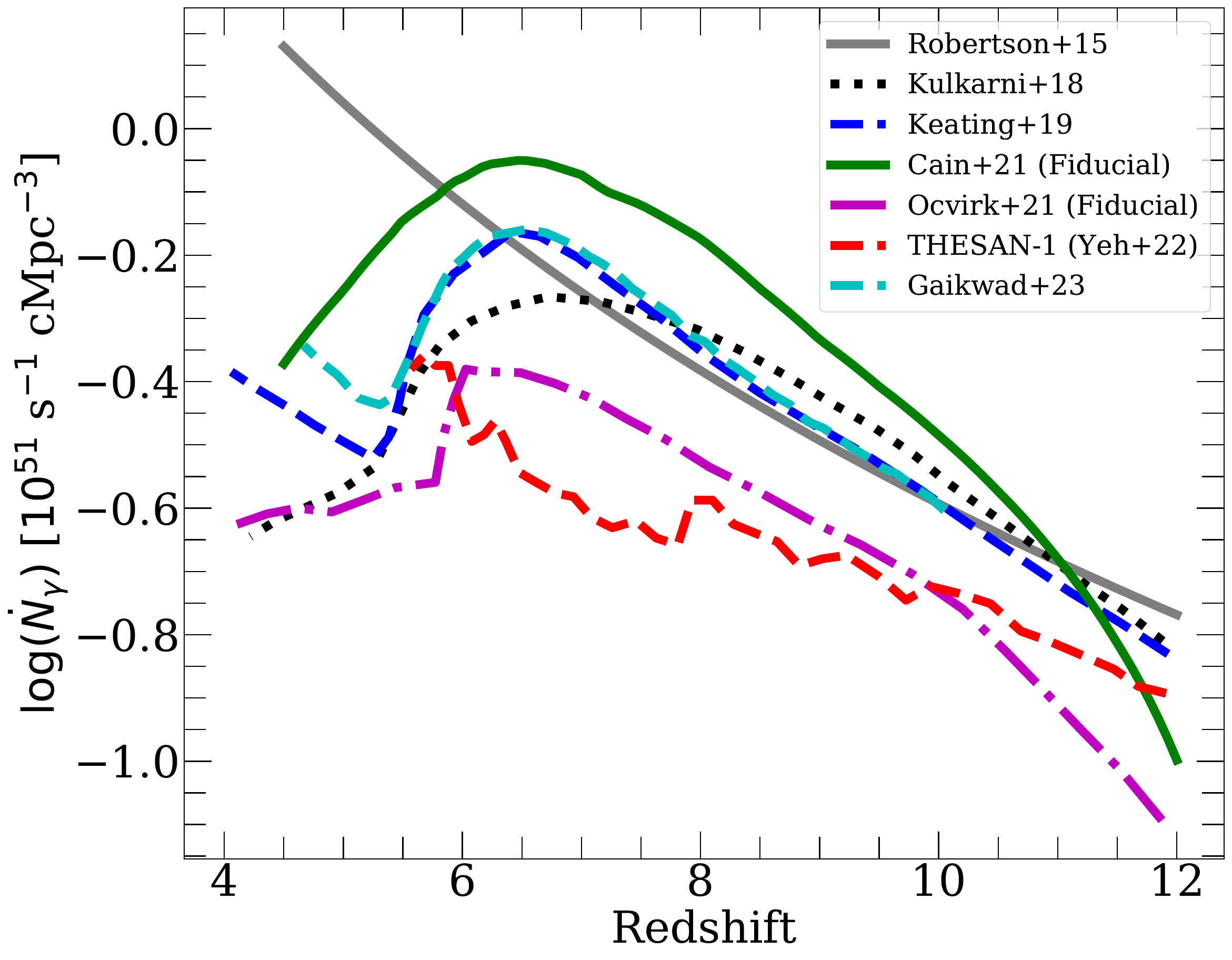}
    \caption{Ionizing emissivity vs. redshift for several models and simulations in the literature.  The faded grey curve shows the emissivity computed assuming $f_{\rm esc} = 0.2$ and $\log(\xi_{\rm ion}) = 25.2$, as in~\citet{Robertson2015}.  The remaining curves come from simulations by \citet{Kulkarni2019,Keating2019,Cain2021,Ocvirk2021,Yeh2022,Gaikwad2023}, which reproduce QSO observations at $5 < z < 6$.  With the exception of THESAN-1~\citep{Yeh2022}, all the simulations find a drop in $\dot{N}_{\gamma}$ starting at $z \sim 6.5$ is required to reproduce these observations.  Our goal is to determine whether this drop is likely to be real, or an artifact of one or more modeling deficiencies in simulations.  } 
    \label{fig:em_lit}
\end{figure}

If galaxies drove reionization, the total ionizing photon emissivity is given by
\begin{equation}
    \label{eq:ndot}
    \dot{N}_{\gamma} \equiv \langle f_{\rm esc} \xi_{\rm ion}\rangle \rho_{\rm UV}
\end{equation}
where $\rho_{\rm UV}$ is the integrated galaxy UVLF, $\xi_{\rm ion}$ is the ionizing efficiency of galaxies, and $f_{\rm esc}$ is their ionizing photon escape fraction, and the average of $f_{\rm esc} \xi_{\rm ion}$ is over the galaxy population and weighted by the luminosity $L_{\rm UV}$.  The UVLF at $M_{\rm UV} \lesssim -17$ has been measured out to $z \approx 10$ with HST~\citep[][]{Finkelstein2019,Bouwens2021} and at higher redshifts with JWST~\citep[e.g.][]{Adams2023}, and most recently even fainter galaxies are being probed using strong lensing up to $z \sim 8$~\citep{Atek2023}.  Assuming a constant $\log(\xi_{\rm ion}) = 25.2$, \cite{Robertson2013} found that a constant $f_{\rm esc} = 0.2$ is sufficient for galaxies to re-ionize the universe by $z \sim 6$ given reasonable assumptions about the faint end of the UVLF.  The grey curve in Figure~\ref{fig:em_lit} shows $\dot{N}_{\gamma}$ from~\citet{Robertson2015}.  It grows rapidly with cosmic time (decreasing redshift), tracing the growth of the UVLF and the underlying halo mass function.  

However, QSO observations at the tail end of reionization may be telling a different story.  Figure~\ref{fig:em_lit} compares $\dot{N}_{\gamma}$ from \cite{Robertson2015} to that from several recent simulations in the literature.  All these simulations reproduce reasonably well (or have been tuned to reproduce) the $5 < z < 6$ Ly$\alpha$ forest and/or MFP.  All of these models, except THESAN-1~\citep{Kannan2022,Yeh2022}, display a {\it drop} in $\dot{N}_{\gamma}$, starting at $z \sim 6 - 6.5$.  The drop is necessary in these models to prevent the mean Ly$\alpha$ forest transmission from over-shooting the measurements at $z < 6$.  Its magnitude ranges from a factor of $1.5-2.5$, and happens within $300$ Myr or less.  This behavior is only explainable by a factor of several evolution in $\langle f_{\rm esc} \xi_{\rm ion}\rangle$ over this relatively short time period.  

There are several reasons why $f_{\rm esc}$ and/or $\xi_{\rm ion}$ may decrease near the end of reionization.  If $f_{\rm esc}$ and/or $\xi_{\rm ion}$ depends strongly on galaxy host halo mass~\citep{Finkelstein2019,Yeh2022,Rosdahl2022}, the evolution of the HMF could drive a decrease in $\langle f_{\rm esc} \xi_{\rm ion}\rangle$.  The ionizing properties of galaxies (particularly faint ones) may be affected by feedback from IGM photo-heating and/or supernovae~\citep{Shapiro1994,Kimm2014,Wu2019b,Ocvirk2021}.  There may be evolution and/or variation in the intrinsic ionizing properties of high-redshift stellar populations~\citep{Maseda2020,Atek2022}.  Dust may reduce the ionizing photon output of the most massive galaxies~\citep{Kostyuk2023,Lewis2023}.

For the most part, the studies referenced in Figure~\ref{fig:em_lit} make little attempt to explain the physical mechanism(s) underlying the drop in $\dot{N}_{\gamma}$.  The $\dot{N}_{\gamma}$ evolution in \cite{Kulkarni2019,Keating2019,Cain2021,Gaikwad2023} is not a prediction of any underlying galaxy model, but is instead tuned to reproduce Ly$\alpha$ forest and/or MFP observations.  The fiducial simulation of \cite{Ocvirk2021}, which has a drop, does include star formation.  In their case, the drop is driven by a combination of photo-heating feedback, which suppresses gas accretion onto galaxies, and supernova feedback, which further disrupts star formation.  However, they account for these effects by assuming a sharp temperature threshold for star formation, and it is unclear how much of the effect results from this choice.  In THESAN-1, which uses the IllustrisTNG galaxy formation model~\citep{Vogelsberger2014,Weinberger2016,Pillepich2017}, the global escape fraction declines by a factor of $2-3$ during reionization~\cite{Yeh2022}, in part due to feedback effects~\citep{Garaldi2022}.  However, this is not enough to produce a decrease in $\dot{N}_{\gamma}$, nor do they find that one is necessary to reproduce QSO observations.  However, as we will see in \S\ref{subsec:clight}, their agreement with QSO observations may be in part due to their use of the reduced speed of light.  

Another possibility is that the drop in $\dot{N}_{\gamma}$ is an artifact of inaccurate or incomplete modeling of the IGM.  \cite{Keating2019} speculated that it may result from an inaccurate treatment of the IGM thermal history. Simulations may also be missing ionizing photon absorbers that would otherwise regulate the growth of the UV background~\citep{Cain2021}.  These absorbers could be missing due to a lack of spatial resolution, which is a problem for most reionization simulations in representative cosmological volumes~\citep{Emberson2013}. 
Absorption in and around massive star-forming halos may be sensitive to details of galaxy dynamics and evolution (see Appendix C of~\cite{Wu2019}).  Simulations that model these details approximately (or not at all) may also be under-estimating this source of IGM opacity.  These, and several other potential IGM modeling considerations, including the spectrum of the ionizing radiation and behavior of ionizing recombination radiation, affect the relationship between the Ly$\alpha$ forest and $\dot{N}_{\gamma}$.  Our goal is to determine whether one or more of these issues, if resolved, might alleviate the need for a drop in $\dot{N}_{\gamma}$.  

This work is organized as follows.  In \S\ref{sec:phys}, we discuss several physical and numerical modeling effects that influence the relationship between $\dot{N}_{\gamma}$ and the Ly$\alpha$ forest.  \S\ref{sec:methods} describes our numerical methods.  In \S\ref{sec:results_IGM} we study each modeling effect individually.  We consider their combined effects, and implications for the properties of ionizing sources, in \S\ref{sec:sources}.  In \S\ref{sec:num_effects}, we consider numerical (non-physical) modeling effects.  We conclude in \S\ref{sec:conc}.  Throughout this work, we assume the following cosmological parameters: $\Omega_m = 0.305$, $\Omega_{\Lambda} = 1 - \Omega_m$, $\Omega_b = 0.048$, $h = 0.68$, $n_s = 0.9667$ and $\sigma_8 = 0.82$, consistent with~\citet{Planck2018} results. All distances are quoted in co-moving units unless otherwise specified. 

\section{Relevant Modeling Considerations}
\label{sec:phys}

In this section, we will discuss several physical effects that affect the relationship between $\dot{N}_{\gamma}$ and the forest in simulations.  We will also discuss how each of these might affect the need for a drop.  

\subsection{Spectrum of the ionizing radiation}
\label{subsec:alpha}

The ionizing photon spectrum emitted by galaxies is often parameterized as a power law of the form
\begin{equation}
    \label{eq:ndotnu}
    \frac{d\dot{N}_{\gamma}}{d\nu} \propto \frac{J_{\nu}}{h_{\rm p} \nu} \propto \nu^{-\alpha-1}
\end{equation}
where $J_{\nu}$ is the galaxy's spectral energy distribution (SED) above $1$ Ryd, $h_{\rm p}$ is Planck's constant, $\nu$ is frequency, and $\alpha$ is the spectral index.  Typical values for $\alpha$ range from $0.5$ to $2.5$, depending assumptions about the ionizing properties of stars~\citep[][]{Bressan2012,Choi2017} and spectral hardening by the interstellar and circumgalactic medium (ISM \& CGM)~\citep{Madau1995,Giguere2009,Haardt2012}.  Ionizing radiation may be further hardened by the IGM after emerging from galaxies.  

The Ly$\alpha$ forest is sensitive to $\alpha$ through the latter's effect on the photo-ionization rate $\Gamma_{\rm HI}$ and temperature $T$.  In optically thin gas, 
\begin{equation}
    \label{eq:gammaHI}
    \Gamma_{\rm HI} = \int_{\nu_0}^{4 \nu_0} d\nu \frac{dN_{\gamma}}{d\nu} c \sigma_{\rm HI}(\nu)  = N_{\gamma} c \langle \sigma_{\rm HI} \rangle
\end{equation}
where $N_{\gamma}$ is the local ionizing photon number density, $c$ is the speed of light, $\langle \sigma_{\rm HI} \rangle$ is the spectrum-averaged HI cross-section, and $h_p \nu_0 \equiv 13.6\text{ eV}$ is the ionization energy of HI\footnote{The cutoff at $4$ Ryd owes to absorption of He II ionizing photons within galaxies.  }.  For $\alpha = 2.5$ ($0.5$), $\langle \sigma_{\rm HI} \rangle = 3.11 \times 10^{-18}$ cm$^{2}$ ($1.93 \times 10^{-18}$ cm$^{2}$), a factor of $1.6$ difference.  The temperature is sensitive to $\alpha$ through the photo-heating rate, 
\begin{equation}
    \label{eq:photoheating}
    \mathcal{H} = \int_{\nu_0}^{4 \nu_0} d\nu  \frac{dN_{\gamma}}{d\nu} c \sigma_{\rm HI}(\nu) (h_{\rm p}\nu - h_{\rm p} \nu_0) \propto N_{\gamma}\langle E_{\rm ion,HI} \rangle
\end{equation}
where $\langle E_{\rm ion,HI} \rangle$ is the average energy injection per ionization per HI atom.  Since harder photons carry more energy, smaller $\alpha$ will lead to larger $\langle E_{\rm ion,HI} \rangle$ and a hotter IGM.  For $\alpha = 2.5$ ($0.5$), $\langle E_{\rm ion,HI} \rangle = 3.16$ ($5.38$) eV, a factor of $1.7$ difference in heating rate\footnote{To get these values, we first assume that the ionizing photon number spectrum is given by Eq.~\ref{eq:ndotnu}, with a sharp cutoff at $4$ Ryd (the HeII ionization threshold).  To get the average energy injection per ionization over the whole spectrum, we evaluate $\langle E_{\rm ion,HI} \rangle = \frac{\int_{\nu_0}^{4 \nu_0} d\nu \frac{d\dot{N}_{\gamma}}{d\nu} \sigma_{\rm HI}^{\nu}(h \nu - 13.6{\rm eV})}{\int_{\nu_0}^{4 \nu_0} d\nu \frac{d\dot{N}_{\gamma}}{d\nu} \sigma_{\rm HI}^{\nu}}$ which assumes the optically thin limit for absorption. } The temperature the gas is heated to initially by reionization, $T_{\rm reion}$, also depends on $\alpha$, although~\cite{DAloisio2019} found that sensitivity to be modest.  

The opacity of the IGM to Ly$\alpha$ photons scales with $\Gamma_{\rm HI}$ and $T$ as
\begin{equation}
    \label{eq:taulya}
    \tau_{\rm Ly\alpha} \propto n_{\rm HI} \propto \frac{\alpha_{\rm A}(T)}{\Gamma_{\rm HI}} \propto T^{-0.7}\Gamma_{\rm HI}^{-1}
\end{equation}
where $\alpha_{\rm A}$ is the case A recombination coefficient of HII.  Because $\Gamma_{\rm HI}$ and $T$ scale oppositely with $\alpha$, the effect of changing $\alpha$ will partially cancel in Eq.~\ref{eq:taulya}.  The fact that $\tau_{\rm Ly\alpha}$ scales more strongly with $\Gamma_{\rm HI}$ than with $T$ suggests that it should increase for harder spectra (smaller $\alpha$), resulting in less Ly$\alpha$ transmission at fixed $N_{\gamma}$.  This would allow $\dot{N}_{\gamma}$ to be higher at fixed Ly$\alpha$ transmission, possibly alleviating the need for a drop in $\dot{N}_{\gamma}$.  

\subsection{Ionizing Recombination Radiation}
\label{subsec:recomb}

Ionizing recombination photons (IRPs) - produced by HI recombinations to the ground state - also affect the relationship between $\dot{N}_{\gamma}$ and the Ly$\alpha$ forest.  These photons are produced mainly within dense gas clumps with high recombination rates, the same structures that act as sinks (see next section).  Modeling IRPs self-consistently requires resolving the structures that create them, a challenging computational task.  Thus, IRPs are often treated approximately assuming one of two limiting cases (although some codes include the full treatment, e.g.~\citet{Rosdahl2013,Kannan2019}).  They are assumed to (1) have no effect on the local IGM, or (2) get re-absorbed immediately by neighboring HI after being emitted (the so-called ``on-the-spot'' approximation).  These approximations are referred to as case A and B, respectively.   Case A (B) is most accurate in low (high) density gas where the local MFP to IRPs is long (short).  

Transmission in the $5 < z < 6$ Ly$\alpha$ forest is set by under-dense gas, where it is safe to assume that the case A approximation is valid locally.  However, the fate of IRPs produced in dense gas clumps is less clear.  The fraction of IRPs escaping these clumps, and their effect on the IGM opacity, are sensitive to the IGM column density distribution~\citep{FaucherGiguere2009,McQuinn2011,Altay2011}.  If a significant fraction of IRPs escape, they may affect the properties of the sinks and the ionizing background.  It is possible that they could affect the $\dot{N}_{\gamma}$-forest relationship and the need for a drop in $\dot{N}_{\gamma}$.  

\subsection{Ionizing photon sinks}
\label{subsec:ionphotsinks}

Ionizing photon sinks - dense gas clumps with high recombination rates - set the ionizing photon MFP and the reionization photon budget.  They also regulate the ionizing background at the end of reionization. In~\citet{Cain2021}, we demonstrated that sinks could prevent the MFP from over-shooting the measurements at $z \sim 5$ without a drop in $\dot{N}_{\gamma}$.  

It is possible that the smallest, most abundant sinks are missing from simulations. These occupy non-star-forming ``mini-halos'' with dark matter (DM) masses of $10^4 - 10^8$ $M_{\odot}$.  They are photo-evaporated over a $\sim 100$ Myr timescale after the IGM surrounding them is re-ionized~\citep{Shapiro2004,Iliev2005,Chan2023}.  Despite being short-lived, mini-halos contribute significantly to the IGM opacity and significantly increase the photon budget~\citep{Park2016,DAloisio2020,Nasir2021,Chan2023}.  Mini-halos are difficult to resolve in simulations because they can be as small as the pre-reionization baryon Jeans scale, which can be a kpc or less depending on the uncertain IGM thermal history.  Their response to reionization is sensitive to the interplay between self-shielding and hydrodynamics, requiring fully coupled RT/hydro to model accurately.  
Another possibility is that simulations are under-estimating the opacity from the most massive sinks, which occupy halos massive enough ($> 10^9$ $M_{\odot}$) to self-shield long after reionization ends.  The opacity after reionization is dominated by these objects~\citep{Prochaska2010}, and they may be important during reionization~\citep{Munoz2016}.  Since halos this massive usually host star-forming galaxies~\citep{Okamoto2008,Finlator2017}, their opacity may be sensitive to how galaxies form and evolve.  For example,~\citet{Wu2019b} found that how they modeled feedback in their galaxies significantly affected clumpiness of the surrounding IGM (their Appendix C).  While several reionization simulations include galaxy formation models~\citep[e.g.][]{Ocvirk2016,Rosdahl2018,Kannan2022}, these differ in how they handle processes like AGN feedback and star formation.  Many reionization simulations do not include galaxy physics~\citep[e.g.][]{Keating2019,Cain2021}.  

\subsection{Clustering of the ionizing sources}
\label{subsec:source_clustering}

The sources of reionization (galaxies) are clustered on spatial scales of $10$s-$100$s of Mpc.  If reionization was driven by the brightest galaxies~\citep{Naidu2020,Matthee2022}, the sources would be more highly biased/clustered than if faint galaxies dominated~\citep{Finkelstein2019,Atek2023}.  Several recent studies, observational and theoretical, suggest fainter galaxies may have higher average $f_{\rm esc}$ and/or $\xi_{\rm ion}$~\citep{Rosdahl2022,Begley2022,Saldana-Lopez2023,Atek2023}.  However, \citet{Naidu2022,Matthee2022} recently argued that the brightest Lyman$\alpha$ emitters (LAEs) may have been the primary drivers.  Their model may be supported by observations of bright LAEs at $z \geq 7$, which hint at the presence of large, highly ionized regions surrounding them~\citep{Mason2018,Endsley2022b}.  

The clustering properties of the sources can affect the Ly$\alpha$ forest.  The more biased the sources are, the more the ionizing background will fluctuate on large scales during reionization, resulting in regions with enhanced Ly$\alpha$ transmission near the brightest sources.  This effect should be important as long as the MFP is less than or comparable to the clustering scale of the sources, which may be true even after reionization~\citep{Davies2016}.  
 The ionizing background surrounding the brightest sources may also affect the response of the sinks to reionization~\citep{Park2016,DAloisio2020,Chan2023}.  

\section{Numerical Methodology}
\label{sec:methods}

In this section, we describe the numerical methods used in this work.  \S\ref{subsec:large_rt}-\ref{subsec:modsinks} review the methodology discussed in~\citet{Cain2022b}, and the remaining sections introduce new features in our framework.  

\subsection{Large-Scale Radiative Transfer}
\label{subsec:large_rt}

We ran radiative transfer (RT) simulations of reionization using the ray-tracing RT code first introduced in~\cite{Cain2021} and further described in~\cite{Cain2022b}, which we refer to as FlexRT (Flexible Radiative Transfer) hereafter.    

FlexRT solves the RT equation in post-processing on a time-series of cosmological density fields and with a uniform RT grid.  Ionizing sources (see next section) are binned to their nearest RT cells and rays are cast from the centers of source cells.  As rays travel, the optical depth through each intersected cell is computed, and photons are deposited accordingly.  Rays can adaptively split and merge using the HealPix formalism~\citep[][]{Gorski1999,Abel2002,Trac2007} to maintain a desired angular resolution.  In this work we track $12$ directions in the radiation field, with a maximum number of rays per cell of $28$.  Rays are deleted when their photon count is attenuated by a factor of $10^{10}$.  

Instead of solving for the HI number density ($n_{\rm HI}$) in each cell, FlexRT adopts a more general approach to compute ionizing photon opacities.  The optical depth at frequency $\nu$ encountered by ray $j$ intersecting cell $i$ is\footnote{To get this relationship between $\tau$ and $x_{\rm ion}$, we assume that I-fronts travel along one cell axis and that rays travel along this direction and enter at the ionized side of the cell opposite the I-front.  In this limit, the fraction of the cell traversed by the ray before reaching the I-front is $x_{\rm ion}$.  We note that our results are not very sensitive to this choice, since $x_{\rm ion}$ cancels out in Eq. 7 in the limit that $\tau << 1$.  }
\begin{equation}
    \label{eq:tauij}
    \tau_{ij}^{\nu} = \frac{x_{\rm ion}^{i}\Delta s^{ij}}{\lambda_{i}^{\nu}}
\end{equation}
where $\Delta s^{ij}$ is the distance traveled by ray $j$ through cell $i$, $\lambda_{i}^{\nu}$ is the mean free path at frequency $\nu$, and $x_{\rm ion}^{i}$ is the ionized fraction.  The photo-ionization rate is given by
\begin{equation}
    \label{eq:gammah1_subgrid}
    \Gamma_{\rm HI}^{i} = \frac{\langle \sigma_{\rm HI} \rangle_{\nu}^{i} \langle \lambda \rangle_{\nu}^{i}}{x_{\rm ion}^{i} V_{\rm cell}^{i} \Delta t}\sum_{j=1}^{N_{\rm rays}} \sum_{\nu=1}^{N_{\rm freq}}N_{\gamma,0}^{ij,\nu}  [1 - \exp(-\tau_{ij}^{\nu})]
\end{equation}
where $N_{\gamma,0}^{ij,\nu}$ is the number of photons initially in frequency bin $\nu$ of ray $j$, $\langle \sigma_{\rm HI} \rangle_{\nu}^{i}$ and $\langle \lambda \rangle_{\nu}^{i}$ are the HI cross-section and mean free path averaged over the ionizing spectrum incident on cell $i$, and $N_{\rm freq}$ is the number of frequency bins.  The inclusion of multi-frequency RT is a new feature in FlexRT, and is further discussed in \S\ref{subsec:multifreq}.  The other quantities have the same meaning as in Eq. 1 of~\citet{Cain2022b}.  We will describe in \S\ref{subsec:modsinks}-\ref{subsec:multifreq} how we calculate $\langle \lambda \rangle_{\nu}$. Appendix~\ref{app:mfreq_gamma} gives the complete derivation of Eq.~\ref{eq:gammah1_subgrid}\footnote{Appendix A of~\citet{Cain2022b} derives only the mono-chromatic version of Eq.~\ref{eq:gammah1_subgrid} (their Eq. 1).  }.  

We model the propagation of sub-resolved I-fronts in the ``moving screen'' approximation, which assumes a sharp boundary between ionized and neutral gas.  The I-front speed is given by
\begin{equation}
    \label{eq:vif}
    v_{\rm IF} = \frac{F_{\gamma}^{\rm inc}}{(1+\chi)n_{\rm H}}
\end{equation}
where $F_{\gamma}^{\rm inc}$ is the ionizing flux incident on the neutral part of the cell (after absorption by the intervening ionized part) and the factor of $1 + \chi = 1.082$ accounts for single ionization of He.  The gas temperature behind the I-front, $T_{\rm reion}$, is estimated using the flux-based method prescribed in~\cite{DAloisio2019}, and the temperature evolution thereafter is calculated using their Eq. 6.  

\subsection{Sources \& Density Fields}
\label{subsec:source_density}

Our sources are halos taken from a DM-only simulations run with the N-body code used in~\cite{Trac2015} with a box size of $200~h^{-1}$Mpc.  We used $N_{\rm dm} = 3600^3$ particles, resulting in a complete halo mass function (to within $10\%$) down to a minimum halo mass of $3 \times 10^{9} h^{-1}M_{\odot}$ ($\approx 200$ DM particles).  Halos were saved every $10$ Myr from $z = 12$ to $4.8$.  We assigned UV luminosities to halos by abundance-matching to the UVLF of~\cite{Finkelstein2019}.  The global (integrated) ionizing emissivity, $\dot{N}_{\gamma}$, is a free function of redshift that we tune to match observations - we describe how we do this in \S\ref{sec:results_IGM}.  Ionizing photons are distributed between halos according to
\begin{equation}
\label{eq:ndot_gam}
\dot{n}_{\gamma} \propto
\begin{cases}
     L_{\rm UV}^{\beta} & M_{\rm halo} \geq M_{\min}\\
    0 & M_{\rm halo} < M_{\min}
\end{cases}
\end{equation}
where $\dot{n}_{\gamma}$ is the emissivity of a given halo, $L_{\rm UV}$ is its UV luminosity, $M_{\min}$ is the minimum mass of halos that produce ionizing photons, and $\beta$ parameterizes how the emissivity is distributed between bright and faint sources.  We will use $M_{\min} = 3 \times 10^{9} h^{-1}M_{\odot}$ and $\beta = 1$ as our fiducial values, and will vary both in \S\ref{subsec:ionsource}.  Density fields are taken from an Eulerian hydro-dynamical simulation with $N = 1024^3$ gas cells run with the code of~\cite{Trac2004}, with the same box size and large-scale initial conditions used in the N-body run.  We re-binned the density fields to a coarse-grained resolution of $N_{\rm rt} = 200^3$ for our FlexRT runs.  

\subsection{Sub-grid Models for $\lambda_{\nu}$}
\label{subsec:modsinks}

Our prescription for $\lambda_{\nu}$ in Eq.~\ref{eq:gammah1_subgrid} follows the formalism outlined in~\citep{Cain2022b}, with the improvement that we now account for the frequency dependence of $\lambda_{\nu}$ (see next section).  We use a suite of high-resolution hydro/RT simulations in $1 h^{-1}$Mpc volumes (like those in~\citet{DAloisio2020}) run with a modified version of the code of~\cite{Trac2007}.  These simulations sample a range of box-scale over-densities~\citep[using DC modes,][]{Gnedin2011}, $\Gamma_{\rm HI}$, and reionization redshifts ($z_{\rm reion}$) to calibrate a sub-grid model for $\lambda_{\nu}$, which evolves in our simulations according to
\begin{equation}
    \label{eq:lambdamaster}
    \frac{d \lambda_{\nu}}{dt} = \pd{\lambda_{\nu}}{t}\Big|_{\Gamma_{\rm HI}} + \pd{\lambda_{\nu}}{\Gamma_{\rm HI}} \Big|_{t}\frac{d\Gamma_{\rm HI}}{dt} - \frac{\lambda_{\nu} - \lambda_{\nu,0}}{t_{\rm relax}},
\end{equation}
The first term on the RHS gives the time evolution of $\lambda_{\nu}$ at fixed $\Gamma_{\rm HI}$ and the second gives the instantaneous dependence on $\Gamma_{\rm HI}$.  We interpolate the first term from our simulation suite and assume $\lambda_{\nu} \propto \Gamma_{\rm HI}^{2/3}$ to calculate the second~\citep[motivated by theoretic expectations for an ionized IGM, e.g.][]{Furlanetto2005}.  The last term accounts for the evolution of $\lambda_{\nu}$ towards its constant-$\Gamma_{\rm HI}$ limit\footnote{That is, the limit in which $\Gamma_{\rm HI}$ has not changed for a long time.  }, $\lambda_{\nu,0}$, which we also interpolate from our simulations.  We assume $t_{\rm relax} = 100$ Myr for the timescale over which the IGM loses memory of its previous $\Gamma_{\rm HI}$ history.  Further details can be found in the text and appendices of~\cite{Cain2022b}.  

In this work, we use two versions of our sub-grid model, both of which are described in~\cite{Cain2022b}: 
\begin{itemize}
    \item {\bf Full Sinks}: This model uses the full, self-consistent evolution of $\lambda_{\nu}$ predicted by the our small-volume simulations suite combined with Eq.~\ref{eq:gammah1_subgrid} and~\ref{eq:lambdamaster}.  It accounts for the pressure smoothing of the IGM after ionization and the effect of un-resolved self-shielded systems. 
    
    \item {\bf Relaxed Limit}: For this model, we extrapolate the low-redshift $\lambda_{\nu}$ in our $z_{\rm reion} = 12$ sub-grid simulations to lower redshift assuming a power law.  This treats the IGM as if it has been ionized for a long time, and is in a pressure smoothed equilibrium, at all times.  In this model, the effects of small, short-lived sinks that are sensitive to pressure smoothing and photo-evaporation (see \S\ref{subsec:ionphotsinks}) are neglected. 
\end{itemize}

\subsection{Multi-Frequency RT}
\label{subsec:multifreq}

We have extended our sub-grid formalism to allow for a multi-frequency RT treatment.  Unfortunately, our sub-grid simulations only saved the MFP averaged over the ionizing spectrum, and the Lyman Continuum (LyC, $912$ $\text{\AA}$) MFP.  To approximate the full frequency dependence, we assume
\begin{equation}
    \label{eq:kappanu}
    \kappa_{\nu} = \kappa_{912} \left(\frac{\sigma_{\rm HI}(\nu)}{\sigma_{\rm HI}^{912}}\right)^{\beta_{\rm N} - 1}
\end{equation}
where $\kappa_{\nu} \equiv 1/\lambda_{\nu}$ is the absorption coefficient at frequency $\nu$ and $\kappa_{912}$ is the LyC absorption coefficient.  This form follows from Eq. 5 of~\cite{Nasir2021} (see also~\citet{Prochaska2009}) assuming that the HI column density distribution is a power law of the form $f(N_{\rm HI}) \propto N_{\rm HI}^{-\beta_{\rm N}}$.  Our sub-grid simulations assume a power law ionizing spectrum of the form $J_{\nu} \propto \nu^{-\alpha}$ with $\alpha = 1.5$.  We can then find $\beta_{\rm N}$ using
\begin{equation}
    \label{eq:betaN}
    \frac{\langle \kappa_\nu \rangle_{\alpha=1.5}}{\kappa_{912}} = \frac{\alpha[1 - 4^{-\alpha-2.75(\beta_{\rm
 N}-1)}]}{[\alpha+2.75(\beta_{\rm N}-1)](1-4^{-\alpha})},
\end{equation}
where $\langle \kappa_\nu \rangle_{\alpha=1.5}$ is the frequency-averaged absorption coefficient in the sub-grid simulations.  Then we calculate $\kappa_{\nu}$ in each frequency bin in FlexRT using Eq.~\ref{eq:kappanu}.  Note that $\langle \sigma_{\rm HI} \rangle_{\nu}^{i}$ and $\langle \lambda \rangle_{\nu}^{i}$ in Eq.~\ref{eq:gammah1_subgrid} are averaged over the spectrum incident on cell $i$ in FlexRT, not the $\alpha = 1.5$ spectrum used in the sub-grid simulations.  This assumes that $\lambda_{\nu}$ can be estimated for any incident spectrum using results from an $\alpha = 1.5$ simulation.  In FlexRT, we use $5$ frequency bins with energies $14.44, 16.64, 19.91, 25.47$ and $37.6$ eV, each containing the same number of photons\footnote{This particular binning is chosen to give approximately the correct average HI cross-section for an $\alpha = 1.5$ spectrum.  The are also approximately the same frequency bins used in the high-resolution simulations.  }.  We test this procedure and give more details in Appendix~\ref{app:kappa_freq}.  We find that our approach is accurate to within $20\%$ even for negative values of $\alpha$.  

\subsection{Recombination Radiation}
\label{subsec:recrad}

We have added an approximate treatment of recombination radiation to FlexRT.  We assume that a fraction $f_{\rm esc}^{\rm rec}$ of IRPs escape from the dense clumps that produce them, and the rest are absorbed locally.  The effective recombination coefficient can be written as
\begin{equation}
    \label{eq:alphaT}
    \alpha(T) =  f_{\rm esc}^{\rm rec} \alpha_{\rm A}(T) + (1 - f_{\rm esc}^{\rm rec})\alpha_{\rm B}(T)
\end{equation}
where $T$ is temperature and $\alpha_{\rm A}$ and $\alpha_{\rm B}$ are the case A and B recombination coefficients, respectively.  We assume Eq.~\ref{eq:alphaT} holds when modeling the ionizing photon opacity. For Ly$\alpha$ forest calculations (described in \S\ref{subsec:lyaforest}), we use $\alpha_{\rm A}$ for the reasons explained in \S\ref{subsec:recomb}.  

We assume that the ionized gas in each cell is in photo-ionization equilibrium\footnote{This is not true in our small volume simulations, which include non-equilibrium effects such as photo-evaporation of self-shielded systems.  However, it is likely a good approximation in the Relaxed Limit sub-grid model, since such processes are largely neglected there. 
 }.  Then, we can write the recombination rate as a function of the ionizing absorption coefficient~\citep{Emberson2013}, with $\kappa \propto \alpha(T)$ at fixed density and $\Gamma_{\rm HI}$.  Our small-volume simulations assume the case B recombination coefficient, so we can re-scale the absorption coefficient from those simulations like
\begin{equation}
    \label{eq:kappa_rec}
    \langle\kappa\rangle_{\nu} = \langle\kappa_{\rm B}\rangle_{\nu}\frac{\alpha(T)}{\alpha_{\rm B}(T)}
\end{equation}
where $\langle\kappa_{\rm B}\rangle_{\nu}$ is the sub-grid model prediction.    

Under these approximations, the emissivity of IRPs in cell $i$ is
\begin{equation}
    \label{eq:ngam_rec}
    \dot{n}_{\gamma}^{\rm rec,i} =  x_{\rm ion}^{\rm i}f_{\rm esc}^{\rm rec}\frac{\Gamma_{\rm HI}^{\rm i}}{\langle\sigma_{\rm HI}\rangle_{\nu}^{i}}\left(\langle \kappa_{\rm A}^{i} \rangle_{\nu} - \langle \kappa_{\rm B}^{i} \rangle_{\nu}\right)(1+\chi)
\end{equation}
where $\langle\kappa_{\rm A}^{i}\rangle_{\nu}$ is Eq.~\ref{eq:kappa_rec} evaluated for $f_{\rm esc} = 1$, and the factor of $1+\chi$ crudely accounts for ground state recombinations from HeII\footnote{This is appropriate because the recombination coefficients for HII and HeII are very similar in the relevant temperature range.  }.  These photons are added to the emissivity from halos in each cell.  The energy of IRPs is
\begin{equation}
    \label{eq:Erec}
    E_{\rm rec} = 13.6 \text{ eV} + \frac{1}{2}m_{\rm e} v^2
\end{equation}
where $m_{\rm p}$ is the electron mass and $v$ is the relative velocity between the recombining electron and proton.  The average kinetic energy is $\approx \frac{3}{2} k T$, where $T$ is the gas temperature.  For $10^4$ K gas this is $1.3$ eV, yielding $E_{\rm rec} = 14.9$ eV and an average HI cross-section of $\langle \sigma_{\rm HI} \rangle_{\nu} = 4.93 \times 10^{-18}$ cm$^{2}$, a factor of $2$ larger than the $\langle \sigma_{\rm HI} \rangle_{\nu} = 2.55 \times 10^{-18}$ cm$^{2}$ for an $\alpha = 1.5$ spectrum.  In our multi-frequency FlexRT runs, we assign IRPs to our lowest energy bin ($14.44$ eV). 
 We provide a complete derivation of Eq.~\ref{eq:alphaT}-\ref{eq:ngam_rec} in Appendix~\ref{app:recomb_tests}.  

\subsection{Modeling Halos as Absorbers}
\label{subsec:galabs}

Lastly, we added a simple prescription for the opacity from halos hosting ionizing sources.  This allows us to assess how missing opacity from massive halos would affect the drop in $\dot{N}_{\gamma}$ (see \S\ref{subsec:ionphotsinks}).  We treat halos as spherically symmetric, optically thick absorbers with cross-section
\begin{equation}
    \label{eq:sigmahalo}
    \sigma_{\rm halo}(M_{\rm halo}, z, \Gamma_{\rm HI}) = \pi R_{\rm halo}^2(M_{\rm halo}, z, \Gamma_{\rm HI})
\end{equation}
where $R_{\rm halo}$ is the radius to which the halo is opaque.  We write this as
\begin{equation}
    \label{eq:Rhalo}
    R_{\rm halo}(M_{\rm halo}, z, \Gamma_{\rm HI}) = f_{200}(M_{\rm halo}, z, \Gamma_{\rm HI}) R_{200}(M_{\rm halo}, z)
\end{equation}
where $R_{200}$ is the radius within which the mean density is $200\times$ the cosmic mean, and $f_{200}$ parameterizes the opacity of the halo (including its sensitivity to $\Gamma_{\rm HI}$).  Halo masses in our N-body simulation are given by the mass enclosed within $R_{200}$, which is approximately the halo virial radius~\citep{Trac2015}.  We will describe momentarily how we estimate $f_{200}$.  

The total opacity from halos in cell $i$ is given by
\begin{equation}
    \label{eq:kappa_halo}
    \kappa_{\rm halo}^{i} = \frac{1}{V_{\rm cell}^{i}}\sum_{j = 1}^{N_{\rm halo}} \sigma_{\rm halo}(M_{ij}, z)
\end{equation}
where the sum runs over all halos occupying cell $i$.  In what follows, we assume that all halos resolved by our N-body simulations (with $M_{\rm halo} \geq 3 \times 10^9$ $h^{-1}M_{\odot}$) contribute extra opacity.  The total opacity in cell $i$ is given by adding $\kappa_{\rm halo}$ to Eq.~\ref{eq:tauij}:\footnote{Note that the extra opacity from halos does not contribute to $\langle \lambda \rangle_{\nu}$ in the numerator of Eq.~\ref{eq:gammah1_subgrid}.  This is because the expression $\langle \lambda \rangle_{\nu} \langle \sigma_{\rm HI} \rangle_{\nu}$ in the numerator of that equation is equal to the inverse of the residual HI number density in the {\it ionized} IGM, but the halos are treated here as if they are completely self-shielding (neutral).  Also, note that $\kappa_{\rm halo}$ is also multiplied by $x_{\rm ion}$ because not doing so would spuriously count opacity from halos still within the neutral part of the cell.  }
\begin{equation}
    \label{eq:tauij_halo}
    \tau_{ij}^{\nu} = x_{\rm ion}^{i}\Delta s^{ij}\left(\frac{1}{\lambda_{i}^{\nu}} + \kappa_{\rm halo}^{i}\right)
\end{equation}

Finally, we need a prescription for $f_{200}$, for which we use an analytical model similar to that described in \S2 of~\cite{Munoz2016} (see also~\cite{Theuns2021,Theuns2023}).  We assume all halos have a spherically symmetric power law density profile
\begin{equation}
    \label{eq:rho_gas}
    \rho_{\rm gas}(r) \propto r^{-\epsilon}
\end{equation}
The radius $r_{\rm abs}$ to which the halo is opaque to ionizing photons can be approximated by the condition
\begin{equation}
    \label{eq:tauion}
    \tau_{\rm ion} \approx n_{\rm HI}(r_{\rm abs}) \langle \sigma_{\rm HI} \rangle r_{\rm abs}  = 1
\end{equation}
where $\tau_{\rm ion}$ is approximately the optical depth encountered by a sightline intersecting the edge of the halo\footnote{The analogous derivation in \S2 of~\citet{Munoz2016} identified the absorber radius as the distance at which the opacity approaching the center of the halo from infinity reaches $1$.  This condition yields the same scaling relations found here, but gives an absorber radius about $70\%$ of that given by Eq.~\ref{eq:tauion}.  This is an acceptable difference given the highly approximate nature of the model.  }.  We show in Appendix~\ref{app:winds} that, assuming photo-ionization equilibrium, $f_{200}$ scales like
\begin{equation}
    \label{eq:f200}
     f_{200} \propto \Gamma_{\rm HI}^{-\frac{1}{2\epsilon-1}}M_{\rm halo}^{\frac{1}{3(2\epsilon-1)}}(1+z)^{\frac{5}{2\epsilon-1}}
\end{equation}
where $\Gamma_{\rm HI}$ is the photo-ionization rate in the IGM surrounding the halo.  The positive scaling with redshift reflects the fact that halos collapsing at higher $z$ have larger physical densities at their virial radii. More details are given in Appendix~\ref{app:winds}.  

\subsection{Modeling the Ly$\alpha$ Forest}
\label{subsec:lyaforest}

We compute Ly$\alpha$ forest statistics on a $N = 2048^3$ version of the same hydro run used to get our density fields.  We traced $4000$ sightlines through the box and mapped on the $\Gamma_{\rm HI}$, $T$, and neutral fraction from FlexRT.  In ionized RT cells (defined as those with $x_{\rm HI} < 0.5$), we compute the residual neutral fraction assuming photo-ionization equilibrium.  The spatial resolution of our hydro simulation is $97.7$ $h^{-1}$kpc, which~\citet{DAloisio2018} found to be un-converged in the mean Ly$\alpha$ transmission at the $50\%$ level.  We correct for this using the approach outlined in Appendix A of~\citet{DAloisio2018}, and the correction factors from their Table A1.   With their correction, our effective spatial resolution for the forest calculation is $12.2$ $h^{-1}$kpc, close to the convergence criteria found by~\citet{Doughty2023}.  

One problem with mapping the FlexRT fields directly to the hydro density grids is that correlations with density on scales smaller than the RT cell size ($1~h^{-1}$Mpc) are missed.  This is a problem for temperature, which correlates strongly with density on small scales.  Hydro cells less dense than the coarse-grained RT cells they occupy contribute most of the Ly$\alpha$ transmission, and in general these have temperatures lower than that of their host RT cell.  The IGM temperature is usually assumed to follow a power law in density of the form
\begin{equation}
    \label{eq:TDelta}
    T(\Delta) = T_0 \Delta^{\gamma-1}
\end{equation}
where $\Delta$ is gas density in units of the mean, $T_0$ is the temperature at mean density, and the power law index $\gamma$ is mainly a function of how recently the gas was ionized.  Recently ionized gas ($z \approx z_{\rm reion}$) is nearly isothermal ($\gamma = 1$), while for $z << z_{\rm reion}$, $\gamma \rightarrow 5/3$.  We estimate $\gamma(z,z_{\rm reion})$ using the analytical solution for the IGM temperature given by~\citet{McQuinn2016}.  Then we re-scale the temperatures in each hydro cell according to
\begin{equation}
    \label{eq:Thydro}
    T_{\rm hydro}^{i} = T_{\rm RT}^{j} \left(\frac{\Delta_{\rm hydro}^{i}}{\Delta_{\rm RT}^{j}}\right)^{\gamma(z,z_{\rm reion}^{j})-1}
\end{equation}
where $T_{\rm RT}^{j}$ is the temperature in RT cell $j$ that contains hydro cell $i$, $\Delta_{\rm hydro}^{i}$ is the density in hydro cell $i$, $\Delta_{\rm RT}^{j}$ is the coarse-grained density in RT cell $j$, and $z_{\rm reion}^{j}$ is the reionization redshift of cell $j$.  We find that this procedure results in a $\sim 10\%$ correction to the forest mean flux.  We test this procedure and give more details in Appendix~\ref{app:TDrelation}.  

\section{Effect of individual modeling considerations}
\label{sec:results_IGM}

\subsection{Reference model}
\label{subsec:ref_model}

In the following sections, we will study the IGM modeling effects described in \S\ref{sec:phys}.  Starting from a reference model (described below), we will vary each modeling choice one at a time and assess how this changes the evolution of $\dot{N}_{\gamma}$ required by the forest.  In {\it all} our FlexRT simulations, $\dot{N}_{\gamma}$ is a non-parametric function of redshift tuned such that the mean Ly$\alpha$ transmission, $\langle F_{\rm Ly\alpha} \rangle$, agrees as closely as possible with recent measurements by~\citet{Bosman2021} using the new XQR-30 QSO sample~\citep{DOdorico2023}.  This is similar to the approach adopted in~\citep{Keating2019,Keating2020}.  Our reference model has the following features: 
\begin{itemize}
    \item A monochromatic spectrum with a $\sigma_{\rm HI} = 2.55\times 10^{-18}$ cm$^{2}$, as in~\citet{Cain2021,Cain2022b}.  This $\sigma_{\rm HI}$ is the average value for a power law spectrum with $\alpha = 1.5$.  
    \item The Relaxed Limit sub-grid model, described in \S\ref{subsec:modsinks}.  We chose this model so that our reference case would be representative of simulations that do not resolve the small, abundant sinks that are sensitive to photo-evaporation and pressure smoothing.  We do not include any extra opacity from halos.  
    \item We use the case B assumption for the ionizing opacity that is native to our sub-grid simulations (i.e. $f_{\rm esc}^{\rm rec} = 0$ in Eq.~\ref{eq:alphaT}).  
\end{itemize}

\begin{figure*}
    \centering
    \includegraphics[scale=0.15]{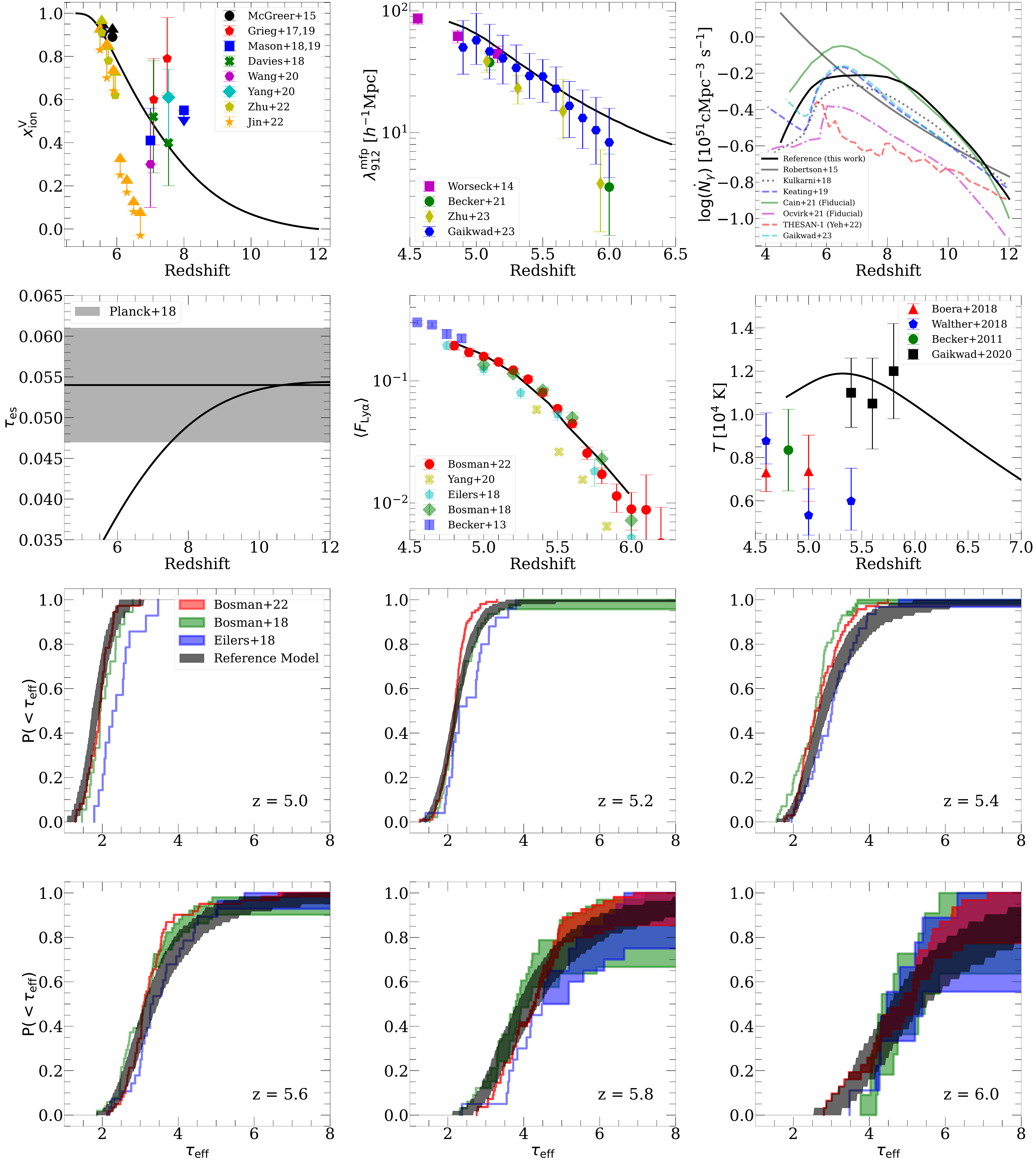}
    \caption{Top two rows, in clockwise order from top left: ionized fraction $x_{\rm ion}$, LyC MFP $\lambda_{912}^{\rm mfp}$, ionizing emissivity $\dot{N}_{\gamma}$, IGM temperature at mean density $T_0$, mean Ly$\alpha$ forest transmission $\langle F_{\rm Ly\alpha}\rangle$, and electron scattering optical depth to the CMB ($\tau_{\rm es}$).  We compare our reference model (black solid curve) to observations from~\citep{McGreer2015,Grieg2016,Grieg2019,Mason2018,Mason2019,Davies2018,Wang2020,Yang2020a,Zhu2022,Jin2023,Worseck2014,Becker2021,Zhu2023,Gaikwad2023,Becker2011,Walther2019,Boera2019,Gaikwad2020,Becker2013,Bosman2018,Eilers2018,Yang2020b,Bosman2021,Planck2018}.  The upper right panel displays the same models shown in Figure~\ref{fig:em_lit} for reference (faded curves).  All the simulations in this work have their emissivity histories calibrated to match the $\langle F_{\rm Ly\alpha} \rangle$ measurements from~\citet{Bosman2021} as closely as possible over $4.8 < z < 6$.  The bottom two rows show the CDF of $\tau_{\rm eff}$, computed over $50$ $h^{-1}$Mpc segments of the Ly$\alpha$ forest, compared to the measurements of~\citet{Bosman2018,Eilers2018,Bosman2021}.  See text for details.  }
    \label{fig:ref_summary}
\end{figure*}

Figure~\ref{fig:ref_summary} compares our reference model with a wide range of observations and other simulations.  The top two rows show, clockwise from the top left, the ionized fraction $x_{\rm ion}$, Lyman Continuum (LyC) MFP $\lambda_{912}^{\rm mfp}$, $\dot{N}_{\gamma}$, IGM temperature at mean density $T_0$, mean Ly$\alpha$ forest transmission $\langle F_{\rm Ly\alpha}\rangle$, and electron scattering optical depth to the CMB, $\tau_{\rm es}$.  We estimate $\lambda_{912}^{\rm mfp}$ by stacking mock spectra calculated from random starting points and fitting to the model of~\citet{Prochaska2009}, as in~\citet{Cain2021}\footnote{\citet{Roth2023} find that this estimator for the MFP agrees well with the measurements using QSOs, even at $z = 6$ when reionization is ongoing. 
 (see also~\citet{Satyavolu2023}).  }.  The upper right panel shows $\dot{N}_{\gamma}$ for the reference model (black solid) and the same curves from Figure~\ref{fig:em_lit}.  The other panels show a collection of recent observational measurements from the literature, referenced in the figure caption.  The red points in the second-from-top middle panel show the most recent measurements of $\langle F_{\rm Ly\alpha}\rangle$ by ~\citet{Bosman2021} at $4.8 < z < 6$, against which we calibrate our models.  

The bottom two rows show the cumulative distribution function (CDF) of effective Ly$\alpha$ optical depth, $\tau_{\rm eff}$, measured over $50$ $h^{-1}$Mpc segments of the forest.  Each panel shows a different redshift between $z = 5$ and $6$.  We compare to measurements by~\citet{Bosman2021} (red),~\citet{Bosman2018} (green), and~\cite{Eilers2018} (blue).  For observations, shaded regions denote limiting assumptions about the value of $\tau_{\rm eff}$ for null detections\footnote{The lower bound assumes $\tau_{\rm eff} = \infty$, while the upper bound assumes $\tau_{\rm eff} = -\ln(2 \sigma)$, where $\sigma$ is the uncertainty of the mean transmission.}.  In each redshift bin, we randomly draw a number of $50$ $h^{-1}$Mpc segments from our simulation sightlines equal to the corresponding number in the~\citet{Bosman2021} data set, and compute the CDF $500$ times.  The shaded gray curves show the $10-90\%$ range from this procedure.  

Our reference model is in broad agreement with the observations in Figure~\ref{fig:ref_summary}.  Our reionization history ends slightly too late for the dark pixel constraints from~\citet{McGreer2015}, but is in agreement with updated limits from~\citet{Jin2023} and recent dark gap constraints from~\citet{Zhu2022}.  The MFP is in reasonable agreement with measurements of~\citet{Worseck2014} and~\citet{Gaikwad2023}, but is slightly too high for those of~\citet{Becker2021,Zhu2023}.  This could owe to the Relaxed Limit model having too few sinks, and/or to our use of the case B approximation for recombination radiation.  Our $\dot{N}_{\gamma}$ is qualitatively similar to those discussed in \S\ref{sec:intro}, with a factor of $\sim 2$ decline between $z = 6$ and the end of the simulation at $z = 4.8$.  Our thermal history is in agreement with the recent measurements of~\citet{Gaikwad2020}, but is slightly too high for measurements at $z \leq 5$.  

Our $\tau_{\rm eff}$ distributions are also in broad agreement with measurements.  Note that our simulations are not explicitly calibrated to match the $\tau_{\rm eff}$ distribution of~\citet{Bosman2021}, only $\langle F_{\rm Ly\alpha}\rangle$.  So, to some extent, our models can be judged by their agreement with $\tau_{\rm eff}$ measurements.  We see that at $5.2 < z < 5.8$, our $\tau_{\rm eff}$ distributions are slightly wider than the  ~\citet{Bosman2021} measurements.  \citet{Bosman2021} found that their $\tau_{\rm eff}$ distributions are in good agreement with simulations in which reionization is completely over by $z = 5.3$.  The fact that our reference model ends reionization slightly later than this ($\sim 5.1$) is likely the reason for this disagreement.  %We will return to this point in subsequent sections.  

\subsection{Ionizing spectrum}
\label{subsec:ionizingspectrum}

We start by varying the properties of the ionizing spectrum.  Our first model, which we label ``Fiducial (Fid.) Multi-Frequency'', adopts the multi-frequency treatment described in \S\ref{subsec:multifreq}, still assuming $\alpha = 1.5$ for sources.  Comparing this model to the reference model (which employs the monochromatic approximation) isolates the effects of IGM filtering\footnote{To maximize the effect of filtering,  we assume $\beta_{\rm N} = 2$ in Eq.~\ref{eq:kappanu} instead of calculating it self-constantly using Eq.~\ref{eq:betaN} in this section.  }.  Our second model, labeled ``Hard Multi-frequency'', also uses multi-frequency RT but assumes $\alpha = 0.5$, on the low end for reasonable stellar population models (see \S\ref{subsec:alpha}).

\begin{figure}
    \centering
    \includegraphics[scale=0.193]{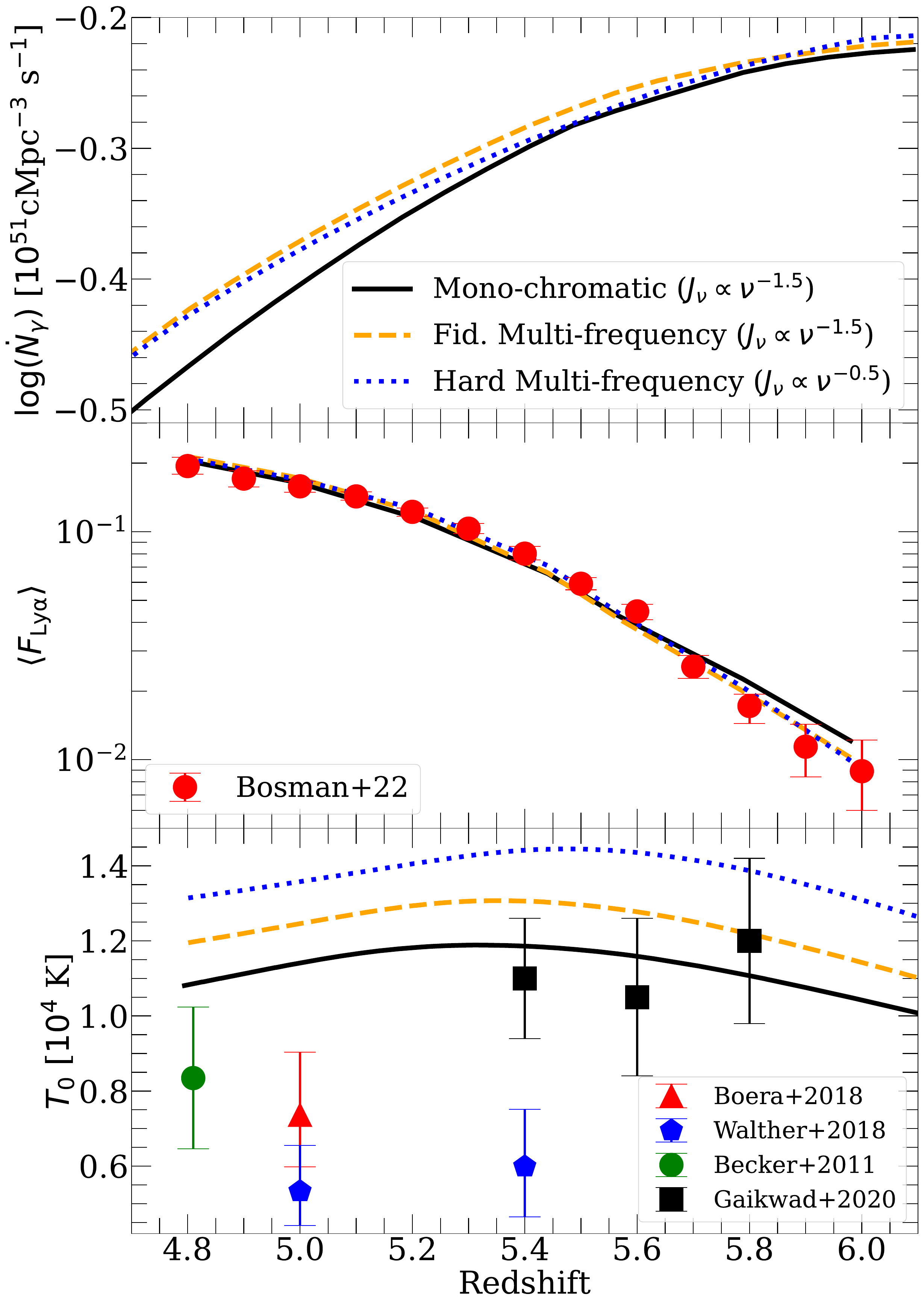}
    \caption{Comparison of our Fid. Multi-Frequency (orange dashed) and Hard Multi-frequency (blue dotted) models to our reference model (solid black).  From top to bottom, the panels show $\dot{N}_{\gamma}$, $\langle F_{\rm Ly\alpha}\rangle$, and $T_0$.  All models are calibrated to match the measurements of $\langle F_{\rm Ly\alpha}\rangle$ by~\citet{Bosman2021}.  Harder spectra work in the direction of making the drop in $\dot{N}_{\gamma}$ shallower, but the effect is small compared to the size of the drop itself.  The Hard Multi-frequency model is also in significant tension with most recent IGM temperature measurements.  }
    \label{fig:ref_vs_mfreq}
\end{figure}

Figure~\ref{fig:ref_vs_mfreq} shows $\dot{N}_{\gamma}$ (top), $\langle F_{\rm Ly\alpha}\rangle$ (middle) and $T_0$ (bottom) for these models (orange dashed and blue dotted, respectively) compared to the reference model (black solid).  By construction, all three models have nearly identical $\langle F_{\rm Ly\alpha}\rangle$, in agreement with~\citet{Bosman2021}.  All three have fairly similar $\dot{N}_{\gamma}$ at $z < 6$, with the Fid. Multi-frequency and Hard Multi-frequency models having a slightly shallower drop than the reference case.  The effects of multi-frequency RT, including a harder spectrum, on $\dot{N}_{\gamma}$ are about $0.05$ dex, much smaller than the $\sim 0.3$ dex drop in the reference model.  The smallness of the effect likely owes to the partial cancellation of $\Gamma_{\rm HI}$ and $T$ effects in the $\langle F_{\rm Ly\alpha} \rangle$, discussed in \S\ref{subsec:alpha}.  Recall that a harder spectrum leads to lower $\Gamma_{\rm HI}$ and higher temperatures, which have opposite effects on the forest transmission.  

We see from the bottom panel that these models also have significantly higher IGM temperatures.  $T_0$ is $\sim 10\%$ ($20\%$) higher in the Fid. Multi-frequency (Hard Multi-frequency) model than in the reference case.  The Hard Multi-frequency case is in $\sim 2 \sigma$ tension with the~\citet{Gaikwad2023} measurements at $z = 5.4$ and $5.6$, and in $\sim 3\sigma$ tension with the~\citet{Boera2019} measurement at $z = 5$.  Even harder spectra, which may allow for a shallower drop, would only worsen this disagreement.  We conclude that although a harder ionizing spectrum works in the direction of making the drop shallower, on its own it is an unlikely explanation.  

\subsection{Ionizing Recombination Radiation}
\label{subsec:recradresults}

\begin{figure}
    \centering
    \includegraphics[scale=0.193]{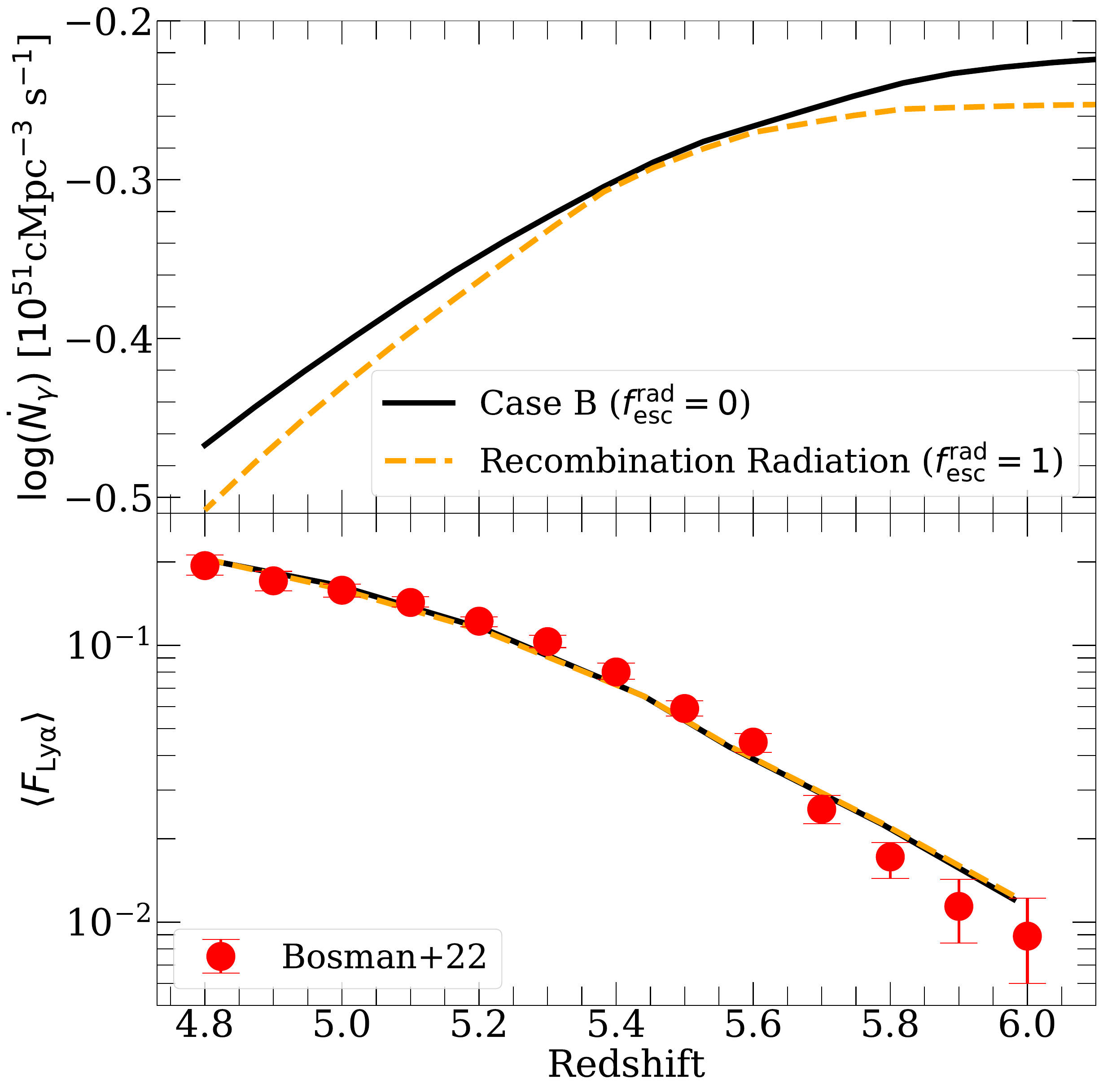}
    \caption{Effect of ionizing recombination photons on the drop in $\dot{N}_{\gamma}$.  Top: $\dot{N}_{\gamma}$ for our recombination model (orange dashed) and reference model (black solid).  Bottom: $\langle F_{\rm Ly\alpha} \rangle$ compared to~\citet{Bosman2021} measurements.  Ionizing recombination radiation has a small effect on the drop, even for parameters that maximize its effect.  See text for details.  }
    \label{fig:ref_vs_recomb}
\end{figure}

Next, we study the effect of ionizing recombination radiation.  Our ``Recombination Radiation'' model assumes $f_{\rm esc}^{\rm rec} = 1$, which corresponds to the case A limit and maximally contrasts the reference model.  To facilitate a fair comparison with our mono-chromatic reference model, here we only include two frequency bins: $h_p \nu = 19.0$ eV for sources (which gives the same $\sigma_{\rm HI}$ assumed in the reference model) and $13.6$ eV for IRPs.  The choice of $13.6$ eV maximizes the ``softening'' effect of IRPs on the ionizing spectrum\footnote{IRPs do not have a net heating or cooling effect, so this choice does not affect the IGM temperature.  }.   Figure~\ref{fig:ref_vs_recomb} shows $\dot{N}_{\gamma}$ (top) and $\langle F_{\rm Ly\alpha} \rangle$ (bottom) for our Recombination Radiation model (orange dashed) and our reference model (black solid).  As in Figure~\ref{fig:ref_vs_mfreq}, the differences in $\dot{N}_{\gamma}$ between models are modest.  The Recombination Radiation model has a slightly flatter $\dot{N}_{\gamma}$ at $z = 6$, with a slightly steeper drop between $z = 5.5$ and $4.8$.  Given that the model in Figure \ref{fig:ref_vs_recomb} likely overestimates the effect of recombination radiation, we conclude that IRPs probably have a small effect on $\dot{N}_{\gamma}$ at fixed $\langle F_{\rm Ly\alpha}\rangle$. 

\begin{figure*}
    \centering
    \includegraphics[scale=0.153]{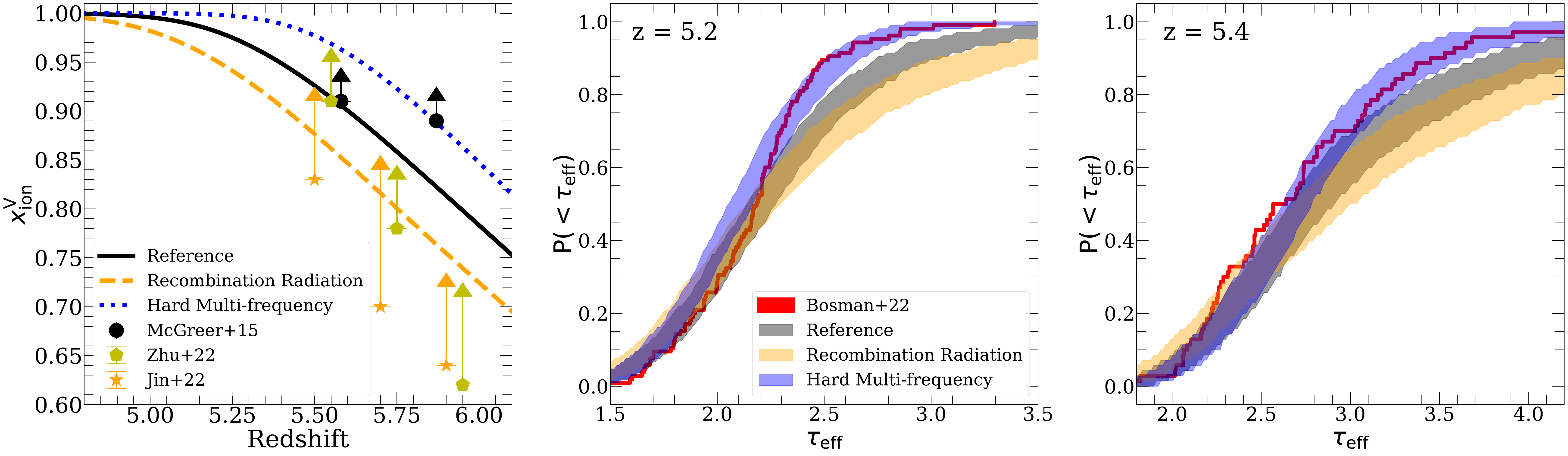}
        \caption{Sensitivity of the reionization history and $\tau_{\rm eff}$ distribution to the ionizing spectrum and recombination radiation.  From left to right, the panels show the reionization history and $\tau_{\rm eff}$ distribution at $z = 5.2$ and $5.4$.  The main effect of recombination radiation is to soften the ionizing spectrum, so we can think of this model as a ``soft spectrum'' scenario.  Models with harder spectra end reionization earlier when calibrated to the same $\langle F_{\rm Ly\alpha} \rangle$ evolution, in better agreement with constraints on the neutral fraction at $z < 6$.  For this reason, they have narrower $\tau_{\rm eff}$ distributions, in better agreement with recent measurements.  }
    \label{fig:xion_taueff_recomb}
\end{figure*}

However, IRPs do have an appreciable effect on the reionization history and the $\tau_{\rm eff}$ distribution.  Figure~\ref{fig:xion_taueff_recomb} shows the $z < 6$ reionization history (left) and $\tau_{\rm eff}$ CDFs at $z = 5.2$ and $5.4$ (middle and right, respectively).  We also show our Hard Multi-frequency model (blue dotted curve and shaded regions) from the previous section.  The Recombination Radiation (Hard Multi-frequency) model ends reionization later (earlier) than the reference case, and displays worse (better) agreement with $\tau_{\rm eff}$ measurements.  The former is in mild tension with the $z = 5.5$ neutral fraction constraint of~\citet{Zhu2022}.  In the Hard Multi-frequency model, at fixed $\langle F_{Ly\alpha}\rangle$,  $\dot{N}_{\gamma}$ is higher and reionization ends earlier, while the opposite is true in the Recombination Radiation model.  

The main reason for later reionization history in the Recombination Radiation model is the soft ionizing spectrum of the IRPs.  Since the spectrum is softer in that model than the reference case, $\Gamma_{\rm HI}$ and $\langle F_{\rm Ly\alpha}\rangle$ are higher at fixed $\dot{N}_{\gamma}$, and so reionization ends later at fixed $\langle F_{\rm Ly\alpha}\rangle$.  We have checked that if the IRPs are assigned the same frequency as the sources, the differences with the Reference model largely disappear.  A key takeaway from Figure~\ref{fig:xion_taueff_recomb} is that harder (softer) ionizing spectra can lead to earlier (later) reionization and better (worse) agreement with measurements of the $\tau_{\rm eff}$ distribution at fixed $\langle F_{\rm Ly\alpha}\rangle$. 

\subsection{Missing Sinks}
\label{subsec:missingsinks}

Ionizing photon sinks regulate the ionizing background, making them promising candidates to explain the drop in $\dot{N}_{\gamma}$.  In this section, we consider whether a population of sinks that are missing from simulations could explain the drop.  

\subsubsection{Missing un-resolved sinks}
\label{subsubsec:shortlived}

\begin{figure*}
    \centering
    \includegraphics[scale=0.166]{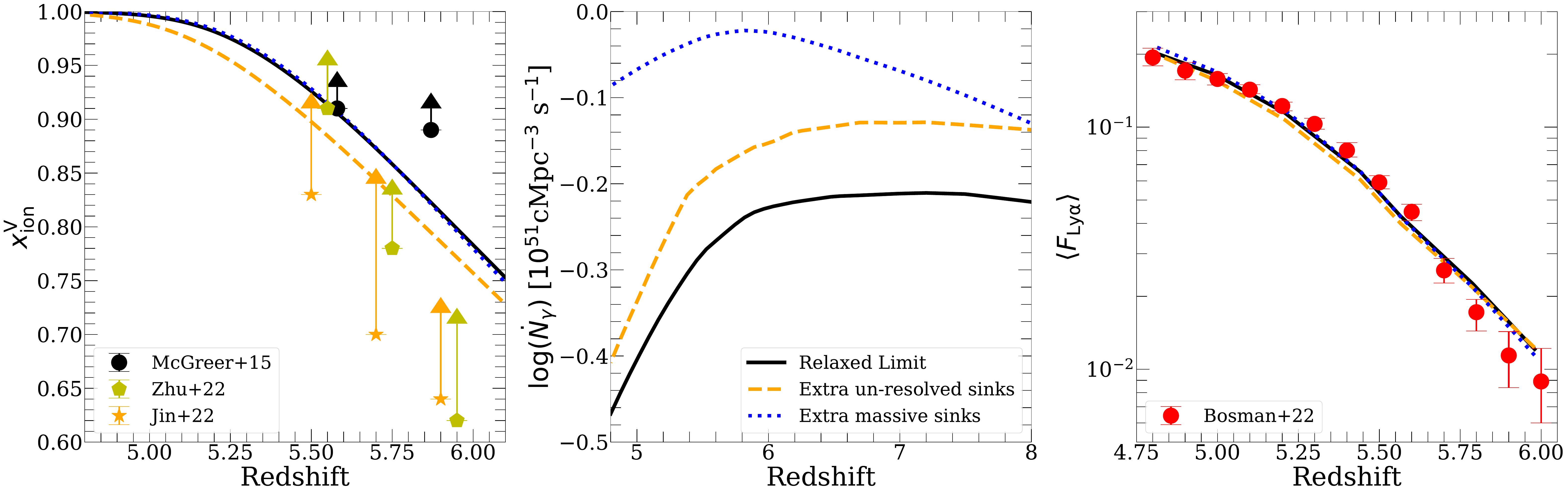}
    \caption{Effect of sinks on the drop in $\dot{N}_{\gamma}$. We show $x_{\rm ion}$, $\dot{N}_{\gamma}$, and $\langle F_{\rm Ly\alpha}\rangle$ (left to right).  We compare our reference model (black solid) to a model that includes the smallest, hardest-to-resolve sinks (orange dashed) and a model with extra opacity from massive, star-forming halos (blue dotted).  The former requires a higher photon budget and ends reionization later, but has an $\sim 0.05$ dex steeper drop than the reference case.  This is because the smallest sinks do not survive very long and cannot help regulate the ionizing background at reionization's end.  Sinks in massive halos are not destroyed by photo-evaporation and pressure smoothing, and rapidly grow in abundance at lower redshifts.  These sinks effectively regulate $\Gamma_{\rm HI}$, reducing the drop to only $\sim 0.05$ dex.  }
    \label{fig:ref_vs_sinks}
\end{figure*}

One reason for sinks to be missing in a simulation is that they are unresolved.  A natural way for us to assess the effect of the smallest sinks is to compare our Relaxed Limit and Full Sinks sub-grid models, described in \S\ref{subsec:modsinks}.  The latter includes the effects of these objects on the IGM opacity, while the former is designed to neglect them.  The orange dashed curve in Figure~\ref{fig:ref_vs_sinks} shows $x_{\rm ion}$ (left), $\dot{N}_{\gamma}$ (middle), and $\langle F_{\rm Ly\alpha} \rangle$ for a simulation that uses the Full Sinks treatment, but is otherwise the same as our reference case (black solid curve).  We emphasize again that this comparison is made at fixed $\langle F_{\rm Ly\alpha} \rangle$, as the right panel shows.  

A surprising result is that including the smallest sinks does not alleviate the need for a drop - in fact, the drop is $\sim 0.05$ dex steeper than in the reference case.  This is because the smallest sinks are destroyed by photo-evaporation and pressures smoothing effects within a few hundred Myr, and only patches of the IGM that ionized recently still have them.  So, although they increase the photon budget (reflected by a later end to reionization and higher $\dot{N}_{\gamma}$), they do not survive long enough to help regulate $\Gamma_{\rm HI}$ after reionization ends.  The result is that an even steeper drop in $\dot{N}_{\gamma}$ is required.  For this reason, attempts to correct the LyC opacity in simulations for missing small-scale power are unlikely to eliminate the need for a drop.  

\subsubsection{Missing massive sinks}
\label{subsubsec:galphys}

It is also possible that simulations are under-predicting the opacity from massive halos.  We have run a simulation that uses the halo absorber model described in \S\ref{subsec:galabs}, with all else the same as in the reference case.  We assume fairly shallow density profiles for our halos, with $\epsilon = 1.5$ (compared to $2$ for isothermal profiles).  This choice is motivated by the possibility that feedback drives significant outflows~\citep[e.g.][]{Weldon2022}, possibly resulting in significant opacity outside the halo virial radius ($f_{200} > 1$).  The blue dotted curve in Figure~\ref{fig:ref_vs_sinks} shows results for our ``Massive Sinks'' model.  Unlike our Full Sinks model, our halo absorber model does make the drop in $\dot{N}_{\gamma}$ shallower.  Moreover, $\dot{N}_{\gamma}$ steadily rises between $z = 6$ and $8$ instead of staying flat.  This is because the abundance of $M \geq 3\times10^9$ $h^{-1}M_{\odot}$ halos grows rapidly throughout reionization, increasing by a factor of $2.6$ ($3.7$) from $z = 8$ to $6$ ($5$)\footnote{The mass contained in these halos grows by a factor of $4.1$ ($7.8$) between $z = 8$ and $6$ ($5)$.  }.  Because these sinks become more abundant rapidly, they can effectively regulate the growth of $\Gamma_{\rm HI}$ and avoid the need for a drop in $\dot{N}_{\gamma}$.  Moreover, these sinks are not evaporated by reionization, so they survive long enough to regulate the post-reionization ionizing background.  

There are a few caveats worth mentioning.  First, our treatment of halos as spherically symmetric, ``billiard-ball'' absorbers with a single density profile is overly simplistic.  Real galaxies have much more complex geometries and dynamics, and as such our model may not capture their opacity very well.  Second, the choice of including extra opacity for $M_{\rm halo} > 3\times10^9$ $h^{-1}M_{\odot}$ is arbitrary, being set by the completeness limit of our N-body simulation.  Lower mass cutoffs would result in more opacity from halos, but would also cause $\kappa_{\rm halo}$ to grow less quickly (or perhaps decline) with cosmic time (see Fig.~\ref{fig:f200} and surrounding discussion).  This would also increase the budget of ionizing photons required to complete reionization.  Finally, our choice of $\epsilon = 1.5$ may not be realistic - a smaller value for $\epsilon$ results in less opacity from halos, all else being the same.  A key ``performance test'' for models like this will be to see if their MFP is consistent with measurements at $5 < z < 6$~\citep[][Davies et. al. in prep.]{Gaikwad2023,Zhu2023}.  We provide such a test in \S\ref{subsec:realistic} and will further address this point in a forthcoming paper.

\section{Implications for the properties of sources}
\label{sec:sources}

\subsection{Realistic \& Optimistic Scenarios}
\label{subsec:realistic}

\begin{figure*}
    \centering
    \includegraphics[scale=0.15]{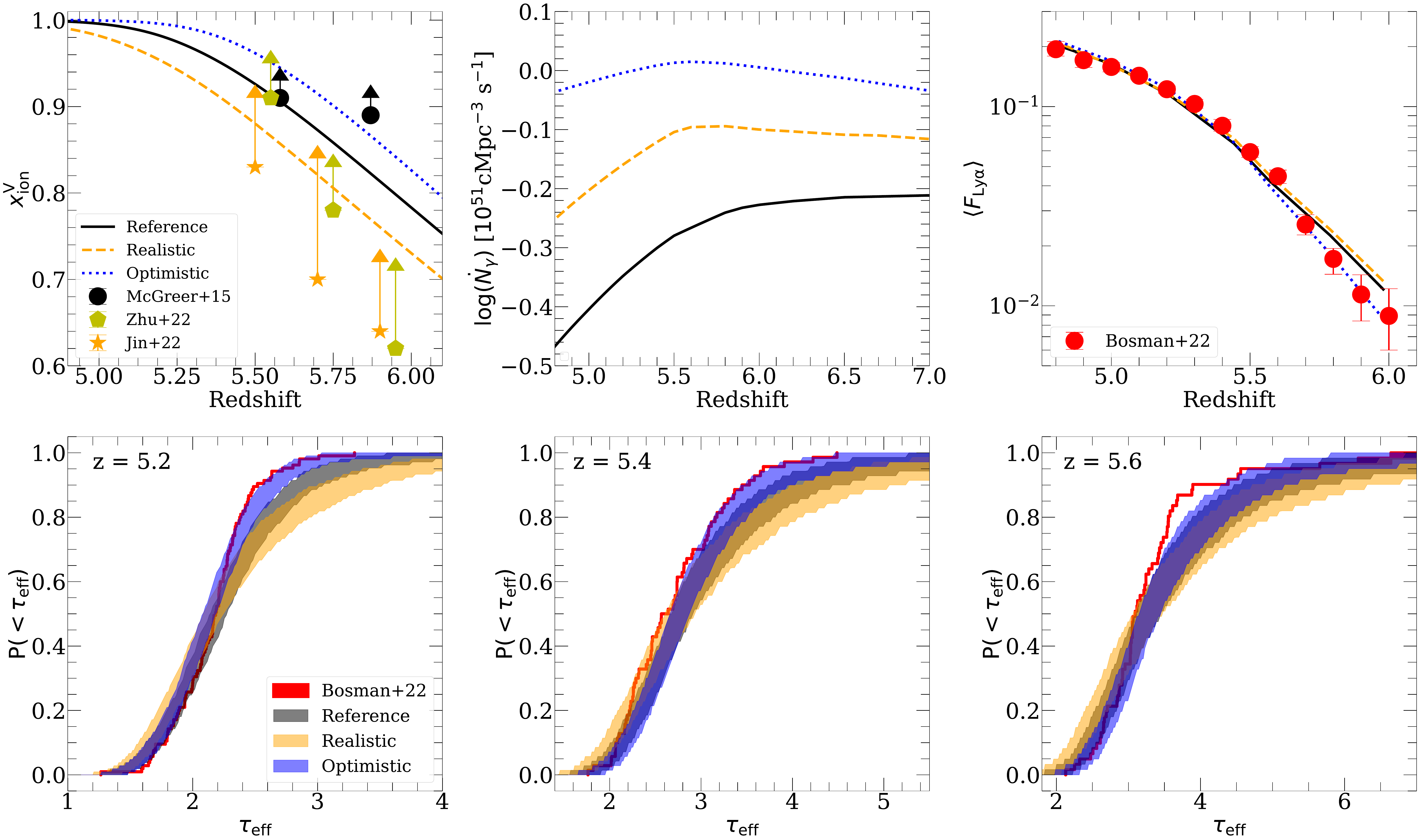}
    \caption{Comparison of Optimistic (blue dotted) and Realistic (orange dashed) models. The top row has the same format as Figure~\ref{fig:ref_vs_sinks}, while the row shows the $\tau_{\rm eff}$ distribution at $z = 5.2$, $5.4$, and $5.6$.  The former includes only effects that make the drop in $\dot{N}_{\gamma}$ smaller and/or the reionization history end earlier.  The Realistic model includes ``intermediate'' treatments of the effects discussed in \S\ref{subsec:ionizingspectrum}-\S\ref{subsec:missingsinks}.   The Optimistic model has almost no drop in $\dot{N}_{\gamma}$ and ends reionization early enough to agree well with the measured $\tau_{\rm eff}$ CDF.  However, even a flat $\dot{N}_{\gamma}$ implies significant evolution of source properties.  Our realistic model requires some drop and ends reionization later than suggested by the $\tau_{\rm eff}$ CDF.  }
    \label{fig:ref_vs_real}
\end{figure*}

In this section, we combine the effects studied in \S\ref{sec:results_IGM}.  Table~\ref{tab:effects} lists the physical effects we studied and describes how each changes the drop in $\dot{N}_{\gamma}$ and the reionization history when $\langle F_{\rm Ly\alpha} \rangle$ is held fixed.  Green text denotes a shallower drop in $\dot{N}_{\gamma}$ or an earlier reionization history, and red the opposite.  We consider two models in this section: a ``Realistic'' model and an ``Optimistic'', shown in Figure~\ref{fig:ref_vs_real} by the orange dashed and blue dotted curves, respectively.  The top row has the same format as Figure~\ref{fig:ref_vs_sinks}, while the bottom row shows the $\tau_{\rm eff}$ CDF at $z = 5.2$, $5.4$, and $5.6$.  

\begin{table}
    \begin{center}
        \begin{tabular}{| c | c c |}
        \hline
        \hline
        Physical Effect & Emissivity Drop & End of Reionization \\ 
        \hline
        \hline
        Harder ionizing spectrum & \textcolor{orange}{$<10\%$ difference} & \textcolor{green}{earlier} \\ \\
        Recombination radiation & \textcolor{orange}{$<10\%$ difference} & \textcolor{red}{later}  \\ \\
        Missing small sinks & \textcolor{red}{steeper} & \textcolor{red}{later} \\ \\
        Missing massive sinks & \textcolor{green}{shallower} & \textcolor{orange}{$\Delta z < 0.1$}\\
        \hline
        \hline
        \end{tabular}
        \end{center}
    \caption{Summary of the modeling considerations studied in \S\ref{sec:results_IGM} and their effects on $\dot{N}_{\gamma}$ and the reionization history.  The second and third columns describe their effect on the drop in $\dot{N}_{\gamma}$ and the reionization history. 
 }
    \label{tab:effects}
\end{table}

The Optimistic model is a combination of our Massive Sinks and Hard Multi-frequency model, and ignores the effect of un-resolved sinks and recombination radiation. This model has almost no drop and an early end to reionization, the latter of which results in good agreement with the measured $\tau_{\rm eff}$ distribution.  However, even this model would require some decline in the ionizing output of galaxies during reionization.  The~\citet{Robertson2015} emissivity, which assumes constant $f_{\rm esc}$ and $\xi_{\rm ion}$, increases by $\sim 0.4$ dex between $z = 8$ and $5$ (Figure~\ref{fig:em_lit}).  So, $\langle f_{\rm esc} \xi_{\rm ion}\rangle$ would have to fall by a similar factor over this redshift range to produce the Optimistic $\dot{N}_{\gamma}$. We note that the Optimistic model has a higher ionizing budget -- $3.2$ photons per H atom -- than the reference model ($2.2$ photons/H atom).  

\begin{figure}
    \centering
    \includegraphics[scale=0.2]{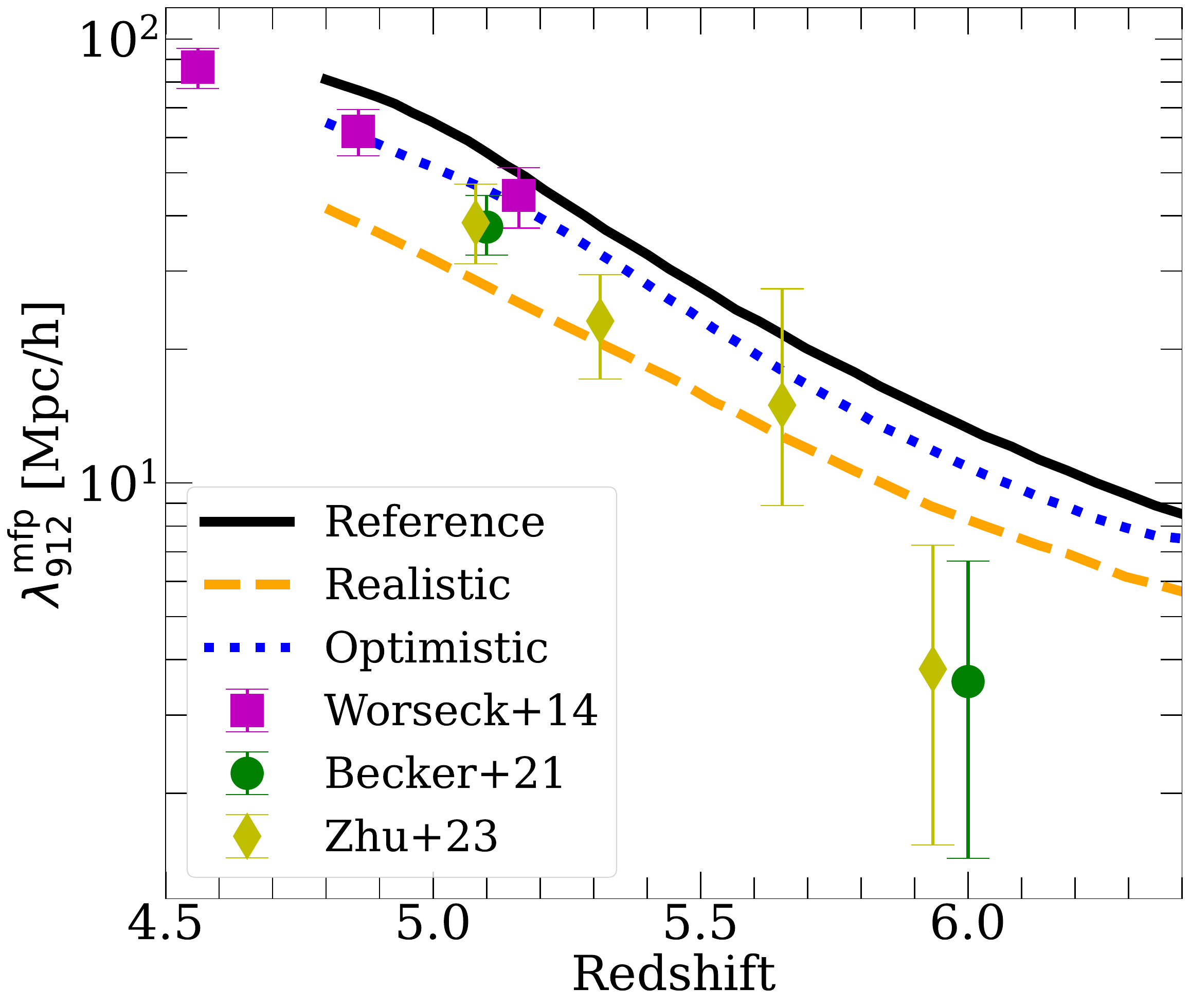}
    \caption{Ionizing photon MFP for the Reference, Optimistic, and Realistic models.  The Optimistic model has a  shorter MFP than the Reference case, and is actually in slightly better agreement with measurements at $z \lesssim 5.5$.  The MFP in the Realistic model is too short for measurements, mainly due to recombination radiation and the inclusion of the smallest sinks.  }
    \label{fig:mfp_plot}
\end{figure}

Our Realistic model includes reasonable prescriptions for all the effects in~\ref{tab:effects}.  We include multi-frequency RT as described in \S\ref{subsec:multifreq}, but keep our fiducial $\alpha = 1.5$ spectrum.  We assume $f_{\rm esc}^{\rm rec} = 0.5$ for recombination radiation, in between case A and B.  We use our Full Sinks sub-grid model to account for the smallest sinks, and include our halo absorber model, but with $\epsilon = 2$ (isothermal density profiles).  The last choice reduces the extra opacity from halos (see Figure~\ref{fig:f200} of Appendix~\ref{app:winds}).  This model has an $\dot{N}_{\gamma}$ between those of the Reference and Optimistic models -- flat until $z = 5.5$, then dropping $\sim 0.15$ dex by $z = 4.8$, suggesting that realistic IGM models will likely require a drop.  However, the details of the drop are sensitive to IGM modeling effects.  This model also ends reionization later than the reference model and displays worse agreement with the $\tau_{\rm eff}$ CDF (similar to that of the Recombination Radiation model in \S\ref{subsec:recradresults}).  This owes to the inclusion of recombination radiation and small sinks (see Figures~\ref{fig:ref_vs_recomb} and~\ref{fig:ref_vs_sinks}).  This may hint that recombination radiation and/or small sinks play a smaller role in reality than they do in our Realistic model -- although see the next section for an alternative solution.  

As mentioned in \S\ref{subsubsec:galphys}, an important check for models that resolve the drop with extra sinks is that they agree with measurements of the MFP.  Figure~\ref{fig:mfp_plot} shows the MFP for our Reference, Optimistic, and Realistic models.  We see that the Optimistic model has a lower MFP than the Reference case, but that it actually agrees better with $z < 5.5$ measurements.  This shows  that it is possible to alleviate the emissivity drop with massive sinks without undershooting MFP measurements.  However, the Realistic model undershoots measurements at $z \leq 5.1$, due to its inclusion of recombination radiation and the smallest sinks\footnote{The former increases the opacity of the gas by a factor of $\sim f_{\rm esc}^{\rm rec}\frac{\alpha_{\rm A} - \alpha_{\rm B}}{\alpha_{\rm B}}  + 1$ (Eq.~\ref{eq:kappa_rec}), but does so without affecting the ionizing photon budget, since IRPs compensate for the extra absorption (Eq.~\ref{eq:ngam_rec})}.   This result highlights the fact that the relationship between the MFP and the forest is sensitive to the effects studied in this work.  

An important caveat is that we have not exhaustively treated everything that could affect the $\dot{N}_{\gamma}$-forest relationship.  For example, the slope of the HI column density distribution, $\beta_{\rm N}$ (Eq.~\ref{eq:betaN}), is not a free parameter in our model, being inferred from our sub-grid simulations.  Treating this as a free parameter would widen the range of possible sinks models.  We also have not fully explored the parameter space of our IGM prescriptions.  For example, our halo density profile parameter $\epsilon$ could have time-dependence, and so could $f_{\rm esc}^{\rm rec}$ or the spectral index $\alpha$.  We will study these points in more detail in a forthcoming paper.  

\subsection{Clustering of the ionizing sources}
\label{subsec:ionsource}

The last physical effect we will study is that of the clustering of ionizing sources (see \S\ref{subsec:source_clustering}).  We will consider three sources models (all assuming our Realistic IGM model): 
\begin{itemize}
    \item Fiducial Sources: This model, which we have assumed so far, has $\beta = 1$ (that is, $\dot{n}_{\gamma} \propto L_{\rm UV}$), and includes halos down to the completeness limit of our N-body simulation, $M_{\rm halo} > 3 \times 10^{9}$ $h^{-1}M_{\odot}$.  
    \item Democratic Sources: We assume $\beta = 0$, which assigns all halos the same ionizing emissivity, independent of $L_{\rm UV}$.  We also include all the halos in the N-body simulation even below the completeness limit\footnote{The smallest halo identified in our N-body simulation has $M_{\rm halo} = 7 \times 10^{8}$ $h^{-1}M_{\odot}$.  Halos below the completeness limit sometimes appear and disappear in subsequent snapshots.  This behavior is un-physical, though it may mimic to some extent the bursty nature of star formation and associated ionizing photon production in low-mass halos~\citep{Emami2019,Dome2023}.  Since our goal is to qualitatively assess the effect of including fainter sources, so this is not a significant concern.  }.  This model maximizes the contribution of the faintest, least clustered ionizing sources to the ionizing budget.  
    \item Oligarchic Sources: our last model assumes $\beta = 1$ and that only $M_{\rm halo}> 2\times 10^{10}$ $h^{-1}M_{\odot}$ halos host ionizing sources, such that only the rarest and most clustered sources contribute to the ionizing budget.  This model is motivated by the bright-galaxy driven models proposed by~\citet{Naidu2020,Matthee2022} and contrasts the Democratic model.  
\end{itemize}

\begin{figure*}
    \centering
    \includegraphics[scale=0.15]{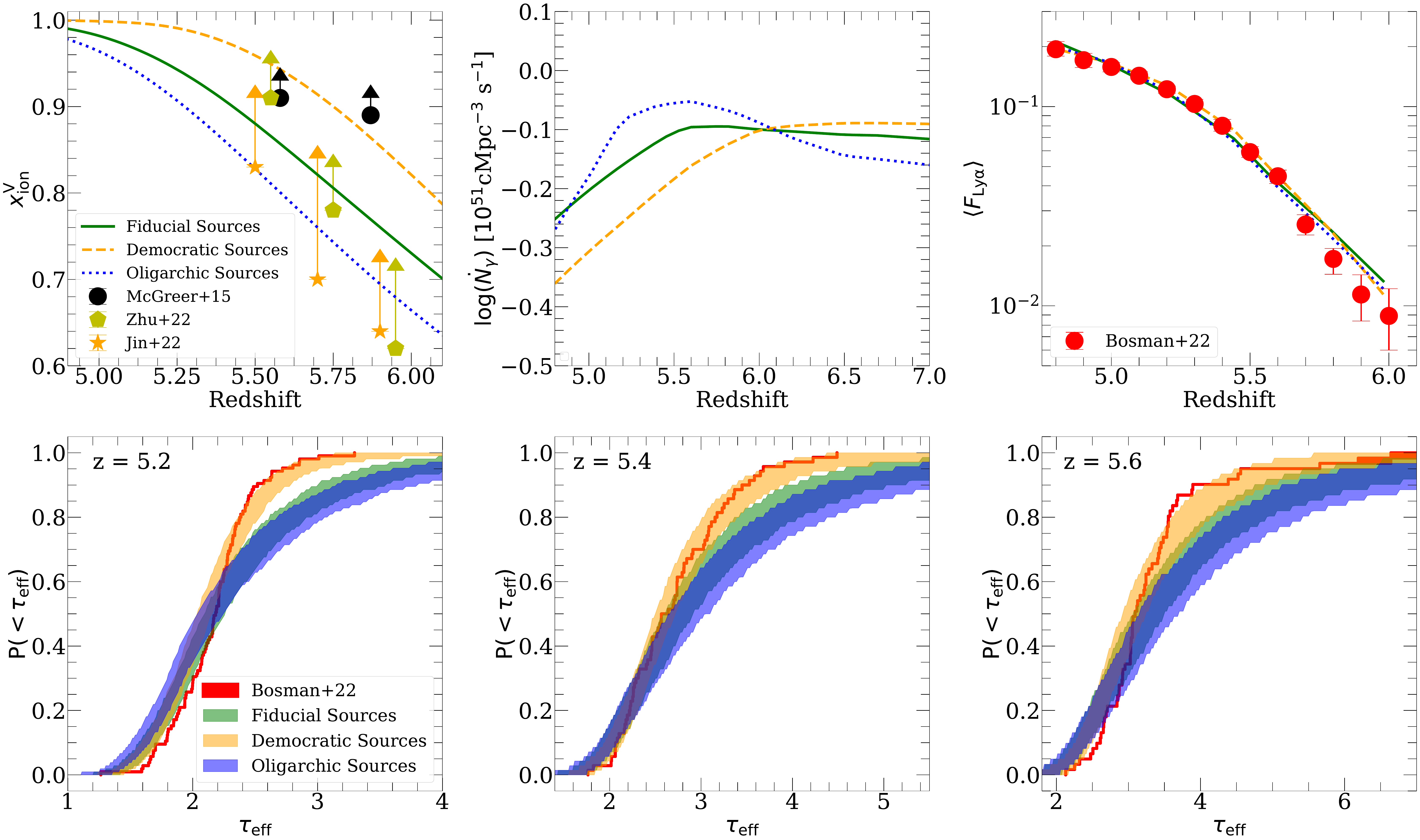}
    \caption{Effect of source clustering on the $\dot{N}_{\gamma}$-forest relationship, in the same format as Figure~\ref{fig:ref_vs_unconv}.  The orange dashed and blue dotted lines compare our Democratic and Oligarchic Sources models to our Fiducial Sources model (green solid).  All three assume the Realistic IGM model from the previous section.  The Democratic (Oligarchic) models end reionization earlier (later) than the Fiducial case and agree better (worse) with $\tau_{\rm eff}$ measurements.  The latter is in some tension with recent neutral fraction constraints from dark gaps and dark pixels.  The Democratic model requires a greater drop in $\dot{N}_{\gamma}$, while $\dot{N}_{\gamma}$ for the Oligarchic model at $5 < z < 6$ is non-monotonic and seems ``artificial''.  }
    \label{fig:compare_sources}
\end{figure*}

Figure~\ref{fig:compare_sources} shows our source models in the same format as Figure~\ref{fig:ref_vs_unconv}.  These models have significantly different reionization histories, with the Democratic (Oligarchic) sources model ending reionization $\Delta z \sim 0.2$ earlier ($\Delta z \sim 0.15$ later) than the Fiducial case.  The Democratic Sources model agrees much better with the $\tau_{\rm eff}$ CDF than the Fiducial or Oligarchic models.  This is because during reionization, this model is less transmissive in biased regions close to sources than the other models at fixed neutral fraction.  So, reionization is allowed to end sooner without overshooting measurements.  However, this model has a larger drop in $\dot{N}_{\gamma}$ than the Fiducial Sources case ($\sim 0.25$ vs. $\sim 0.15$ dex).  The earlier end to reionization in this case leads to a faster buildup of photons in the IGM by $z \sim 5$, and compensating for this requires a larger drop in $\dot{N}_{\gamma}$.   

The Oligarchic model is in significant tension with the $\tau_{\rm eff}$ CDF and some of the neutral fraction constraints from dark gaps/pixels.   Up to $z = 6.5$, $\dot{N}_{\gamma}$ is flat, then increases by $0.1$ dex between $z = 6.5$ and $5.5$ and then rapidly decreases by $0.2$ dex after $5.5$.  This odd behavior is necessary to match $\langle F_{\rm \alpha}\rangle$ measurements.  Taken together with the very high neutral fraction, this ``artificial'' behavior of $\dot{N}_{\gamma}$ seems to suggest that the forest disfavors the Oligarchic Sources scenario.  However, we caution there is at least partial degeneracy between the effects of the source model and the assumed IGM parameters.  

\section{Numerical Effects}
\label{sec:num_effects}

In this section, we will briefly study two numerical effects that affect the $\dot{N}_{\gamma}$-forest relationship - the reduced speed of light approximation and the numerical convergence of the Ly$\alpha$ forest itself.  

\subsection{Reduced Speed of Light Approximation}
\label{subsec:clight}

A self-consistent numerical treatment of reionization requires solving the RT equation.  The maximum time step required to resolve the transport of ionizing radiation in the Eulerian frame is the light-crossing time of the spatial resolution element, $\Delta x_{\rm cell}/c$, where $\Delta x_{\rm cell}$ is the RT cell size and $c$ is the speed of light.  This time scale is usually much shorter than any other timescales in the problem.  Because of this, a commonly used time-saving approach is to decrease the speed of light in the simulation, which increases the time step without compromising the accuracy.  This is called the reduced speed of light approximation (RSLA), and it has been used in a number of reionization studies~\citep[e.g.][]{Katz2017,DAloisio2020,Kannan2022}.  However, the RSLA can cause inaccuracies in large-scale reionization simulations.  Using moment-based M1 RT\footnote{\cite{Wu2021} showed that moment-based RT may itself lead to inaccuracies in the post-reionization ionizing background.  However, based on the arguments in this section, we believe that their result should hold for all RT methods.  }, \citet{Ocvirk2019} demonstrated that the RSLA can lead to a significant under-estimate of the ionizing background near the end of reionization, even if the reionization history is converged\footnote{A related effect is that the RSLA leads to under-estimates of the speed of I-fronts.  This is studied in~\citet{Deparis2019}.  }.  
  
\begin{figure}
    \centering
    \includegraphics[scale=0.22]{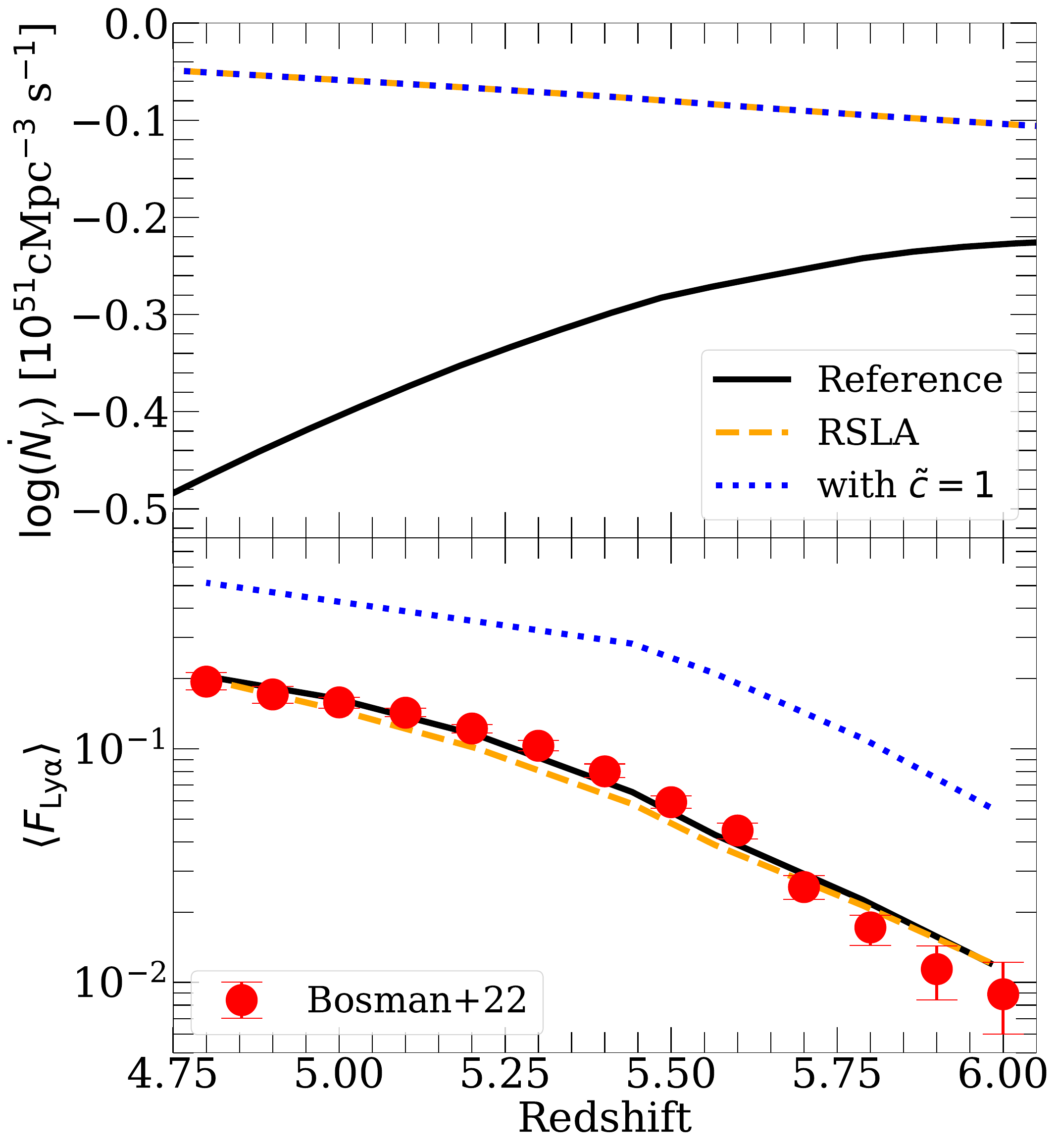}
    \caption{Effect of the RSLA on the $\dot{N}_{\gamma}$-$\langle F_{\rm Ly\alpha}\rangle$ relationship.  The orange dashed curve shows a model calibrated to match the forest, but with $\tilde{c} = 0.2c$.  This model requires no drop in $\dot{N}_{\gamma}$.  The blue dotted curve assumes the same $\dot{N}_{\gamma}$ but sets $\tilde{c} = 1$, causing $\langle F_{\rm Ly\alpha}\rangle$ to overshoot measurements considerably.  }
    \label{fig:compare_redc}
\end{figure}

Figure~\ref{fig:compare_redc} compares two models to the reference case (in the same format as Figure~\ref{fig:ref_vs_recomb}).  The first (the ``RSLA'' model, orange dashed) uses $\tilde{c} = 0.2$ (as in the recent THESAN simulations~\citep{Kannan2022}) and has $\dot{N}_{\gamma}$ re-calibrated to match $\langle F_{\rm Ly\alpha}\rangle$ measurements.  The RSLA model has a similar reionization history, thermal history, and MFP as the reference case.  Strikingly, it has no drop in $\dot{N}_{\gamma}$.  The blue dotted curve shows what happens when we use the $\dot{N}_{\gamma}$ from the RSLA model, but set $\tilde{c} = 1$.  This case recovers the expected result - an overshoot of $\langle F_{\rm Ly\alpha}\rangle$ by a factor of $2.5-6$.  We see that using the RSLA can spuriously lead to the conclusion that no drop is needed.  

The reason for this is as follows.  Early in reionization, $\Gamma_{\rm HI}$ in ionized bubbles is independent of $\tilde{c}$.  When the IGM is still a collection of isolated ionized bubbles, the photon density $N_{\gamma}$ in these bubbles is over-estimated by a factor of $c/\tilde{c}$.  Since $\Gamma_{\rm HI} \propto N_{\gamma}c$ in ionized gas, $\Gamma_{\rm HI}$ is independent of $\tilde{c}$.  Later on, as ionized bubbles overlap, $N_{\gamma}$ becomes similar for simulations with different $\tilde{c}$, and $\Gamma_{\rm HI}$ is under-estimated by a factor of $\approx \tilde{c}/c$. 
 This behavior is illustrated in Figure 3 of~\citet{Ocvirk2019}.  The transition between these regimes happens in the last half of reionization, when $\Gamma_{\rm HI}$ is increasing, such that the RSLA artificially slows down the growth of $\Gamma_{\rm HI}$.  

 In our RSLA model, the slower growth of $\Gamma_{\rm HI}$ demands a larger $\dot{N}_{\gamma}$ to reproduce the same Ly$\alpha$ forest properties as the reference case.  As a result, instead of declining at $z < 6.5$, $\dot{N}_{\gamma}$ continues to increase.  This is a possible explanation for why the THESAN simulations were able to able to reproduce forest observations reasonably well without any drop in $\dot{N}_{\gamma}$.  However, it is also possible that their treatment of galaxies played a role in reducing the drop, as illustrated in \S\ref{subsubsec:galphys} and \S\ref{subsec:realistic}.  They also use a different RT method than ours (moment-based vs. ray tracing).  It is therefore difficult to assess exactly how large this effect is in THESAN, although Figure~\ref{fig:compare_redc} suggests that it is likely significant.  

\subsection{Numerical Convergence of the Ly$\alpha$ Forest}
\label{subsec:convresult}

Convergence of the forest itself could also affect the $\dot{N}_{\gamma}$-forest relationship.  At $5 < z < 6$, even the mean density IGM is mostly opaque to Ly$\alpha$ photons, so the mean transmission is set under-dense gas.  Capturing the distribution of under-densities to which the forest is sensitive requires high spatial resolution.  It also requires a fairly large volume to include a representative distribution of large-scale voids, where under-dense gas is more common.   Simulations that fail to meet one or both of these requirements under-estimate the mean transmission of the forest.  Recently, \citet{Doughty2023} showed that box sizes of $\geq 20$ $h^{-1}$Mpc and spatial resolution of $\leq 10$ $h^{-1}$kpc are necessary to converge on the mean transmission of the forest in uniform grid simulations.  The requirements are likely more stringent for SPH/adaptive mesh simulations, which have worse resolution in under-densities~\citep{Bolton2009}.  

\begin{figure*}
    \centering
    \includegraphics[scale=0.16]{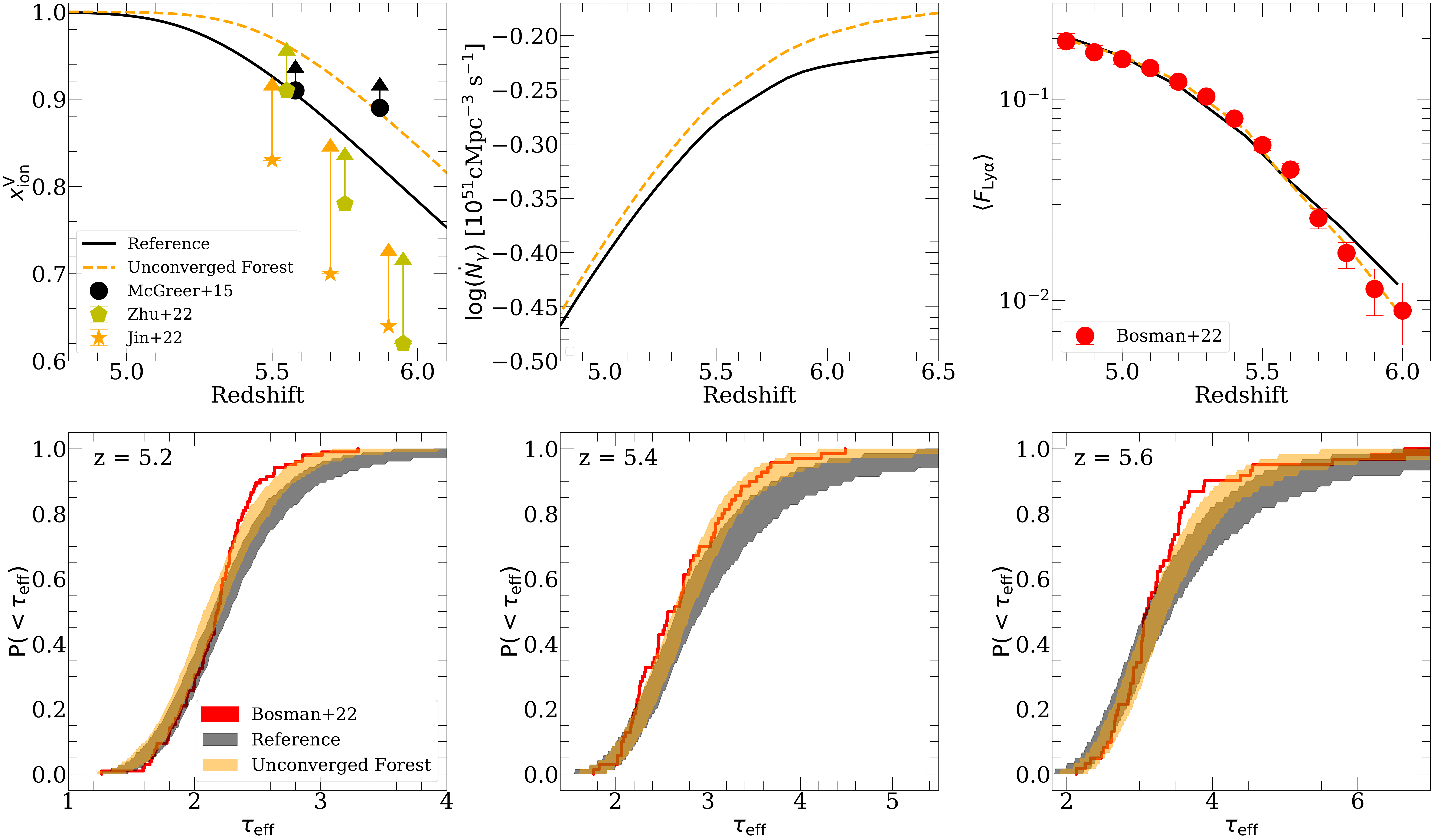}
    \caption{Effect of an under-resolved Ly$\alpha$ forest.  The orange dashed curve shows a model where we remove the spatial resolution correction applied to our forest calculations and re-calibrate to match measurements.  The top row shares the same format as Figure~\ref{fig:ref_vs_sinks}, and the bottom row shows the $\tau_{\rm eff}$ at $z = 5.2$, $5.4$, and $5.6$ (left to right).  This model ends reionization earlier, has a slightly steeper drop in $\dot{N}_{\gamma}$, and agrees better with the measured $\tau_{\rm eff}$ CDF than does the reference model.  This shows that in self-consistent simulations, an un-resolved Ly$\alpha$ forest can bias conclusions about reionization.  }
    \label{fig:ref_vs_unconv}
\end{figure*}

To assess the effect of forest convergence, we calibrated a model like our reference case, but without the spatial resolution correction for the Ly$\alpha$ forest described in \S\ref{subsec:lyaforest}.  This reduced our effective cell size in the forest calculation to the hydro resolution, $\Delta x = 97.6$ $h^{-1}$kpc.  Figure~\ref{fig:ref_vs_unconv} compares this ``Unconverged forest'' model to our reference case, in the same format as Figure~\ref{fig:ref_vs_real}.  At fixed $\langle F_{\rm Ly\alpha}\rangle$, the un-converged model ends reionization $\Delta z \sim 0.2$ earlier than the reference case, displays better overall agreement with $\tau_{\rm eff}$ measurements, and has a slightly steeper drop in $\dot{N}_{\gamma}$ (by $\sim 0.05$ dex).  This is because $\langle F_{\rm Ly\alpha}\rangle$ is under-estimated in this model, allowing for a larger $\Gamma_{\rm HI}$ at fixed $\langle F_{\rm Ly\alpha}\rangle$.  This in turn allows $\dot{N}_{\gamma}$ to be higher and reionization to end sooner.  

This result, like the previous one, is a cautionary tale for modeling the Ly$\alpha$ forest.  Simulations that are calibrated to match measurements of the mean forest transmission may come to incorrect conclusions if they are under-resolved in the forest.  In our case, this caused accidentally good agreement with $\tau_{\rm eff}$ measurements because the reionization history was allowed to end earlier than it did when the forest was converged.  We caution that this is only an issue for simulation that self-consistently model the interplay between $\dot{N}_{\gamma}$, the reionization history, and $\langle F_{\rm Ly\alpha}\rangle$.  We emphasize that these differences arise primarily because the neutral fractions in the models are different.  At fixed neutral fraction {\it and} $\langle F_{\rm Ly\alpha}\rangle$, we find little difference in the shape of the $\tau_{\rm eff}$ distribution between models.  

\section{Conclusions}
\label{sec:conc}

In this work, we have studied the implications of Ly$\alpha$ forest measurements at $5 < z < 6$ for the evolution of the ionizing emissivity at the end of reionization.  Several recent works have found that matching these measurements requires a drop in emissivity in this redshift range, requiring strong evolution in the ionizing properties of galaxies.  This work investigated the possibility that this drop is an artifact of inadequate modeling of the IGM in reionization simulations.  We have done this by comparing reionization simulations run with our radiative transfer code, FlexRT~\citep[][]{Cain2021,Cain2022b} with different IGM modeling assumptions.  Our main conclusions can be summarized as follows: 
\begin{itemize}

    \item At fixed forest transmission, $\dot{N}_{\gamma}$ is fairly insensitive to the hardness of the ionizing spectrum.  Models with hard ionizing spectra ($\alpha \leq 0.5$) result in slightly shallower (by $\sim 0.05$ dex) drops and earlier reionization histories, the latter yielding better agreement with the distribution of Ly$\alpha$ opacities ($\tau_{\rm eff}$).  However, such models also yield higher IGM temperatures than suggested by measurements.  

    \item Accounting for ionizing recombination radiation also has a small effect on the behavior of $\dot{N}_{\gamma}$.  However, the ``spectral softening'' of the radiation field by recombination photons requires a later reionization history to match the forest transmission, resulting in worse agreement with the $\tau_{\rm eff}$ distribution.  

    \item Accounting for the effect of the small, hard-to-resolve ionizing photon sinks on the IGM opacity requires a slightly steeper drop in $\dot{N}_{\gamma}$ and pushes the end of reionization later.  These sinks increase the reionization photon budget significantly, but they are destroyed on a timescale of a few hundred Myr after ionization.  Because these sinks do not survive long after reionization, they do not alleviate the need for a drop in $\dot{N}_{\gamma}$.  

    \item Unlike the small sinks, missing opacity from massive sinks may be able to explain the drop.  These structures are large enough to self-shield against the ionizing background, allowing them to survive past the end of reionization.  They also become much more numerous at lower redshifts.  So, they are able to regulate the ionizing background and in some scenarios eliminate the need for a drop.  

    \item Under optimistic assumptions about the properties of the IGM, agreement with the forest transmission can be achieved without a drop in $\dot{N}_{\gamma}$.  In this scenario, massive sinks contribute significantly to the IGM opacity, the ionizing spectrum is very hard, and the smallest sinks play a minimal role.  However, even models like this require $\dot{N}_{\gamma}$ to grow less steeply than the UVLF, suggesting some evolution in underlying source properties.  Models that include both small and massive sinks and a more reasonable ionizing spectrum still require some drop.  

    \item Scenarios where reionization is driven by lower-mass, less clustered ionizing sources yield earlier reionization histories (in better agreement with $\tau_{\rm eff}$) but also a steeper drop in $\dot{N}_{\gamma}$.  In models driven by bright sources, reionization may have to end later than recent measurements suggest to match forest transmission measurements.  

    \item The reduced speed of light approximation (RSLA) artificially blunts the growth of the ionizing background at the end of reionization.  This can erroneously lead to the conclusion that a drop in $\dot{N}_{\gamma}$ is not required to match forest observations.  

    \item Insufficient spatial resolution in numerical simulations can lead to an under-estimation of the Ly$\alpha$ forest mean transmission.  This can result in reionization ending earlier than it would otherwise. In some cases, this can cause spuriously good agreement with the measured $\tau_{\rm eff}$ distribution.  
    
\end{itemize}

We are led to conclude that either (1) the forest demands at least some evolution in the ionizing properties of sources near reionization's end, or (2) there are other effects setting the relationship between $\dot{N}_\gamma$ and the forest transmission that we have not considered here.  The first possibility would have important implications for models of the high-$z$ galaxy population. JWST has already begun to discover galaxies at $z \geq 6$ that may have exceptionally high ionizing efficiency~\citep{Atek2023,Cameron2023}, hinting at such evolution.  Improved modeling of the IGM will be key to helping to understand the implications of observations like these over the next few years.  Further work should also explore in more detail the possible mechanisms that may drive an evolution in source properties, which we do not address directly in this work.  

\section*{Acknowledgements}

 C.C. acknowledges helpful conversations with Yongda Zhu, George Becker, Sarah Bosman, Bayu Wilson, and Pierre Ocvirk.  C.C. also thanks Pierre Ocvirk for inspiring the title.  A.D.’s group is supported by NASA 19-ATP19-0191, NSF AST-2045600, and JWST-AR-02608.001-A. C.C. was supported by a combination of the above and the Beus Center for Cosmic Foundations while this work was ongoing.  All computations were made possible by NSF XSEDE allocations TG-PHY210041 and TG-PHY230063, and by the NASA HEC Program through the NAS Division at Ames Research Center.  

%%%%%%%%%%%%%%%%%%%%%%%%%%%%%%%%%%%%%%%%%%%%%%%%%%
\section*{Data Availability}

The data underlying this article will be shared upon reasonable request to the corresponding author.



\bibliographystyle{mnras}
\bibliography{references} 

\begin{thebibliography}{}
\makeatletter
\relax
\def\mn@urlcharsother{\let\do\@makeother \do\$\do\&\do\#\do\^\do\_\do\%\do\~}
\def\mn@doi{\begingroup\mn@urlcharsother \@ifnextchar [ {\mn@doi@}
  {\mn@doi@[]}}
\def\mn@doi@[#1]#2{\def\@tempa{#1}\ifx\@tempa\@empty \href
  {http://dx.doi.org/#2} {doi:#2}\else \href {http://dx.doi.org/#2} {#1}\fi
  \endgroup}
\def\mn@eprint#1#2{\mn@eprint@#1:#2::\@nil}
\def\mn@eprint@arXiv#1{\href {http://arxiv.org/abs/#1} {{\tt arXiv:#1}}}
\def\mn@eprint@dblp#1{\href {http://dblp.uni-trier.de/rec/bibtex/#1.xml}
  {dblp:#1}}
\def\mn@eprint@#1:#2:#3:#4\@nil{\def\@tempa {#1}\def\@tempb {#2}\def\@tempc
  {#3}\ifx \@tempc \@empty \let \@tempc \@tempb \let \@tempb \@tempa \fi \ifx
  \@tempb \@empty \def\@tempb {arXiv}\fi \@ifundefined
  {mn@eprint@\@tempb}{\@tempb:\@tempc}{\expandafter \expandafter \csname
  mn@eprint@\@tempb\endcsname \expandafter{\@tempc}}}

\bibitem[\protect\citeauthoryear{{Abel} \& {Wandelt}}{{Abel} \&
  {Wandelt}}{2002}]{Abel2002}
{Abel} T.,  {Wandelt} B.~D.,  2002, \mn@doi [\mnras]
  {10.1046/j.1365-8711.2002.05206.x}, \href
  {https://ui.adsabs.harvard.edu/abs/2002MNRAS.330L..53A} {330, L53}

\bibitem[\protect\citeauthoryear{{Adams} et~al.,}{{Adams}
  et~al.}{2023}]{Adams2023}
{Adams} N.~J.,  et~al., 2023, \mn@doi [arXiv e-prints]
  {10.48550/arXiv.2304.13721}, \href
  {https://ui.adsabs.harvard.edu/abs/2023arXiv230413721A} {p. arXiv:2304.13721}

\bibitem[\protect\citeauthoryear{{Altay}, {Theuns}, {Schaye}, {Crighton}  \&
  {Dalla Vecchia}}{{Altay} et~al.}{2011}]{Altay2011}
{Altay} G.,  {Theuns} T.,  {Schaye} J.,  {Crighton} N. H.~M.,   {Dalla Vecchia}
  C.,  2011, \mn@doi [\apjl] {10.1088/2041-8205/737/2/L37}, \href
  {https://ui.adsabs.harvard.edu/abs/2011ApJ...737L..37A} {737, L37}

\bibitem[\protect\citeauthoryear{{Atek}, {Furtak}, {Oesch}, {van Dokkum},
  {Reddy}, {Contini}, {Illingworth}  \& {Wilkins}}{{Atek}
  et~al.}{2022}]{Atek2022}
{Atek} H.,  {Furtak} L.~J.,  {Oesch} P.,  {van Dokkum} P.,  {Reddy} N.,
  {Contini} T.,  {Illingworth} G.,   {Wilkins} S.,  2022, \mn@doi [\mnras]
  {10.1093/mnras/stac360}, \href
  {https://ui.adsabs.harvard.edu/abs/2022MNRAS.511.4464A} {511, 4464}

\bibitem[\protect\citeauthoryear{{Atek} et~al.,}{{Atek}
  et~al.}{2023}]{Atek2023}
{Atek} H.,  et~al., 2023, \mn@doi [arXiv e-prints] {10.48550/arXiv.2308.08540},
  \href {https://ui.adsabs.harvard.edu/abs/2023arXiv230808540A} {p.
  arXiv:2308.08540}

\bibitem[\protect\citeauthoryear{{Becker} \& {Bolton}}{{Becker} \&
  {Bolton}}{2013}]{Becker2013}
{Becker} G.~D.,  {Bolton} J.~S.,  2013, \mn@doi [\mnras]
  {10.1093/mnras/stt1610}, \href
  {https://ui.adsabs.harvard.edu/abs/2013MNRAS.436.1023B} {436, 1023}

\bibitem[\protect\citeauthoryear{{Becker}, {Bolton}, {Haehnelt}  \&
  {Sargent}}{{Becker} et~al.}{2011}]{Becker2011}
{Becker} G.~D.,  {Bolton} J.~S.,  {Haehnelt} M.~G.,   {Sargent} W. L.~W.,
  2011, \mn@doi [\mnras] {10.1111/j.1365-2966.2010.17507.x}, \href
  {https://ui.adsabs.harvard.edu/abs/2011MNRAS.410.1096B} {410, 1096}

\bibitem[\protect\citeauthoryear{{Becker}, {Bolton}, {Madau}, {Pettini},
  {Ryan-Weber}  \& {Venemans}}{{Becker} et~al.}{2015}]{Becker2015}
{Becker} G.~D.,  {Bolton} J.~S.,  {Madau} P.,  {Pettini} M.,  {Ryan-Weber}
  E.~V.,   {Venemans} B.~P.,  2015, \mn@doi [MNRAS] {10.1093/mnras/stu2646},
  447, 3402

\bibitem[\protect\citeauthoryear{{Becker}, {D'Aloisio}, {Christenson}, {Zhu},
  {Worseck}  \& {Bolton}}{{Becker} et~al.}{2021}]{Becker2021}
{Becker} G.~D.,  {D'Aloisio} A.,  {Christenson} H.~M.,  {Zhu} Y.,  {Worseck}
  G.,   {Bolton} J.~S.,  2021, \mn@doi [\mnras] {10.1093/mnras/stab2696}, \href
  {https://ui.adsabs.harvard.edu/abs/2021MNRAS.508.1853B} {508, 1853}

\bibitem[\protect\citeauthoryear{{Begley} et~al.,}{{Begley}
  et~al.}{2022}]{Begley2022}
{Begley} R.,  et~al., 2022, arXiv e-prints, \href
  {https://ui.adsabs.harvard.edu/abs/2022arXiv220204088B} {p. arXiv:2202.04088}

\bibitem[\protect\citeauthoryear{{Boera}, {Becker}, {Bolton}  \&
  {Nasir}}{{Boera} et~al.}{2019}]{Boera2019}
{Boera} E.,  {Becker} G.~D.,  {Bolton} J.~S.,   {Nasir} F.,  2019, \mn@doi
  [\apj] {10.3847/1538-4357/aafee4}, \href
  {https://ui.adsabs.harvard.edu/abs/2019ApJ...872..101B} {872, 101}

\bibitem[\protect\citeauthoryear{{Bolton} \& {Becker}}{{Bolton} \&
  {Becker}}{2009}]{Bolton2009}
{Bolton} J.~S.,  {Becker} G.~D.,  2009, \mn@doi [\mnras]
  {10.1111/j.1745-3933.2009.00700.x}, \href
  {https://ui.adsabs.harvard.edu/abs/2009MNRAS.398L..26B} {398, L26}

\bibitem[\protect\citeauthoryear{{Bosman}, {Fan}, {Jiang}, {Reed}, {Matsuoka},
  {Becker}  \& {Haehnelt}}{{Bosman} et~al.}{2018}]{Bosman2018}
{Bosman} S. E.~I.,  {Fan} X.,  {Jiang} L.,  {Reed} S.,  {Matsuoka} Y.,
  {Becker} G.,   {Haehnelt} M.,  2018, \mn@doi [\mnras]
  {10.1093/mnras/sty1344}, 479, 1055

\bibitem[\protect\citeauthoryear{{Bosman} et~al.,}{{Bosman}
  et~al.}{2022}]{Bosman2021}
{Bosman} S. E.~I.,  et~al., 2022, \mn@doi [\mnras] {10.1093/mnras/stac1046},
  \href {https://ui.adsabs.harvard.edu/abs/2022MNRAS.514...55B} {514, 55}

\bibitem[\protect\citeauthoryear{{Bouwens}, {Illingworth}, {Oesch}, {Caruana},
  {Holwerda}, {Smit}  \& {Wilkins}}{{Bouwens} et~al.}{2015}]{Bouwens2015}
{Bouwens} R.~J.,  {Illingworth} G.~D.,  {Oesch} P.~A.,  {Caruana} J.,
  {Holwerda} B.,  {Smit} R.,   {Wilkins} S.,  2015, \mn@doi [\apj]
  {10.1088/0004-637X/811/2/140}, \href
  {https://ui.adsabs.harvard.edu/abs/2015ApJ...811..140B} {811, 140}

\bibitem[\protect\citeauthoryear{{Bouwens} et~al.,}{{Bouwens}
  et~al.}{2021}]{Bouwens2021}
{Bouwens} R.~J.,  et~al., 2021, \mn@doi [\aj] {10.3847/1538-3881/abf83e}, \href
  {https://ui.adsabs.harvard.edu/abs/2021AJ....162...47B} {162, 47}

\bibitem[\protect\citeauthoryear{{Bressan}, {Marigo}, {Girardi}, {Salasnich},
  {Dal Cero}, {Rubele}  \& {Nanni}}{{Bressan} et~al.}{2012}]{Bressan2012}
{Bressan} A.,  {Marigo} P.,  {Girardi} L.,  {Salasnich} B.,  {Dal Cero} C.,
  {Rubele} S.,   {Nanni} A.,  2012, \mn@doi [\mnras]
  {10.1111/j.1365-2966.2012.21948.x}, \href
  {https://ui.adsabs.harvard.edu/abs/2012MNRAS.427..127B} {427, 127}

\bibitem[\protect\citeauthoryear{{Cain}, {D'Aloisio}, {Gangolli}  \&
  {Becker}}{{Cain} et~al.}{2021}]{Cain2021}
{Cain} C.,  {D'Aloisio} A.,  {Gangolli} N.,   {Becker} G.~D.,  2021, \mn@doi
  [\apjl] {10.3847/2041-8213/ac1ace}, \href
  {https://ui.adsabs.harvard.edu/abs/2021ApJ...917L..37C} {917, L37}

\bibitem[\protect\citeauthoryear{{Cain}, {D'Aloisio}, {Gangolli}  \&
  {McQuinn}}{{Cain} et~al.}{2022}]{Cain2022b}
{Cain} C.,  {D'Aloisio} A.,  {Gangolli} N.,   {McQuinn} M.,  2022, \mn@doi
  [arXiv e-prints] {10.48550/arXiv.2207.11266}, \href
  {https://ui.adsabs.harvard.edu/abs/2022arXiv220711266C} {p. arXiv:2207.11266}

\bibitem[\protect\citeauthoryear{{Cameron}, {Katz}, {Witten}, {Saxena},
  {Laporte}  \& {Bunker}}{{Cameron} et~al.}{2023}]{Cameron2023}
{Cameron} A.~J.,  {Katz} H.,  {Witten} C.,  {Saxena} A.,  {Laporte} N.,
  {Bunker} A.~J.,  2023, \mn@doi [arXiv e-prints] {10.48550/arXiv.2311.02051},
  \href {https://ui.adsabs.harvard.edu/abs/2023arXiv231102051C} {p.
  arXiv:2311.02051}

\bibitem[\protect\citeauthoryear{{Chan}, {Benitez-Llambay}, {Theuns}, {Frenk}
  \& {Bower}}{{Chan} et~al.}{2023}]{Chan2023}
{Chan} T.~K.,  {Benitez-Llambay} A.,  {Theuns} T.,  {Frenk} C.,   {Bower} R.,
  2023, \mn@doi [arXiv e-prints] {10.48550/arXiv.2305.04959}, \href
  {https://ui.adsabs.harvard.edu/abs/2023arXiv230504959C} {p. arXiv:2305.04959}

\bibitem[\protect\citeauthoryear{Choi, Conroy  \& Byler}{Choi
  et~al.}{2017}]{Choi2017}
Choi J.,  Conroy C.,   Byler N.,  2017, \mn@doi [The Astrophysical Journal]
  {10.3847/1538-4357/aa679f}, 838, 159

\bibitem[\protect\citeauthoryear{{D'Aloisio}, {Upton Sanderbeck}, {McQuinn},
  {Trac}  \& {Shapiro}}{{D'Aloisio} et~al.}{2017}]{Daloisio2017}
{D'Aloisio} A.,  {Upton Sanderbeck} P.~R.,  {McQuinn} M.,  {Trac} H.,
  {Shapiro} P.~R.,  2017, \mn@doi [\mnras] {10.1093/mnras/stx711}, \href
  {https://ui.adsabs.harvard.edu/abs/2017MNRAS.468.4691D} {468, 4691}

\bibitem[\protect\citeauthoryear{{D'Aloisio}, {McQuinn}, {Davies}  \&
  {Furlanetto}}{{D'Aloisio} et~al.}{2018}]{DAloisio2018}
{D'Aloisio} A.,  {McQuinn} M.,  {Davies} F.~B.,   {Furlanetto} S.~R.,  2018,
  \mn@doi [\mnras] {10.1093/mnras/stx2341}, 473, 560

\bibitem[\protect\citeauthoryear{{D'Aloisio}, {McQuinn}, {Maupin}, {Davies},
  {Trac}, {Fuller}  \& {Upton Sanderbeck}}{{D'Aloisio}
  et~al.}{2019}]{DAloisio2019}
{D'Aloisio} A.,  {McQuinn} M.,  {Maupin} O.,  {Davies} F.~B.,  {Trac} H.,
  {Fuller} S.,   {Upton Sanderbeck} P.~R.,  2019, \mn@doi [\apj]
  {10.3847/1538-4357/ab0d83}, \href
  {https://ui.adsabs.harvard.edu/abs/2019ApJ...874..154D} {874, 154}

\bibitem[\protect\citeauthoryear{D'Aloisio, McQuinn, Trac, Cain  \&
  Mesinger}{D'Aloisio et~al.}{2020}]{DAloisio2020}
D'Aloisio A.,  McQuinn M.,  Trac H.,  Cain C.,   Mesinger A.,  2020, \mn@doi
  [The Astrophysical Journal] {10.3847/1538-4357/ab9f2f}, 898, 149

\bibitem[\protect\citeauthoryear{{D'Odorico} et~al.,}{{D'Odorico}
  et~al.}{2023}]{DOdorico2023}
{D'Odorico} V.,  et~al., 2023, \mn@doi [\mnras] {10.1093/mnras/stad1468}, \href
  {https://ui.adsabs.harvard.edu/abs/2023MNRAS.523.1399D} {523, 1399}

\bibitem[\protect\citeauthoryear{{Davies} \& {Furlanetto}}{{Davies} \&
  {Furlanetto}}{2016}]{Davies2016}
{Davies} F.~B.,  {Furlanetto} S.~R.,  2016, \mn@doi [\mnras]
  {10.1093/mnras/stw931}, 460, 1328

\bibitem[\protect\citeauthoryear{Davies et~al.,}{Davies
  et~al.}{2018}]{Davies2018}
Davies F.~B.,  et~al., 2018, \mn@doi [The Astrophysical Journal]
  {10.3847/1538-4357/aad6dc}, 864, 142

\bibitem[\protect\citeauthoryear{{Davies}, {Bosman}, {Furlanetto}, {Becker}  \&
  {D'Aloisio}}{{Davies} et~al.}{2021}]{Davies2021b}
{Davies} F.~B.,  {Bosman} S. E.~I.,  {Furlanetto} S.~R.,  {Becker} G.~D.,
  {D'Aloisio} A.,  2021, \mn@doi [\apjl] {10.3847/2041-8213/ac1ffb}, \href
  {https://ui.adsabs.harvard.edu/abs/2021ApJ...918L..35D} {918, L35}

\bibitem[\protect\citeauthoryear{{Deparis}, {Aubert}, {Ocvirk}, {Chardin}  \&
  {Lewis}}{{Deparis} et~al.}{2019}]{Deparis2019}
{Deparis} N.,  {Aubert} D.,  {Ocvirk} P.,  {Chardin} J.,   {Lewis} J.,  2019,
  \mn@doi [\aap] {10.1051/0004-6361/201832889}, \href
  {https://ui.adsabs.harvard.edu/abs/2019A&A...622A.142D} {622, A142}

\bibitem[\protect\citeauthoryear{{Dome}, {Tacchella}, {Fialkov}, {Dekel},
  {Ginzburg}, {Lapiner}  \& {Looser}}{{Dome} et~al.}{2023}]{Dome2023}
{Dome} T.,  {Tacchella} S.,  {Fialkov} A.,  {Dekel} A.,  {Ginzburg} O.,
  {Lapiner} S.,   {Looser} T.~J.,  2023, \mn@doi [arXiv e-prints]
  {10.48550/arXiv.2305.07066}, \href
  {https://ui.adsabs.harvard.edu/abs/2023arXiv230507066D} {p. arXiv:2305.07066}

\bibitem[\protect\citeauthoryear{{Doughty}, {Hennawi}, {Davies}, {Luki{\'c}}
  \& {O{\~n}orbe}}{{Doughty} et~al.}{2023}]{Doughty2023}
{Doughty} C.~C.,  {Hennawi} J.~F.,  {Davies} F.~B.,  {Luki{\'c}} Z.,
  {O{\~n}orbe} J.,  2023, \mn@doi [arXiv e-prints] {10.48550/arXiv.2305.16200},
  \href {https://ui.adsabs.harvard.edu/abs/2023arXiv230516200D} {p.
  arXiv:2305.16200}

\bibitem[\protect\citeauthoryear{{Eilers}, {Davies}  \& {Hennawi}}{{Eilers}
  et~al.}{2018}]{Eilers2018}
{Eilers} A.-C.,  {Davies} F.~B.,   {Hennawi} J.~F.,  2018, \mn@doi [\apj]
  {10.3847/1538-4357/aad4fd}, 864, 53

\bibitem[\protect\citeauthoryear{{Emami}, {Siana}, {Weisz}, {Johnson}, {Ma}  \&
  {El-Badry}}{{Emami} et~al.}{2019}]{Emami2019}
{Emami} N.,  {Siana} B.,  {Weisz} D.~R.,  {Johnson} B.~D.,  {Ma} X.,
  {El-Badry} K.,  2019, \mn@doi [\apj] {10.3847/1538-4357/ab211a}, \href
  {https://ui.adsabs.harvard.edu/abs/2019ApJ...881...71E} {881, 71}

\bibitem[\protect\citeauthoryear{Emberson, Thomas  \& Alvarez}{Emberson
  et~al.}{2013}]{Emberson2013}
Emberson J.~D.,  Thomas R.~M.,   Alvarez M.~A.,  2013, \mn@doi [The
  Astrophysical Journal] {10.1088/0004-637x/763/2/146}, 763, 146

\bibitem[\protect\citeauthoryear{{Endsley} \& {Stark}}{{Endsley} \&
  {Stark}}{2022}]{Endsley2022b}
{Endsley} R.,  {Stark} D.~P.,  2022, \mn@doi [\mnras] {10.1093/mnras/stac524},
  \href {https://ui.adsabs.harvard.edu/abs/2022MNRAS.511.6042E} {511, 6042}

\bibitem[\protect\citeauthoryear{{Fan} et~al.,}{{Fan} et~al.}{2006}]{Fan2006}
{Fan} X.,  et~al., 2006, \mn@doi [\aj] {10.1086/504836}, 132, 117

\bibitem[\protect\citeauthoryear{{Faucher-Gigu{\`e}re}, {Lidz}, {Zaldarriaga}
  \& {Hernquist}}{{Faucher-Gigu{\`e}re} et~al.}{2009a}]{Giguere2009}
{Faucher-Gigu{\`e}re} C.-A.,  {Lidz} A.,  {Zaldarriaga} M.,   {Hernquist} L.,
  2009a, \mn@doi [\apj] {10.1088/0004-637X/703/2/1416}, \href
  {https://ui.adsabs.harvard.edu/abs/2009ApJ...703.1416F} {703, 1416}

\bibitem[\protect\citeauthoryear{{Faucher-Gigu{\`e}re}, {Lidz}, {Zaldarriaga}
  \& {Hernquist}}{{Faucher-Gigu{\`e}re} et~al.}{2009b}]{FaucherGiguere2009}
{Faucher-Gigu{\`e}re} C.-A.,  {Lidz} A.,  {Zaldarriaga} M.,   {Hernquist} L.,
  2009b, \mn@doi [\apj] {10.1088/0004-637X/703/2/1416}, \href
  {https://ui.adsabs.harvard.edu/abs/2009ApJ...703.1416F} {703, 1416}

\bibitem[\protect\citeauthoryear{{Finkelstein} et~al.,}{{Finkelstein}
  et~al.}{2019}]{Finkelstein2019}
{Finkelstein} S.~L.,  et~al., 2019, \mn@doi [\apj] {10.3847/1538-4357/ab1ea8},
  \href {https://ui.adsabs.harvard.edu/abs/2019ApJ...879...36F} {879, 36}

\bibitem[\protect\citeauthoryear{{Finlator} et~al.,}{{Finlator}
  et~al.}{2017}]{Finlator2017}
{Finlator} K.,  et~al., 2017, \mn@doi [\mnras] {10.1093/mnras/stw2433}, \href
  {https://ui.adsabs.harvard.edu/abs/2017MNRAS.464.1633F} {464, 1633}

\bibitem[\protect\citeauthoryear{Furlanetto \& Oh}{Furlanetto \&
  Oh}{2005}]{Furlanetto2005}
Furlanetto S.~R.,  Oh S.~P.,  2005, \mn@doi [Monthly Notices of the Royal
  Astronomical Society] {10.1111/j.1365-2966.2005.09505.x}, 363, 1031

\bibitem[\protect\citeauthoryear{{Gaikwad} et~al.,}{{Gaikwad}
  et~al.}{2020}]{Gaikwad2020}
{Gaikwad} P.,  et~al., 2020, \mn@doi [\mnras] {10.1093/mnras/staa907}, \href
  {https://ui.adsabs.harvard.edu/abs/2020MNRAS.494.5091G} {494, 5091}

\bibitem[\protect\citeauthoryear{{Gaikwad} et~al.,}{{Gaikwad}
  et~al.}{2023}]{Gaikwad2023}
{Gaikwad} P.,  et~al., 2023, \mn@doi [arXiv e-prints]
  {10.48550/arXiv.2304.02038}, \href
  {https://ui.adsabs.harvard.edu/abs/2023arXiv230402038G} {p. arXiv:2304.02038}

\bibitem[\protect\citeauthoryear{{Garaldi}, {Kannan}, {Smith}, {Springel},
  {Pakmor}, {Vogelsberger}  \& {Hernquist}}{{Garaldi}
  et~al.}{2022}]{Garaldi2022}
{Garaldi} E.,  {Kannan} R.,  {Smith} A.,  {Springel} V.,  {Pakmor} R.,
  {Vogelsberger} M.,   {Hernquist} L.,  2022, \mn@doi [\mnras]
  {10.1093/mnras/stac257}, \href
  {https://ui.adsabs.harvard.edu/abs/2022MNRAS.tmp..401G} {}

\bibitem[\protect\citeauthoryear{{Georgakakis} et~al.,}{{Georgakakis}
  et~al.}{2015}]{Georgakakis2015}
{Georgakakis} A.,  et~al., 2015, \mn@doi [\mnras] {10.1093/mnras/stv1703},
  \href {https://ui.adsabs.harvard.edu/abs/2015MNRAS.453.1946G} {453, 1946}

\bibitem[\protect\citeauthoryear{{Gnedin}, {Kravtsov}  \& {Rudd}}{{Gnedin}
  et~al.}{2011}]{Gnedin2011}
{Gnedin} N.~Y.,  {Kravtsov} A.~V.,   {Rudd} D.~H.,  2011, \mn@doi [\apjs]
  {10.1088/0067-0049/194/2/46}, 194, 46

\bibitem[\protect\citeauthoryear{{Gorski}, {Wandelt}, {Hansen}, {Hivon}  \&
  {Banday}}{{Gorski} et~al.}{1999}]{Gorski1999}
{Gorski} K.~M.,  {Wandelt} B.~D.,  {Hansen} F.~K.,  {Hivon} E.,   {Banday}
  A.~J.,  1999, arXiv e-prints, \href
  {https://ui.adsabs.harvard.edu/abs/1999astro.ph..5275G} {pp
  astro--ph/9905275}

\bibitem[\protect\citeauthoryear{Greig, Mesinger, Haiman  \& Simcoe}{Greig
  et~al.}{2016}]{Grieg2016}
Greig B.,  Mesinger A.,  Haiman Z.,   Simcoe R.~A.,  2016, \mn@doi [Monthly
  Notices of the Royal Astronomical Society] {10.1093/mnras/stw3351}, 466, 4239

\bibitem[\protect\citeauthoryear{{Greig}, {Mesinger}  \& {Ba{\~n}ados}}{{Greig}
  et~al.}{2019}]{Grieg2019}
{Greig} B.,  {Mesinger} A.,   {Ba{\~n}ados} E.,  2019, \mn@doi [\mnras]
  {10.1093/mnras/stz230}, \href
  {https://ui.adsabs.harvard.edu/abs/2019MNRAS.484.5094G} {484, 5094}

\bibitem[\protect\citeauthoryear{{Haardt} \& {Madau}}{{Haardt} \&
  {Madau}}{2012}]{Haardt2012}
{Haardt} F.,  {Madau} P.,  2012, \mn@doi [\apj] {10.1088/0004-637X/746/2/125},
  \href {https://ui.adsabs.harvard.edu/abs/2012ApJ...746..125H} {746, 125}

\bibitem[\protect\citeauthoryear{{Hu} et~al.,}{{Hu} et~al.}{2019}]{Hu2019}
{Hu} W.,  et~al., 2019, \mn@doi [\apj] {10.3847/1538-4357/ab4cf4}, \href
  {https://ui.adsabs.harvard.edu/abs/2019ApJ...886...90H} {886, 90}

\bibitem[\protect\citeauthoryear{Iliev, Shapiro  \& Raga}{Iliev
  et~al.}{2005}]{Iliev2005}
Iliev I.~T.,  Shapiro P.~R.,   Raga A.~C.,  2005, \mn@doi [Monthly Notices of
  the Royal Astronomical Society] {10.1111/j.1365-2966.2005.09155.x}, 361, 405

\bibitem[\protect\citeauthoryear{{Jin} et~al.,}{{Jin} et~al.}{2023}]{Jin2023}
{Jin} X.,  et~al., 2023, \mn@doi [\apj] {10.3847/1538-4357/aca678}, \href
  {https://ui.adsabs.harvard.edu/abs/2023ApJ...942...59J} {942, 59}

\bibitem[\protect\citeauthoryear{{Kannan}, {Vogelsberger}, {Marinacci},
  {McKinnon}, {Pakmor}  \& {Springel}}{{Kannan} et~al.}{2019}]{Kannan2019}
{Kannan} R.,  {Vogelsberger} M.,  {Marinacci} F.,  {McKinnon} R.,  {Pakmor} R.,
    {Springel} V.,  2019, \mn@doi [\mnras] {10.1093/mnras/stz287}, \href
  {https://ui.adsabs.harvard.edu/abs/2019MNRAS.485..117K} {485, 117}

\bibitem[\protect\citeauthoryear{{Kannan}, {Garaldi}, {Smith}, {Pakmor},
  {Springel}, {Vogelsberger}  \& {Hernquist}}{{Kannan}
  et~al.}{2022}]{Kannan2022}
{Kannan} R.,  {Garaldi} E.,  {Smith} A.,  {Pakmor} R.,  {Springel} V.,
  {Vogelsberger} M.,   {Hernquist} L.,  2022, \mn@doi [\mnras]
  {10.1093/mnras/stab3710}, \href
  {https://ui.adsabs.harvard.edu/abs/2022MNRAS.511.4005K} {511, 4005}

\bibitem[\protect\citeauthoryear{{Katz}, {Kimm}, {Sijacki}  \&
  {Haehnelt}}{{Katz} et~al.}{2017}]{Katz2017}
{Katz} H.,  {Kimm} T.,  {Sijacki} D.,   {Haehnelt} M.~G.,  2017, \mn@doi
  [\mnras] {10.1093/mnras/stx608}, \href
  {https://ui.adsabs.harvard.edu/abs/2017MNRAS.468.4831K} {468, 4831}

\bibitem[\protect\citeauthoryear{{Keating}, {Weinberger}, {Kulkarni},
  {Haehnelt}, {Chardin}  \& {Aubert}}{{Keating} et~al.}{2020a}]{Keating2019}
{Keating} L.~C.,  {Weinberger} L.~H.,  {Kulkarni} G.,  {Haehnelt} M.~G.,
  {Chardin} J.,   {Aubert} D.,  2020a, \mn@doi [\mnras]
  {10.1093/mnras/stz3083}, \href
  {https://ui.adsabs.harvard.edu/abs/2020MNRAS.491.1736K} {491, 1736}

\bibitem[\protect\citeauthoryear{{Keating}, {Kulkarni}, {Haehnelt}, {Chardin}
  \& {Aubert}}{{Keating} et~al.}{2020b}]{Keating2020}
{Keating} L.~C.,  {Kulkarni} G.,  {Haehnelt} M.~G.,  {Chardin} J.,   {Aubert}
  D.,  2020b, \mn@doi [\mnras] {10.1093/mnras/staa1909}, \href
  {https://ui.adsabs.harvard.edu/abs/2020MNRAS.497..906K} {497, 906}

\bibitem[\protect\citeauthoryear{{Kimm} \& {Cen}}{{Kimm} \&
  {Cen}}{2014}]{Kimm2014}
{Kimm} T.,  {Cen} R.,  2014, \mn@doi [\apj] {10.1088/0004-637X/788/2/121},
  \href {https://ui.adsabs.harvard.edu/abs/2014ApJ...788..121K} {788, 121}

\bibitem[\protect\citeauthoryear{{Kostyuk}, {Nelson}, {Ciardi}, {Glatzle}  \&
  {Pillepich}}{{Kostyuk} et~al.}{2023}]{Kostyuk2023}
{Kostyuk} I.,  {Nelson} D.,  {Ciardi} B.,  {Glatzle} M.,   {Pillepich} A.,
  2023, \mn@doi [\mnras] {10.1093/mnras/stad677}, \href
  {https://ui.adsabs.harvard.edu/abs/2023MNRAS.521.3077K} {521, 3077}

\bibitem[\protect\citeauthoryear{{Kulkarni}, {Keating}, {Haehnelt}, {Bosman},
  {Puchwein}, {Chardin}  \& {Aubert}}{{Kulkarni} et~al.}{2019}]{Kulkarni2019}
{Kulkarni} G.,  {Keating} L.~C.,  {Haehnelt} M.~G.,  {Bosman} S. E.~I.,
  {Puchwein} E.,  {Chardin} J.,   {Aubert} D.,  2019, \mn@doi [\mnras]
  {10.1093/mnrasl/slz025}, \href
  {https://ui.adsabs.harvard.edu/abs/2019MNRAS.485L..24K} {485, L24}

\bibitem[\protect\citeauthoryear{{Lewis}, {Ocvirk}, {Dubois}, {Aubert},
  {Chardin}, {Gillet}  \& {Th{\'e}lie}}{{Lewis} et~al.}{2023}]{Lewis2023}
{Lewis} J. S.~W.,  {Ocvirk} P.,  {Dubois} Y.,  {Aubert} D.,  {Chardin} J.,
  {Gillet} N.,   {Th{\'e}lie} {\'E}.,  2023, \mn@doi [\mnras]
  {10.1093/mnras/stad081}, \href
  {https://ui.adsabs.harvard.edu/abs/2023MNRAS.519.5987L} {519, 5987}

\bibitem[\protect\citeauthoryear{{Madau}}{{Madau}}{1995}]{Madau1995}
{Madau} P.,  1995, \mn@doi [\apj] {10.1086/175332}, \href
  {https://ui.adsabs.harvard.edu/abs/1995ApJ...441...18M} {441, 18}

\bibitem[\protect\citeauthoryear{{Maseda} et~al.,}{{Maseda}
  et~al.}{2020}]{Maseda2020}
{Maseda} M.~V.,  et~al., 2020, \mn@doi [\mnras] {10.1093/mnras/staa622}, \href
  {https://ui.adsabs.harvard.edu/abs/2020MNRAS.493.5120M} {493, 5120}

\bibitem[\protect\citeauthoryear{{Mason} et~al.,}{{Mason}
  et~al.}{2018}]{Mason2018}
{Mason} C.~A.,  et~al., 2018, \mn@doi [\apjl] {10.3847/2041-8213/aabbab}, \href
  {https://ui.adsabs.harvard.edu/abs/2018ApJ...857L..11M} {857, L11}

\bibitem[\protect\citeauthoryear{Mason et~al.,}{Mason et~al.}{2019}]{Mason2019}
Mason C.~A.,  et~al., 2019, \mn@doi [Monthly Notices of the Royal Astronomical
  Society] {10.1093/mnras/stz632}, 485, 3947

\bibitem[\protect\citeauthoryear{{Matthee} et~al.,}{{Matthee}
  et~al.}{2022}]{Matthee2022}
{Matthee} J.,  et~al., 2022, \mn@doi [\mnras] {10.1093/mnras/stac801}, \href
  {https://ui.adsabs.harvard.edu/abs/2022MNRAS.tmp..794M} {}

\bibitem[\protect\citeauthoryear{{Matthee} et~al.,}{{Matthee}
  et~al.}{2023}]{Matthee2023}
{Matthee} J.,  et~al., 2023, \mn@doi [arXiv e-prints]
  {10.48550/arXiv.2306.05448}, \href
  {https://ui.adsabs.harvard.edu/abs/2023arXiv230605448M} {p. arXiv:2306.05448}

\bibitem[\protect\citeauthoryear{{McGreer} et~al.,}{{McGreer}
  et~al.}{2013}]{McGreer2013}
{McGreer} I.~D.,  et~al., 2013, \mn@doi [\apj] {10.1088/0004-637X/768/2/105},
  \href {https://ui.adsabs.harvard.edu/abs/2013ApJ...768..105M} {768, 105}

\bibitem[\protect\citeauthoryear{{McGreer}, {Mesinger}  \&
  {D'Odorico}}{{McGreer} et~al.}{2015}]{McGreer2015}
{McGreer} I.~D.,  {Mesinger} A.,   {D'Odorico} V.,  2015, \mn@doi [\mnras]
  {10.1093/mnras/stu2449}, \href
  {https://ui.adsabs.harvard.edu/abs/2015MNRAS.447..499M} {447, 499}

\bibitem[\protect\citeauthoryear{{McQuinn} \& {Upton Sanderbeck}}{{McQuinn} \&
  {Upton Sanderbeck}}{2016}]{McQuinn2016}
{McQuinn} M.,  {Upton Sanderbeck} P.~R.,  2016, \mn@doi [\mnras]
  {10.1093/mnras/stv2675}, \href
  {https://ui.adsabs.harvard.edu/abs/2016MNRAS.456...47M} {456, 47}

\bibitem[\protect\citeauthoryear{McQuinn, Oh  \& Faucher-Gigu{\`{e}}re}{McQuinn
  et~al.}{2011}]{McQuinn2011}
McQuinn M.,  Oh S.~P.,   Faucher-Gigu{\`{e}}re C.-A.,  2011, \mn@doi [The
  Astrophysical Journal] {10.1088/0004-637x/743/1/82}, 743, 82

\bibitem[\protect\citeauthoryear{{Mu{\~n}oz}, {Oh}, {Davies}  \&
  {Furlanetto}}{{Mu{\~n}oz} et~al.}{2016}]{Munoz2016}
{Mu{\~n}oz} J.~A.,  {Oh} S.~P.,  {Davies} F.~B.,   {Furlanetto} S.~R.,  2016,
  \mn@doi [\mnras] {10.1093/mnras/stv2355}, \href
  {https://ui.adsabs.harvard.edu/abs/2016MNRAS.455.1385M} {455, 1385}

\bibitem[\protect\citeauthoryear{{Naidu}, {Tacchella}, {Mason}, {Bose}, {Oesch}
   \& {Conroy}}{{Naidu} et~al.}{2020}]{Naidu2020}
{Naidu} R.~P.,  {Tacchella} S.,  {Mason} C.~A.,  {Bose} S.,  {Oesch} P.~A.,
  {Conroy} C.,  2020, \mn@doi [\apj] {10.3847/1538-4357/ab7cc9}, \href
  {https://ui.adsabs.harvard.edu/abs/2020ApJ...892..109N} {892, 109}

\bibitem[\protect\citeauthoryear{{Naidu} et~al.,}{{Naidu}
  et~al.}{2022}]{Naidu2022}
{Naidu} R.~P.,  et~al., 2022, \mn@doi [\mnras] {10.1093/mnras/stab3601}, \href
  {https://ui.adsabs.harvard.edu/abs/2022MNRAS.510.4582N} {510, 4582}

\bibitem[\protect\citeauthoryear{Nasir \& D’Aloisio}{Nasir \&
  D’Aloisio}{2020}]{Nasir2020}
Nasir F.,  D’Aloisio A.,  2020, \mn@doi [Monthly Notices of the Royal
  Astronomical Society] {10.1093/mnras/staa894}, 494, 3080–3094

\bibitem[\protect\citeauthoryear{{Nasir}, {Cain}, {D'Aloisio}, {Gangolli}  \&
  {McQuinn}}{{Nasir} et~al.}{2021}]{Nasir2021}
{Nasir} F.,  {Cain} C.,  {D'Aloisio} A.,  {Gangolli} N.,   {McQuinn} M.,  2021,
  \mn@doi [\apj] {10.3847/1538-4357/ac2eb9}, \href
  {https://ui.adsabs.harvard.edu/abs/2021ApJ...923..161N} {923, 161}

\bibitem[\protect\citeauthoryear{{Ocvirk} et~al.,}{{Ocvirk}
  et~al.}{2016}]{Ocvirk2016}
{Ocvirk} P.,  et~al., 2016, \mn@doi [\mnras] {10.1093/mnras/stw2036}, \href
  {https://ui.adsabs.harvard.edu/abs/2016MNRAS.463.1462O} {463, 1462}

\bibitem[\protect\citeauthoryear{{Ocvirk}, {Aubert}, {Chardin}, {Deparis}  \&
  {Lewis}}{{Ocvirk} et~al.}{2019}]{Ocvirk2019}
{Ocvirk} P.,  {Aubert} D.,  {Chardin} J.,  {Deparis} N.,   {Lewis} J.,  2019,
  \mn@doi [\aap] {10.1051/0004-6361/201832923}, \href
  {https://ui.adsabs.harvard.edu/abs/2019A&A...626A..77O} {626, A77}

\bibitem[\protect\citeauthoryear{{Ocvirk} et~al.,}{{Ocvirk}
  et~al.}{2020}]{Ocvirk2018}
{Ocvirk} P.,  et~al., 2020, \mn@doi [\mnras] {10.1093/mnras/staa1266}, \href
  {https://ui.adsabs.harvard.edu/abs/2020MNRAS.496.4087O} {496, 4087}

\bibitem[\protect\citeauthoryear{{Ocvirk}, {Lewis}, {Gillet}, {Chardin},
  {Aubert}, {Deparis}  \& {Th{\'e}lie}}{{Ocvirk} et~al.}{2021}]{Ocvirk2021}
{Ocvirk} P.,  {Lewis} J. S.~W.,  {Gillet} N.,  {Chardin} J.,  {Aubert} D.,
  {Deparis} N.,   {Th{\'e}lie} {\'E}.,  2021, \mn@doi [\mnras]
  {10.1093/mnras/stab2502}, \href
  {https://ui.adsabs.harvard.edu/abs/2021MNRAS.507.6108O} {507, 6108}

\bibitem[\protect\citeauthoryear{Okamoto, Gao  \& Theuns}{Okamoto
  et~al.}{2008}]{Okamoto2008}
Okamoto T.,  Gao L.,   Theuns T.,  2008, \mn@doi [Monthly Notices of the Royal
  Astronomical Society] {10.1111/j.1365-2966.2008.13830.x}, 390, 920

\bibitem[\protect\citeauthoryear{{Ouchi} et~al.,}{{Ouchi}
  et~al.}{2018}]{Ouchi2018}
{Ouchi} M.,  et~al., 2018, \mn@doi [\pasj] {10.1093/pasj/psx074}, \href
  {https://ui.adsabs.harvard.edu/abs/2018PASJ...70S..13O} {70, S13}

\bibitem[\protect\citeauthoryear{{Park}, {Shapiro}, {Choi}, {Yoshida}, {Hirano}
   \& {Ahn}}{{Park} et~al.}{2016}]{Park2016}
{Park} H.,  {Shapiro} P.~R.,  {Choi} J.-h.,  {Yoshida} N.,  {Hirano} S.,
  {Ahn} K.,  2016, \mn@doi [ApJ] {10.3847/0004-637X/831/1/86}, 831, 86

\bibitem[\protect\citeauthoryear{Pillepich et~al.,}{Pillepich
  et~al.}{2017}]{Pillepich2017}
Pillepich A.,  et~al., 2017, \mn@doi [Monthly Notices of the Royal Astronomical
  Society] {10.1093/mnras/stx2656}, 473, 4077

\bibitem[\protect\citeauthoryear{{Planck Collaboration} et~al.,}{{Planck
  Collaboration} et~al.}{2020}]{Planck2018}
{Planck Collaboration} et~al., 2020, \mn@doi [\aap]
  {10.1051/0004-6361/201833910}, \href
  {https://ui.adsabs.harvard.edu/abs/2020A&A...641A...6P} {641, A6}

\bibitem[\protect\citeauthoryear{{Prochaska}, {Worseck}  \&
  {O'Meara}}{{Prochaska} et~al.}{2009}]{Prochaska2009}
{Prochaska} J.~X.,  {Worseck} G.,   {O'Meara} J.~M.,  2009, \mn@doi [\apjl]
  {10.1088/0004-637X/705/2/L113}, \href
  {https://ui.adsabs.harvard.edu/abs/2009ApJ...705L.113P} {705, L113}

\bibitem[\protect\citeauthoryear{{Prochaska}, {O'Meara}  \&
  {Worseck}}{{Prochaska} et~al.}{2010}]{Prochaska2010}
{Prochaska} J.~X.,  {O'Meara} J.~M.,   {Worseck} G.,  2010, \mn@doi [\apj]
  {10.1088/0004-637X/718/1/392}, \href
  {https://ui.adsabs.harvard.edu/abs/2010ApJ...718..392P} {718, 392}

\bibitem[\protect\citeauthoryear{Qin, Mesinger, Bosman  \& Viel}{Qin
  et~al.}{2021}]{Qin2021}
Qin Y.,  Mesinger A.,  Bosman S. E.~I.,   Viel M.,  2021, \mn@doi [Monthly
  Notices of the Royal Astronomical Society] {10.1093/mnras/stab1833}, 506,
  2390

\bibitem[\protect\citeauthoryear{{Robertson} et~al.,}{{Robertson}
  et~al.}{2013}]{Robertson2013}
{Robertson} B.~E.,  et~al., 2013, \mn@doi [\apj] {10.1088/0004-637X/768/1/71},
  \href {https://ui.adsabs.harvard.edu/abs/2013ApJ...768...71R} {768, 71}

\bibitem[\protect\citeauthoryear{{Robertson}, {Ellis}, {Furlanetto}  \&
  {Dunlop}}{{Robertson} et~al.}{2015}]{Robertson2015}
{Robertson} B.~E.,  {Ellis} R.~S.,  {Furlanetto} S.~R.,   {Dunlop} J.~S.,
  2015, \mn@doi [\apjl] {10.1088/2041-8205/802/2/L19}, 802, L19

\bibitem[\protect\citeauthoryear{{Rosdahl}, {Blaizot}, {Aubert}, {Stranex}  \&
  {Teyssier}}{{Rosdahl} et~al.}{2013}]{Rosdahl2013}
{Rosdahl} J.,  {Blaizot} J.,  {Aubert} D.,  {Stranex} T.,   {Teyssier} R.,
  2013, \mn@doi [\mnras] {10.1093/mnras/stt1722}, \href
  {https://ui.adsabs.harvard.edu/abs/2013MNRAS.436.2188R} {436, 2188}

\bibitem[\protect\citeauthoryear{{Rosdahl} et~al.,}{{Rosdahl}
  et~al.}{2018}]{Rosdahl2018}
{Rosdahl} J.,  et~al., 2018, \mn@doi [\mnras] {10.1093/mnras/sty1655}, \href
  {https://ui.adsabs.harvard.edu/abs/2018MNRAS.479..994R} {479, 994}

\bibitem[\protect\citeauthoryear{{Rosdahl} et~al.,}{{Rosdahl}
  et~al.}{2022}]{Rosdahl2022}
{Rosdahl} J.,  et~al., 2022, \mn@doi [\mnras] {10.1093/mnras/stac1942}, \href
  {https://ui.adsabs.harvard.edu/abs/2022MNRAS.515.2386R} {515, 2386}

\bibitem[\protect\citeauthoryear{{Roth}, {D'Aloisio}, {Cain}, {Wilson}, {Zhu}
  \& {Becker}}{{Roth} et~al.}{2023}]{Roth2023}
{Roth} J.~T.,  {D'Aloisio} A.,  {Cain} C.,  {Wilson} B.,  {Zhu} Y.,   {Becker}
  G.~D.,  2023, \mn@doi [arXiv e-prints] {10.48550/arXiv.2311.06348}, \href
  {https://ui.adsabs.harvard.edu/abs/2023arXiv231106348R} {p. arXiv:2311.06348}

\bibitem[\protect\citeauthoryear{{Saldana-Lopez} et~al.,}{{Saldana-Lopez}
  et~al.}{2023}]{Saldana-Lopez2023}
{Saldana-Lopez} A.,  et~al., 2023, \mn@doi [\mnras] {10.1093/mnras/stad1283},
  \href {https://ui.adsabs.harvard.edu/abs/2023MNRAS.522.6295S} {522, 6295}

\bibitem[\protect\citeauthoryear{{Satyavolu}, {Kulkarni}, {Keating}  \&
  {Haehnelt}}{{Satyavolu} et~al.}{2023}]{Satyavolu2023}
{Satyavolu} S.,  {Kulkarni} G.,  {Keating} L.~C.,   {Haehnelt} M.~G.,  2023,
  \mn@doi [arXiv e-prints] {10.48550/arXiv.2311.06344}, \href
  {https://ui.adsabs.harvard.edu/abs/2023arXiv231106344S} {p. arXiv:2311.06344}

\bibitem[\protect\citeauthoryear{{Shapiro}, {Giroux}  \& {Babul}}{{Shapiro}
  et~al.}{1994}]{Shapiro1994}
{Shapiro} P.~R.,  {Giroux} M.~L.,   {Babul} A.,  1994, \mn@doi [\apj]
  {10.1086/174120}, \href
  {https://ui.adsabs.harvard.edu/abs/1994ApJ...427...25S} {427, 25}

\bibitem[\protect\citeauthoryear{Shapiro, Iliev  \& Raga}{Shapiro
  et~al.}{2004}]{Shapiro2004}
Shapiro P.~R.,  Iliev I.~T.,   Raga A.~C.,  2004, \mn@doi [Monthly Notices of
  the Royal Astronomical Society] {10.1111/j.1365-2966.2004.07364.x}, 348, 753

\bibitem[\protect\citeauthoryear{{Theuns}}{{Theuns}}{2021}]{Theuns2021}
{Theuns} T.,  2021, \mn@doi [\mnras] {10.1093/mnras/staa3412}, \href
  {https://ui.adsabs.harvard.edu/abs/2021MNRAS.500.2741T} {500, 2741}

\bibitem[\protect\citeauthoryear{{Theuns} \& {Chan}}{{Theuns} \&
  {Chan}}{2024}]{Theuns2023}
{Theuns} T.,  {Chan} T.~K.,  2024, \mn@doi [\mnras] {10.1093/mnras/stad3176},
  \href {https://ui.adsabs.harvard.edu/abs/2024MNRAS.527..689T} {527, 689}

\bibitem[\protect\citeauthoryear{Trac \& Cen}{Trac \& Cen}{2007}]{Trac2007}
Trac H.,  Cen R.,  2007, \mn@doi [The Astrophysical Journal] {10.1086/522566},
  671, 1

\bibitem[\protect\citeauthoryear{{Trac} \& {Pen}}{{Trac} \&
  {Pen}}{2004}]{Trac2004}
{Trac} H.,  {Pen} U.-L.,  2004, \mn@doi [\na] {10.1016/j.newast.2004.02.002},
  9, 443

\bibitem[\protect\citeauthoryear{{Trac}, {Cen}  \& {Mansfield}}{{Trac}
  et~al.}{2015}]{Trac2015}
{Trac} H.,  {Cen} R.,   {Mansfield} P.,  2015, \mn@doi [\apj]
  {10.1088/0004-637X/813/1/54}, 813, 54

\bibitem[\protect\citeauthoryear{{Trebitsch} et~al.,}{{Trebitsch}
  et~al.}{2021}]{Trebitsch2021}
{Trebitsch} M.,  et~al., 2021, \mn@doi [\aap] {10.1051/0004-6361/202037698},
  \href {https://ui.adsabs.harvard.edu/abs/2021A&A...653A.154T} {653, A154}

\bibitem[\protect\citeauthoryear{Vogelsberger et~al.,}{Vogelsberger
  et~al.}{2014}]{Vogelsberger2014}
Vogelsberger M.,  et~al., 2014, \mn@doi [Monthly Notices of the Royal
  Astronomical Society] {10.1093/mnras/stu1536}, 444, 1518

\bibitem[\protect\citeauthoryear{{Walther}, {O{\~n}orbe}, {Hennawi}  \&
  {Luki{\'c}}}{{Walther} et~al.}{2019}]{Walther2019}
{Walther} M.,  {O{\~n}orbe} J.,  {Hennawi} J.~F.,   {Luki{\'c}} Z.,  2019,
  \mn@doi [\apj] {10.3847/1538-4357/aafad1}, \href
  {https://ui.adsabs.harvard.edu/abs/2019ApJ...872...13W} {872, 13}

\bibitem[\protect\citeauthoryear{{Wang} et~al.,}{{Wang}
  et~al.}{2020}]{Wang2020}
{Wang} F.,  et~al., 2020, \mn@doi [\apj] {10.3847/1538-4357/ab8c45}, \href
  {https://ui.adsabs.harvard.edu/abs/2020ApJ...896...23W} {896, 23}

\bibitem[\protect\citeauthoryear{Weinberger et~al.,}{Weinberger
  et~al.}{2016}]{Weinberger2016}
Weinberger R.,  et~al., 2016, \mn@doi [Monthly Notices of the Royal
  Astronomical Society] {10.1093/mnras/stw2944}, 465, 3291

\bibitem[\protect\citeauthoryear{{Weldon} et~al.,}{{Weldon}
  et~al.}{2022}]{Weldon2022}
{Weldon} A.,  et~al., 2022, \mn@doi [\mnras] {10.1093/mnras/stac1822}, \href
  {https://ui.adsabs.harvard.edu/abs/2022MNRAS.515..841W} {515, 841}

\bibitem[\protect\citeauthoryear{{Willott} et~al.,}{{Willott}
  et~al.}{2010}]{Willott2010}
{Willott} C.~J.,  et~al., 2010, \mn@doi [\aj] {10.1088/0004-6256/139/3/906},
  \href {https://ui.adsabs.harvard.edu/abs/2010AJ....139..906W} {139, 906}

\bibitem[\protect\citeauthoryear{{Worseck} et~al.,}{{Worseck}
  et~al.}{2014}]{Worseck2014}
{Worseck} G.,  et~al., 2014, \mn@doi [\mnras] {10.1093/mnras/stu1827}, \href
  {https://ui.adsabs.harvard.edu/abs/2014MNRAS.445.1745W} {445, 1745}

\bibitem[\protect\citeauthoryear{{Wu}, {Kannan}, {Marinacci}, {Vogelsberger}
  \& {Hernquist}}{{Wu} et~al.}{2019a}]{Wu2019b}
{Wu} X.,  {Kannan} R.,  {Marinacci} F.,  {Vogelsberger} M.,   {Hernquist} L.,
  2019a, \mn@doi [\mnras] {10.1093/mnras/stz1726}, \href
  {https://ui.adsabs.harvard.edu/abs/2019MNRAS.488..419W} {488, 419}

\bibitem[\protect\citeauthoryear{Wu, McQuinn, Kannan, D’Aloisio, Bird,
  Marinacci, Davé  \& Hernquist}{Wu et~al.}{2019b}]{Wu2019}
Wu X.,  McQuinn M.,  Kannan R.,  D’Aloisio A.,  Bird S.,  Marinacci F.,
  Davé R.,   Hernquist L.,  2019b, \mn@doi [Monthly Notices of the Royal
  Astronomical Society] {10.1093/mnras/stz2807}, 490, 3177

\bibitem[\protect\citeauthoryear{{Wu}, {McQuinn}  \& {Eisenstein}}{{Wu}
  et~al.}{2021}]{Wu2021}
{Wu} X.,  {McQuinn} M.,   {Eisenstein} D.,  2021, \mn@doi [\jcap]
  {10.1088/1475-7516/2021/02/042}, \href
  {https://ui.adsabs.harvard.edu/abs/2021JCAP...02..042W} {2021, 042}

\bibitem[\protect\citeauthoryear{{Yang} et~al.,}{{Yang}
  et~al.}{2020a}]{Yang2020a}
{Yang} J.,  et~al., 2020a, \mn@doi [\apjl] {10.3847/2041-8213/ab9c26}, \href
  {https://ui.adsabs.harvard.edu/abs/2020ApJ...897L..14Y} {897, L14}

\bibitem[\protect\citeauthoryear{{Yang} et~al.,}{{Yang}
  et~al.}{2020b}]{Yang2020b}
{Yang} J.,  et~al., 2020b, \mn@doi [\apj] {10.3847/1538-4357/abbc1b}, \href
  {https://ui.adsabs.harvard.edu/abs/2020ApJ...904...26Y} {904, 26}

\bibitem[\protect\citeauthoryear{{Yeh} et~al.,}{{Yeh} et~al.}{2022}]{Yeh2022}
{Yeh} J. Y.~C.,  et~al., 2022, arXiv e-prints, \href
  {https://ui.adsabs.harvard.edu/abs/2022arXiv220502238Y} {p. arXiv:2205.02238}

\bibitem[\protect\citeauthoryear{{Zhu} et~al.,}{{Zhu} et~al.}{2022}]{Zhu2022}
{Zhu} Y.,  et~al., 2022, \mn@doi [\apj] {10.3847/1538-4357/ac6e60}, \href
  {https://ui.adsabs.harvard.edu/abs/2022ApJ...932...76Z} {932, 76}

\bibitem[\protect\citeauthoryear{{Zhu} et~al.,}{{Zhu} et~al.}{2023}]{Zhu2023}
{Zhu} Y.,  et~al., 2023, \mn@doi [arXiv e-prints] {10.48550/arXiv.2308.04614},
  \href {https://ui.adsabs.harvard.edu/abs/2023arXiv230804614Z} {p.
  arXiv:2308.04614}

\makeatother
\end{thebibliography}

%%%%%%%%%%%%%%%%% APPENDICES %%%%%%%%%%%%%%%%%%%%%

\appendix

\section{Multi-frequency generalization for $\Gamma_{\rm HI}$ (Eq. 7)}
\label{app:mfreq_gamma}

Consider an infinitely sharp I-front traveling along one axis of cell $i$.  Then ray $j$ intersecting cell $i$ will travel a distance $x_{\rm ion}^{i} \Delta s^{ij}$ (recall $\Delta s^{ij}$ is the total path length of ray $j$ through cell $i$) before reaching neutral gas.  The number of photons absorbed over this distance is
\begin{equation}
    \label{eq:B1}
    N_{\rm abs}^{i} = \sum_{j=1}^{N_{\rm rays}} 
 \sum_{\nu}^{}N_{0,\nu}^{ij} \left(1 - \exp\left[\frac{-x_{\rm ion}^{i} \Delta s^{ij}}{\lambda_{\nu}^{i}}\right]\right)
\end{equation}
where the outer sum runs over all rays $j$ intersecting cell $i$ and the inner sum runs over all frequency bins of ray $j$.  Here, $N_{0,\nu}^{ij}$ is the number of photons in ray $j$ entering cell $i$ at frequency $\nu$ and $\lambda_{\nu}^{i}$ is the frequency-dependent mean free path.  During a time step $\Delta t$, $\Gamma_{\rm HI}$ in ionized gas behind the I-front is
\begin{equation}
    \label{eq:B2}
    \Gamma_{\rm HI}^{i} = \frac{\text{\# of photons absorbed per time}}{\text{\# of HI atoms in ionized gas}} = \frac{N_{\rm abs}^{i}/\Delta t}{n_{\rm HI}^{\Gamma} x_{\rm ion}^{i} V_{\rm cell}}
\end{equation}
where $x_{\rm ion}^{i} V_{\rm cell}$ is the ionized volume of cell $i$ and
\begin{equation}
    \label{eq:B3}
    n_{\rm HI}^{\Gamma} \equiv \frac{\langle \Gamma_{\rm HI} n_{\rm HI}\rangle_{\rm V}}{\langle \Gamma_{\rm HI} \rangle_{\rm V}}
\end{equation}
is the $\Gamma_{\rm HI}$-weighted HI number density (the V sub-script denotes a volume average). In appendix C of~\cite{Cain2022b}, we showed that the frequency-averaged MFP in our sub-grid simulations is given by
\begin{equation}
    \label{eq:lambda_cain22}
    \langle \lambda_{\nu}^{-1} \rangle_{\nu}^{i} = \frac{\langle n_{\rm HI}\Gamma_{\rm HI} \rangle_{\rm V}}{F_{\gamma}} = \frac{1}{F_{\gamma}}\int_{\nu_{\rm HI}}^{4\nu_{\rm HI}} d\nu \frac{I_{\nu}}{h\nu}\lambda_{\nu}^{-1}
\end{equation}
where $I_{\nu}$ and $F_{\gamma}$ are the specific intensity and ionizing photon flux at the source planes in the sub-grid simulations, respectively.  Using this result, we can write
\begin{equation}
    \label{eq:B4}
    \frac{n_{\rm HI}^{\Gamma, i}}{F_{\gamma}} = \frac{1}{\langle \Gamma_{\rm HI} \rangle_{\rm V}}\int_{\nu_{\rm HI}}^{4\nu_{\rm HI}} d\nu \frac{1}{F_{\gamma}}\frac{I_{\nu}}{h\nu}\lambda_{\nu}^{-1} \approx \frac{\langle \lambda_{\nu}^{-1} \rangle_{\nu}^{i}}{\langle \Gamma_{\rm HI} \rangle_{\rm V}}
\end{equation}
where $\langle \lambda_{\nu}^{-1} \rangle_{\nu}^{i}$ is the opacity averaged over the spectrum incident on cell $i$.  
 Eq.~\ref{eq:B4} ignores IGM filtering over distances smaller than the cell size, but this approximation holds as long as $\lambda_{912}^{\rm mfp} >> \Delta x_{\rm cell}$. 
Combining Eq.~\ref{eq:B1}-\ref{eq:B4} yields
\begin{equation}
    \label{eq:B5}
    \Gamma_{\rm HI}^{i} = \frac{\sum_{j=1}^{N_{\rm rays}} 
 \sum_{\nu}^{}N_{0,\nu}^{ij} \left(1 - \exp\left[\frac{-x_{\rm ion}^{i} \Delta s^{ij}}{\lambda_{\nu}^{i}}\right]\right)}{ (\langle\lambda_{\nu}^{-1} \rangle_{\nu}F_{\gamma}/\langle \Gamma_{\rm HI} \rangle_{\rm V}) x_{\rm ion}^{i} V_{\rm cell} \Delta t}
\end{equation}
The ratio $F_{\gamma}/\langle \Gamma_{\rm HI} \rangle_{\rm V}$ can be simplified as long as $F_{\gamma}\langle \sigma_{\rm HI}\rangle_{\nu}^{i} \approx \langle \Gamma_{\rm HI} \rangle_{\rm V}$, where $\langle \sigma_{\rm HI}\rangle_{\nu}^{i}$ is the HI cross-section averaged over the spectrum incident on cell $i$.  Under this approximation, 
\begin{equation}
    \label{eq:B6}
    \Gamma_{\rm HI}^{i} = \frac{\langle \sigma_{\rm HI}\rangle_{\nu}^{i}\sum_{j=1}^{N_{\rm rays}} 
 \sum_{\nu}^{}N_{0,\nu}^{ij} \left(1 - \exp\left[\frac{-x_{\rm ion}^{i} \Delta s^{ij}}{\lambda_{\nu}^{i}}\right]\right)}{\langle \lambda_{\nu}^{-1} \rangle_{\nu}^{i}  x_{\rm ion}^{i} V_{\rm cell}\Delta t}
\end{equation}
If we define $\langle \lambda_{\nu} \rangle_{\nu} \equiv \langle \lambda_{\nu}^{-1} \rangle_{\nu}^{-1}$, then Eq.~\ref{eq:B6} can be written in a form similar to Eq. 1 of~\cite{Cain2021},
\begin{equation}
    \label{eq:B6}
    \Gamma_{\rm HI}^{i} = \frac{\langle \lambda_{\nu} \rangle_{\nu}^{i}\langle\sigma_{\rm HI}\rangle_{\nu}^{i}\sum_{j=1}^{N_{\rm rays}} 
 \sum_{\nu}^{}N_{0,\nu}^{ij}  \left(1 - \exp\left[\frac{-x_{\rm ion}^{i} \Delta s^{ij}}{\lambda_{\nu}^{i}}\right]\right)}{   x_{\rm ion}^{i} V_{\rm cell}\Delta t}
\end{equation}

\section{Testing the multi-frequency model \& effect IGM Filtering}
\label{app:kappa_freq}

In this appendix, we validate our multi-frequency RT treatment described in \S\ref{subsec:multifreq}.  We tested this procedure using a set of small-volume hydro/RT simulations similar to the ones used to calibrate our sub-grid model.  We have run three simulations in $0.512~h^{-1}$Mpc boxes with $N = 512^3$ gas/RT cells, each with $\Gamma_{\rm HI} = 3 \times 10^{-13}$, $z_{\rm reion} = 8$, and with the box-scale mean density equal to the cosmic mean.  Our simulations have different power law spectra with indices $\alpha = 1.5$ (as used in our sub-grid simulations), $0.5$, and $-0.5$, each with $5$ frequency bins spanning $1-4$ Ryd.  

\begin{figure}
    \centering
    \includegraphics[scale=0.27]{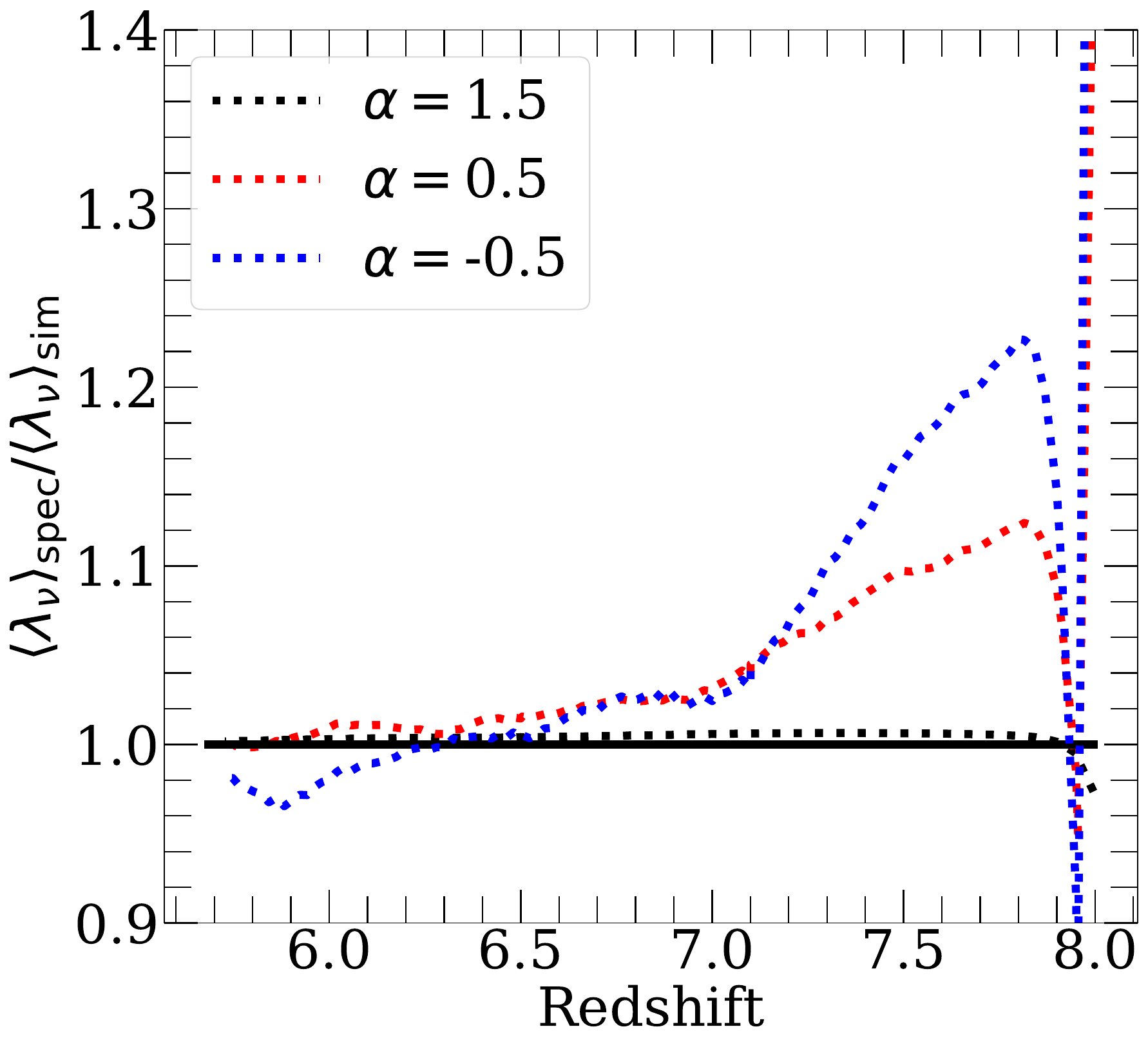}
    \caption{Direct test of our multi-frequency procedure described in \S\ref{subsec:multifreq}.  We have run small-scale simulations with $\alpha = 1.5$, $0.5$, and $-0.5$, and used the procedure in that section to compute $\langle \lambda_{\nu} \rangle_{\rm spec}$, which is the frequency-averaged MFP estimated using only information from the $\alpha = 1.5$ simulation.  Here we show the ratio between $\langle \lambda_{\nu} \rangle_{\rm spec}$ and the true value extracted from each simulation $\langle \lambda \rangle_{\rm sim}$.  The agreement is within $1\%$ for $\alpha = 1.5$ and always within $20\%$ for the others.  }
    \label{fig:mfp_alpha_test}
\end{figure}

\begin{figure*}
    \centering
    \includegraphics[scale=0.28]{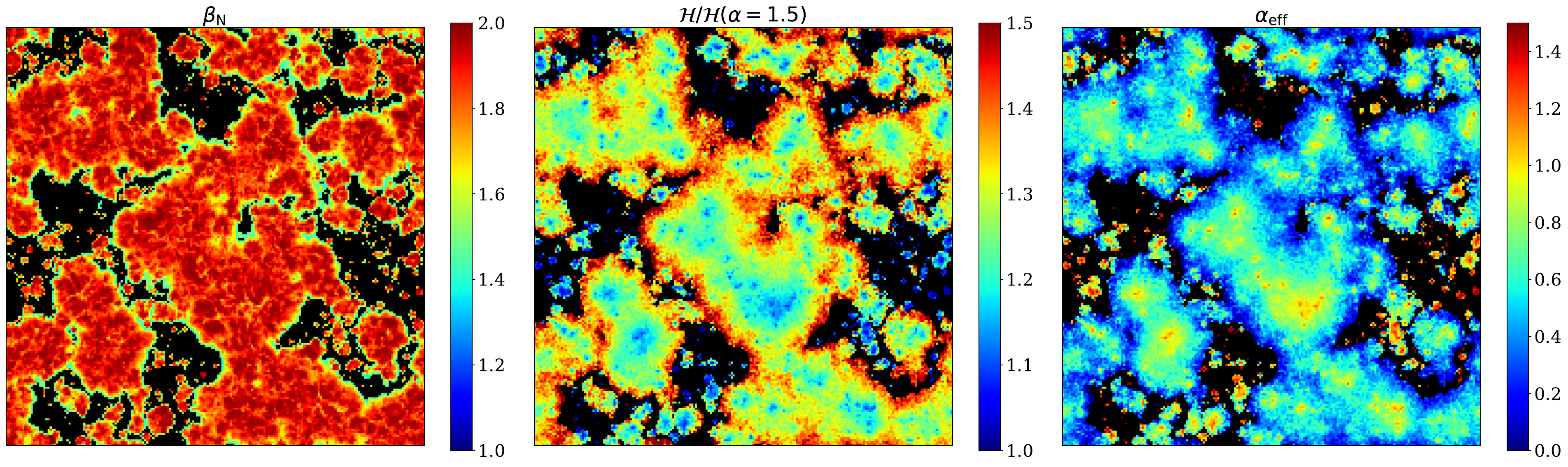}
    \caption{Visualization of the effect of IGM filtering at $z = 6$ in our multi-frequency treatment described in \S\ref{subsec:multifreq}.  This simulation assumes $\alpha = 1.5$ for sources and uses our Full Sinks sub-grid model.  Left: $\beta_{\rm N}$ estimated using Eq.~\ref{eq:betaN}.  We find $\beta_{\rm N} \approx 1.6-1.9$ in most ionized gas, with smaller values ($\approx 1.4-1.5$) close to I-fronts.  Middle: ratio of the IGM heating rate with the value expected for an $\alpha = 1.5$ spectrum.  IGM heating rates are enhanced by as much as a factor of $1.5$ close to ionization fronts, where the effect of IGM filtering is greatest.  Right: spectral index ($\alpha_{\rm eff}$) that would reproduce the heating rates shown in the middle panel.  We see that $\alpha_{\rm eff}$ can be as small as $0$ close to ionization fronts.  }
    \label{fig:betaN}
\end{figure*}

For each $\alpha$, we use the result of the $\alpha = 1.5$ simulation to estimate the frequency-averaged MFP, which we call $\langle \lambda_{\nu} \rangle_{\rm spec}$, following the procedure described in \S\ref{subsec:multifreq}.  Next, we calculate the actual average MFP in each simulation, $\langle \lambda_{\nu} \rangle_{\rm sim}$.  Figure~\ref{fig:mfp_alpha_test} shows the ratio of these quantities for each value of $\alpha$.  As expected, we find very close agreement for $\alpha = 1.5$.  For $\alpha = 0.5$ ($-0.5$), the re-constructed $\langle \lambda_{\nu} \rangle_{\rm spec}$ differs from the truth by at most $10\%$ ($20\%$), with the agreement improving to better than $5\%$ $\Delta z = 1$ after the gas is ionized.  This disagreement likely owes to differences in the self-shielding properties of the gas in simulations with different $\alpha$, which are are unable to account for with our current approach.   Given that $\langle \lambda_{\nu} \rangle_{\rm sim}$ for the $\alpha = 1.5$ and $-0.5$ simulations are different by almost a factor of $3$, this level of agreement is acceptable for our purposes.  

Next, we illustrate the importance of including multi-frequency RT in our FlexRT simulations for the Ly$\alpha$ forest.  To show this, we have run an illustrative simulation with the same box size and fiducial source prescription as the simulations presented in this paper.  The simulation includes multi-frequency RT with $\alpha = 1.5$ and uses our Full Sinks sub-grid model for the IGM.  In the left panel of In Figure~\ref{fig:betaN}, we show a slice through the column density distribution slope $\beta_{\rm N}$ at $z = 6$, computed using Eq.~\ref{eq:betaN}.  We find $\beta_{\rm N} \approx 1.6-1.9$ in highly ionized gas, with the lowest values in the most over-dense cells, consistent with the findings of~\cite{McQuinn2011}~\footnote{They found $\beta_{\rm N} \approx 1.8$ for optically thin systems.  }.  We find lower values ($\beta_{\rm N} \approx 1.4-1.5$) in the most recently ionized gas close to I-fronts.  This is consistent with the results of~\citet{Nasir2021} - they found that recently ionized gas has a shallower column density slope owing to abundant tiny sinks that have not yet photo-evaporated.  These values are somewhat higher than those typically assumed for the IGM at $z \sim 6$ (e.g.~\citet{Becker2021} assumes $\beta_{\rm N} = 1.3$).  Larger values of $\beta_{\rm N}$ result in a stronger frequency dependence of the ionizing opacity (Eq.~\ref{eq:kappanu}) and more IGM filtering.  

The middle panel of Figure~\ref{fig:betaN} shows the ratio of the photo-heating rate $\mathcal{H}$ in our simulation with that expected for an $\alpha = 1.5$ spectrum (Eq.~\ref{eq:photoheating}).  We see that near the centers of ionized bubbles where the sources are clustered, this ratio is close to $1$.  This is because the radiation spectrum in those regions is on average close to that emitted by the sources, since most of the radiation has not yet been filtered by the IGM.  Near the edges of the ionized bubbles, where the radiation has traveled several mean free paths on average, this ratio gets as large as $1.5$, demonstrating that it has been hardened significantly by the IGM.  The right panel converts this enhanced heating rate into an effective spectral index $\alpha_{\rm eff}$, defined such that $\mathcal{H}(\alpha_{\rm eff}) \equiv \mathcal{H}$ for each cell.  We see that $\alpha_{\rm eff}$ declines from $\approx 1.5$ at the center of ionized bubbles to $\approx 0$ near their edges.  These illustrations highlight the importance of IGM filtering effects for both the averages and spatial fluctuations of the heating rate and photo-ionization rate in the IGM, which are crucial for accurately modeling the Ly$\alpha$ forest.  

\section{Derivation \& Tests of the Recombination Radiation Model (Eq. 13-15)}
\label{app:recomb_tests}

Here, we will derive and test our model for recombination radiation, Eq.~\ref{eq:alphaT}-\ref{eq:ngam_rec}.  We start with Eq.~\ref{eq:alphaT}, which describes the effective recombination coefficient if a fraction $f_{\rm esc}^{\rm rec}$ of ionizing recombination photons escape the dense clumps where they are produced.  Assuming photo-ionization equilibrium, 
\begin{equation}
    \label{eq:photo_eq}
    \langle \Gamma_{\rm HI} n_{\rm HI} \rangle_{\rm V} = \langle \alpha(T) n_e n_{\rm HII} \rangle_{\rm V}
\end{equation}
where the average is over volume.  In Appendix B of~\cite{Cain2022b}, we showed that the left hand size of this equation can be expressed in terms of the frequency-averaged MFP $\langle{\lambda_{\nu}^{-1}}\rangle_{\nu}^{-1}$, 
\begin{equation}
    \label{eq:lambdagamma}
    \langle \Gamma_{\rm HI} n_{\rm HI} \rangle_{\rm V} = \frac{\langle \Gamma_{\rm HI} \rangle_{\rm V}} {\langle{\lambda_{\nu}^{-1}}\rangle_{\nu}^{-1} \langle\sigma_{\rm HI}\rangle_{\nu}}
\end{equation}
Combining these equations yields 
\begin{equation}
    \label{eq:lambda_rec}
    \langle{\lambda_{\nu}^{-1}}\rangle_{\nu}^{-1} = \frac{\Gamma_{\rm HI}}{\langle\sigma_{\rm HI}\rangle_{\nu} \langle \alpha(T) n_e n_{\rm HII} \rangle_{\rm V}}
\end{equation}

Let the MFP extracted from our sub-grid simulations, which assume the case B recombination approximation, be denoted $\langle{\lambda_{\nu,B}^{-1}}\rangle_{\nu}^{-1} \equiv \langle \kappa_{\rm B} \rangle_{\nu}$.  Since Eq.~\ref{eq:lambda_rec} also holds in this case, we have
\begin{equation}
    \label{eq:lambda_B_rec}
    \langle{\lambda_{\nu}^{-1}}\rangle_{\nu}^{-1} \equiv \langle \kappa \rangle_{\nu} = \langle \kappa_{\rm B} \rangle_{\nu}\frac{\langle \alpha(T) n_e n_{\rm HII} \rangle_{\rm V}}{\langle \alpha_{\rm B}(T) n_e n_{\rm HII} \rangle_{\rm V}} \approx \langle \kappa_{\rm B} \rangle_{\nu} \frac{\alpha(T)}{\alpha_{\rm B}(T)}
\end{equation}
which is Eq.~\ref{eq:kappa_rec}.  In the final step, we have assumed that the ratio of the recombination rates in the sub-grid simulations can be approximated by the ratio of the recombination coefficients evaluated at the temperature computed in FlexRT.  

Lastly, we derive Eq.~\ref{eq:ngam_rec}.  The production rate of ionizing recombination photons that are not immediately re-absorbed is given by
\begin{equation}
    \dot{n}_{\gamma}^{\rm rec} \approx x_{\rm ion}(\langle \alpha(T) n_e n_{\rm HII} \rangle_{\rm V} - \langle \alpha_{\rm B}(T) n_e n_{\rm HII} \rangle_{\rm V})(1 + \chi)
\end{equation}
where the factor of $1+\chi$ approximately accounts for ground state recombinations from HeII.  Applying Eq.~\ref{eq:lambda_rec} on both terms inside parentheses yields
\begin{equation}
    \dot{n}_{\gamma}^{\rm rec} =  x_{\rm ion}\left(\frac{\Gamma_{\rm HI}}{\langle \sigma_{\rm HI}\rangle_{\nu} \langle \lambda_{\nu}^{-1}\rangle_{\nu}^{-1}} - \frac{\Gamma_{\rm HI}}{\langle \sigma_{\rm HI}\rangle_{\nu} \langle \lambda_{\nu,B}^{-1}\rangle_{\nu}^{-1}}\right)(1+\chi) 
\end{equation}
\begin{equation*}
    = x_{\rm ion}\frac{\Gamma_{\rm HI}}{\langle \sigma_{\rm HI}\rangle_{\nu}} (\langle \kappa \rangle_{\nu} - \langle \kappa_{\rm B}\rangle_{\nu})(1+\chi)
\end{equation*}

Combining with Eq.~\ref{eq:alphaT}-\ref{eq:kappa_rec} yields
\begin{equation}
\dot{n}_{\gamma}^{\rm rec} =  x_{\rm ion}\frac{\Gamma_{\rm HI}}{\langle \sigma_{\rm HI}\rangle_{\nu}} (f_{\rm esc}^{\rm rec}\langle \kappa_{\rm A}\rangle_{\nu} + (1-f_{\rm esc}^{\rm rec})\langle \kappa_{\rm B}\rangle_{\nu}- \langle \kappa_{\rm B}\rangle_{\nu})(1+\chi)    
\end{equation}
which simplifies to Eq.~\ref{eq:ngam_rec}.  

\section{Derivation of the halo absorber model}
\label{app:winds}

\begin{figure}
    \centering
    \includegraphics[scale=0.33]{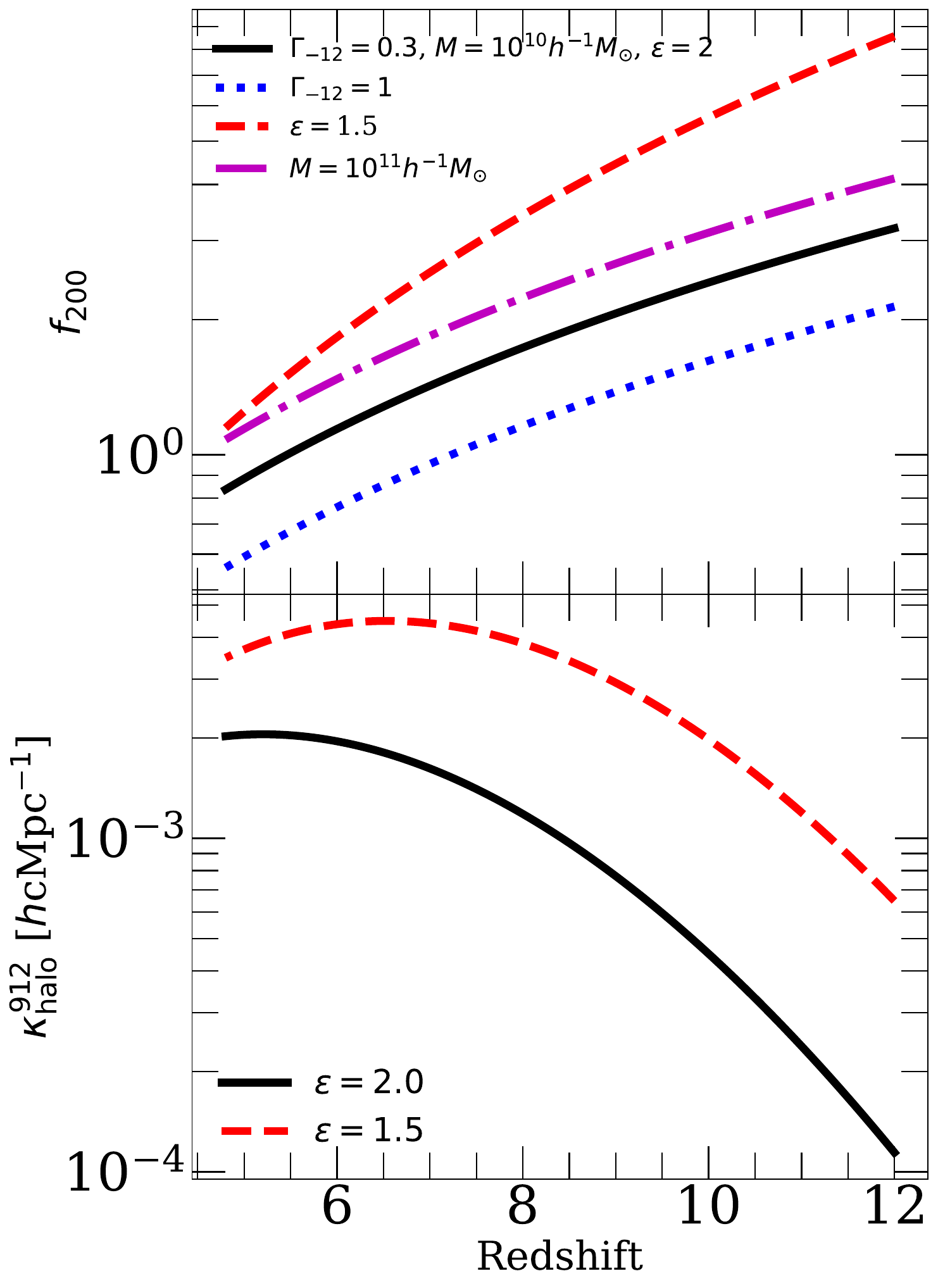}
    \caption{Features of the halo absorber model described in \S\ref{subsec:galabs}.  The black solid curve in the top panel shows $f_{200}$ for a halo with $\Gamma_{-12} = 0.3$, $M_{\rm halo } = 10^{10}$ $h^{-1}M_{\odot}$, and $\epsilon = 2$.  The other curves show the effects of varying these parameters one at a time.  The bottom panel shows the total opacity from halos, integrated over the HMF, for $\epsilon = 2$ and $1.5$ (and $\Gamma_{-12} = 0.3$).  See text for details.  }
    \label{fig:f200}
\end{figure}

In this appendix, we provide a more complete derivation of our halo absorber model described in \S\ref{subsec:galabs}, and explore some of its key features.  First, we can approximately write $M_{200}$ as a function of the gas density profile of the halo, 
\begin{equation}
    \label{eq:appC:M200}
    M_{200} \approx \frac{\Omega_m}{\Omega_b}\int_{0}^{R_{200}} dr (4 \pi r^2) \rho_{\rm gas}(r) = \frac{\Omega_m}{\Omega_b}\int_{0}^{R_{200}} dr (4 \pi r^2) \frac{n_{\rm H}(r)m_p}{X_{\rm Hy}}
\end{equation}
where we have assumed that the gas in the halo traces the dark matter with fraction $\Omega_b/\Omega_m$, and that the halo is spherically symmetric.  In the second equality, $m_p$ is the proton mass and $X_{\rm Hy}$ is the hydrogen fraction.  Combining Eq.~\ref{eq:appC:M200} with Eq.~\ref{eq:rho_gas} and the definition of $R_{200}$ and integrating yields a solution for $n_{\rm H}(r)$: 
\begin{equation}
    \label{eq:appC:nHr}
    n_{\rm H}(r) = \frac{\Omega_b}{\Omega_m}\frac{X_{\rm Hy}}{m_p}(3 - \epsilon) \frac{200 \rho_{\rm crit}(z)}{3} \left(\frac{r}{R_{200}}\right)^{-\epsilon}
\end{equation}
Putting this result into Eq.~\ref{eq:tauion} and assuming photo-ionizational equilibrium gives
\begin{equation}
    \label{eq:app_rabs}
    r_{\rm abs} = \left[\frac{\langle\sigma_{\rm HI}\rangle \alpha_{\rm B}(T_0) (1+\chi)}{\Gamma_{\rm HI}}\right]^{\frac{1}{2 \epsilon - 1}}
\end{equation}
\begin{equation*}
  \times \left[\frac{200\Omega_b X_{\rm Hy} (3 - \epsilon)\rho_{\rm crit}(z)}{3\Omega_m m_p}\right]^{\frac{2}{2 \epsilon - 1}} 
 R_{200}^{\frac{2\epsilon}{2\epsilon-1}}
\end{equation*}
where $f_{200}$ is the right hand side divided by $R_{200}$ (Eq.~\ref{eq:Rhalo}).  We assume $T_0 = 10^4$K for ionized gas in and around halos.  The scaling relations in Eq.~\ref{eq:f200} follow from Eq.~\ref{eq:app_rabs}.  

The top panel of Figure~\ref{fig:f200} shows how $f_{200}$ responds to each of the parameters in our model as a function of redshift.   The black solid curve shows a halo with $\Gamma_{-12} = 0.3$, $M_{\rm halo} = 10^{10}$ $h^{-1}M_{\odot}$, and $\epsilon = 2$ (isothermal density profile).  The blue dotted, red dashed, and magenta dot-dashed curves show the effect of changing $\Gamma_{-12}$, $\epsilon$, and $M_{\rm halo}$, respectively, one at a time (see legend).  Higher $\Gamma_{-12}$ ($M_{\rm halo}$) results in lower (higher) $f_{200}$, but does not affect the shape of its redshift evolution.  Smaller $\epsilon$ leads to both larger $f_{200}$ and much steeper redshift evolution.  

\begin{figure}
    \centering
    \includegraphics[scale=0.18]{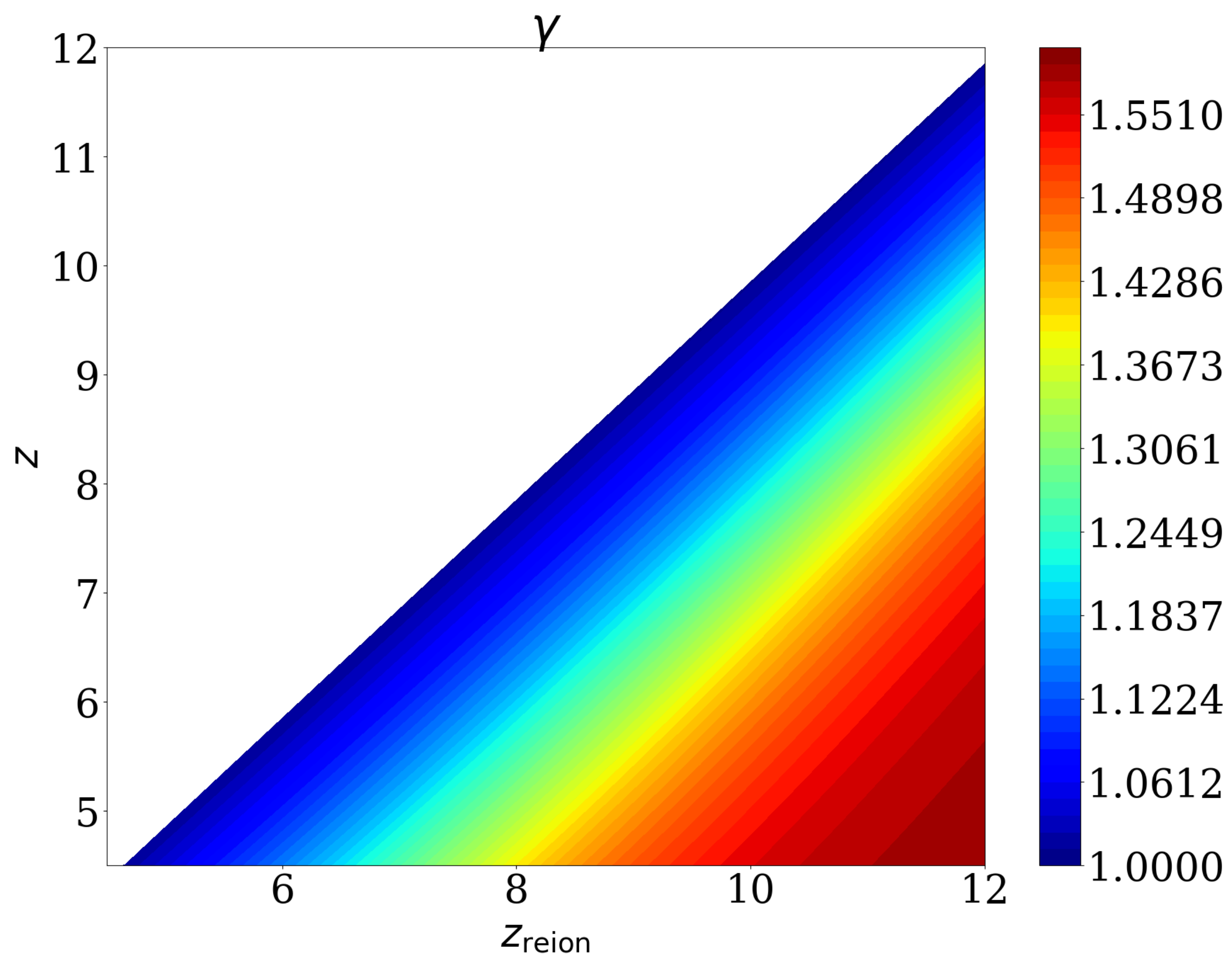}
    \caption{Power law index of the temperature-density relation, $\gamma(z,z_{\rm re})$, used to correct TDR when mapping our coarse-grained RT temperatures onto the high-resolution density grid for our Ly$\alpha$ forest calculations.  $\gamma$ approaches it's limiting value of $5/3$ when $z << z_{\rm reion}$, while when $z = z_{\rm reion}$, $\gamma = 1$ (for isothermal gas).  These values are calculated by fitting the solution of the~\citet{McQuinn2016} IGM temperature model to a power law at $\Delta \leq 1$.  }
    \label{fig:gamma_z_zre}
\end{figure}

\begin{figure*}
    \centering
    \includegraphics[scale=0.24]{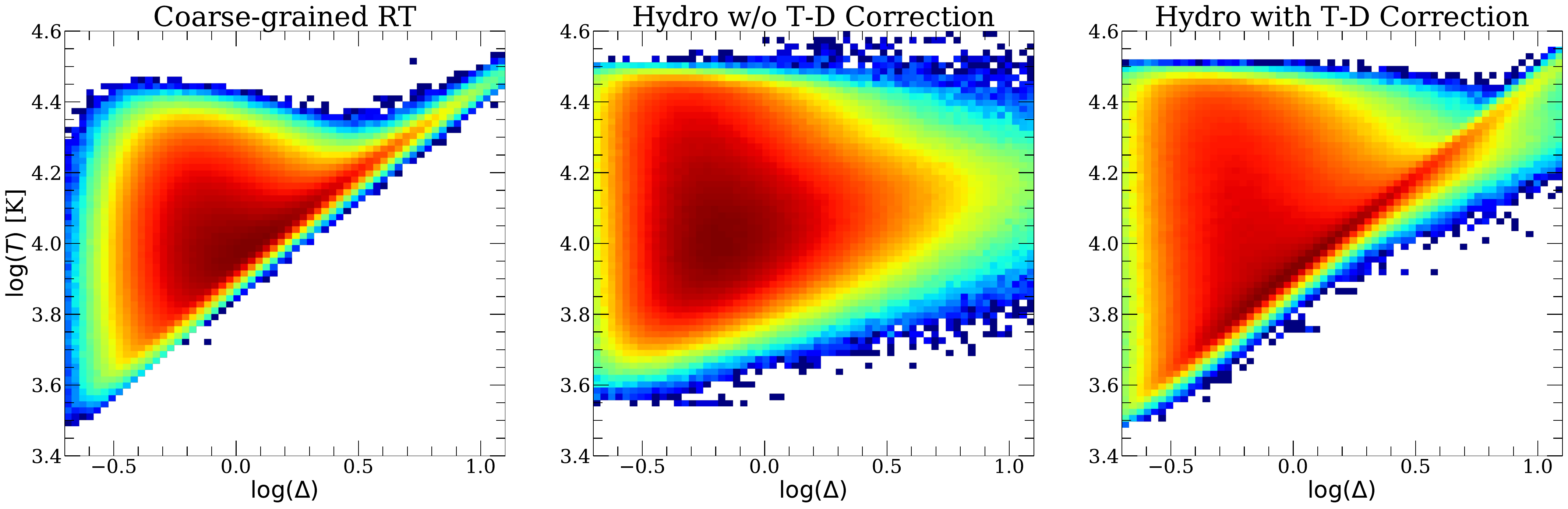}
    \caption{The TDR in our in our Ly$\alpha$ forest calculation before and after the 
    correction described in this section.  Left: the T-D relation in FlexRT at $z = 5.6$ in one of our simulations.  Middle: TDR on the high-resolution density grid using a straight mapping of the FlexRT temperatures.  Right: TDR on the high-res density grid after applying our correction.  We see that this procedure approximately restores the qualitative features seen in the TDR in FlexRT: a tight power law with a population of recently ionized, under-dense cells at much higher temperatures.  }
    \label{fig:td_plot}
\end{figure*}

The bottom panel shows the total Lyman Limit opacity from halos, integrated over the HMF, for the black and red curves in the top panel.  We see that the model with $\epsilon = 1.5$ (our Optimistic model) results in $2.5-5\times$ more opacity from halos than $\epsilon = 2$ (our Realistic model).  However, the opacity from halos begins to turn over at $z = 7$ in the former, as effect of $f_{200}$ decreasing outpaces that of the rapidly growing HMF.  This turnover is steeper in the actual simulations because $\Gamma_{\rm HI}$ increases rapidly at $z < 6$.  This decline in halo opacity at the end of reionization may be one reason why our Optimistic models allows for only a flat ionizing emissivity rather than the rapidly rising one predicted by~\citet{Robertson2015}.  In a forthcoming paper, we will explore the parameter space of this halo absorber model in more detail, and determine whether models that do allow for a rapidly rising emissivity are physically reasonable.  

\section{Correcting the IGM Temperature-Density Relation}
\label{app:TDrelation}

In this appendix, we describe in more detail our procedure for correction the IGM temperature-density relation (TDR) in our Ly$\alpha$ forest calculation, described in \S\ref{subsec:lyaforest}.  We describe the expected TDR in each FlexRT cell using the power law index $\gamma(z,z_{\rm reion})$ (see Eq.~\ref{eq:TDelta}).  We estimate $\gamma(z,z_{\rm reion})$ by fitting the solution of the~\cite{McQuinn2016} analytic temperature model to a power law at $\Delta \leq 1$ (the density range that sets high-z forest transmission).  We then assign $\gamma$ values to each cell using the FlexRT estimate of $z_{\rm reion}$, defined as the redshift at which the neutral fraction crosses $0.5$.  Figure~\ref{fig:gamma_z_zre} shows $\gamma$ vs. $z$ vs. $z_{\rm reion}$ for $5 < z, z_{\rm reion} < 12$.  Freshly ionized gas ($z = z_{\rm reion}$) is isothermal ($\gamma = 1$), while for $z << z_{\rm reion}$, $\gamma \rightarrow 5/3$.  

In Figure~\ref{fig:td_plot}, we show the effect of applying this correction to the TDR in our high-resolution simulation used to calculate forest statistics.  The left panel shows the TDR in FlexRT, which shows a clear power law with a population of lower low density cells at higher temperatures.  These hotter under-dense cells are those most recently re-ionized.  Mapping these temperatures onto the high-resolution hydro grid used for the forest almost completely erases these features, as the middle panel shows.  The right panel shows the TDR on the hydro grid after applying our procedure, which recovers the qualitative behavior seen in the left panel.  The correction also reduces the total forest transmission by $10-20\%$ at $5 < z < 6$, since the lowest-density cells with the highest transmission end up colder than they would be otherwise.

%%%%%%%%%%%%%%%%%%%%%%%%%%%%%%%%%%%%%%%%%%%%%%%%%%

% Don't change these lines
\bsp	% typesetting comment
\label{lastpage}
\end{document}